\documentclass[11pt]{article}
\usepackage[english]{babel}
\usepackage[latin1]{inputenc}
\usepackage{amsmath,latexsym,amssymb}
\usepackage{amsfonts}
\usepackage{amsthm}
\usepackage{slashed}
\usepackage{color}


\newtheorem{thm}{Theorem}[section]

\newtheorem{defi}[thm]{Definition}

\newtheorem{lem}[thm]{Lemma}
\newtheorem{prop}[thm]{Proposition}
\newtheorem{rem}[thm]{Remark}

\begin{document}

\title{Supersymmetric field equations from momentum space}
\author{Daniel Bennequin\footnotemark[2] \hspace{2mm} and \hspace{1mm} Michel Egeileh\footnotemark[3]}
\date{\footnotemark[2] Institut de Mathématiques de Jussieu, Université Paris Diderot \\
Boîte Courrier 7012, 2, place Jussieu, 75251 Paris
  cedex 05, France  \\  E-mail: {\it bennequin@math.jussieu.fr}\\ \vspace{3mm}
\footnotemark[3] Department of Mathematics, American University of Beirut\\
P.O. Box 11-0236/Mathematics, Riad El-Solh/Beirut 1107 2020, Lebanon  \\  E-mail: {\it me58@aub.edu.lb}}

\maketitle

\begin{abstract}
It is known that every irreducible unitary representation of positive energy of the Poincar\'e group can be realized as a subspace of tensor fields on Minkowski spacetime subjected to suitable partial differential equations. We first describe geometrically the general mechanism that produces, via Fourier transform, the invariant differential operators corresponding to those representations. Then, using a super-version of the Fourier transform, we show explicitly how a massive irreducible unitary representation of the super Poincar\'e group in dimension $(4|4)$ can be realized as a linear sub-supermanifold of suitably constrained superfunctions. In this way, we obtain supersymmetric equations in terms of {\it ordinary} (non-Grassmannian) fields. Finally, using the functor of points, we show how our equations can be related in a natural way to the Wess-Zumino equations for massive chiral superfields.\\
\end{abstract}

\tableofcontents

\section{Introduction}

From the point of view of relativistic quantum mechanics, $1$-particle states of an elementary particle constitute a Hilbert space which, by the requirement of relativistic invariance, must carry an irreducible unitary representation of the Poincar\'e group $G:=V\rtimes\hbox{Spin}(V)$, where $V$ is a Lorentzian vector space.\\

Denote by $\widehat{G}$ the {\it unitary dual} of $G$, that is the set of all isomorphism classes of irreducible unitary representations of $G$. The various types of elementary particles are thus labeled by elements of $\widehat{G}$; their classification is well-known and goes back to Wigner \cite{wig}. For instance, one can apply the Mackey little group method for classifying the irreducible unitary representations of semidirect products. For the Poincar\'e group in signature $(1,3)$, the irreducible representations of physical interest are classified by a nonnegative real number $m$ (the mass), and a half-integer $\sigma\in\{0,\frac{1}{2},1,\frac{3}{2},...\}$ (the spin). \\

On the other hand, if $\mathring{M}$ is {\it Minkowski spacetime} (the affine space directed by $V$), and if $W$ is a
representation of $H:=\hbox{Spin}(V)$, one can construct the {\it space of spin-tensor fields of type $W$ on $\mathring{M}$}
as follows. Viewing $\mathring{M}$ as a homogeneous space for $G$, we have a $G$-equivariant $H$-principal bundle
$G\longrightarrow \mathring{M}$; let $\mathbb{W}$ be the associated vector bundle by the action of $H$ on $W$.
This a $G$-equivariant vector bundle over $\mathring{M}$, and the space of spin-tensor fields is defined as the
space of smooth sections $\Gamma(\mathring{M},\mathbb{W})$. It carries a natural representation of the Poincar\'e group $G$.\\

We address in this paper the following question. Given an irreducible unitary
representation $\mathcal{H}_{(m,\sigma)}$ of $G$, find a space of spin-tensor
fields $\Gamma(\mathring{M},\mathbb{W})$ such that $\mathcal{H}_{(m,\sigma)}$
appears in the decomposition of (a suitable Hilbert space completion of a quotient of) $\Gamma(\mathring{M},\mathbb{W})$ under $G$. If we think of the elements of $\Gamma(\mathring{M},\mathbb{W})$ as the fields of some field theory on $\mathring{M}$, this would mean that the particle corresponding to $\mathcal{H}_{(m,\sigma)}$ belongs to the spectrum of that field theory.\\

For a standard example, consider a massive spinless particle $\mathcal{H}_{(m,0)}$ (with $m>0$). Then one can take $W=\mathbb{C}$ with the trivial representation of $H$. The corresponding space of spin-tensor fields can be identified with $\mathcal{C}^{\infty}(\mathring{M},\mathbb{C})$, and one obtains a realization of $\mathcal{H}_{(m,0)}$ by considering a subspace of solutions of the Klein-Gordon equation $(\square+m^2)\phi=0$.\\

The next standard example is that of a massive spin $\frac{1}{2}$ particle $\mathcal{H}_{(m,\frac{1}{2})}$ (with $m>0$). Then one can take $W=S_{\mathbb{C}}$, the complex four-dimensional spinor representation of $\hbox{Spin}(V)$. The corresponding space of spin-tensor fields can be identified with $\mathcal{C}^{\infty}(\mathring{M},S_{\mathbb{C}})$, and one obtains a realization of $\mathcal{H}_{(m,\frac{1}{2})}$ by considering a subspace of solutions of the Dirac equation $(\slashed D-im)\psi=0$.\\

The differential operators in these equations are of course $G$-equivariant. They can be obtained by constructing first
an explicit realization of the irreducible representation $\mathcal{H}_{(m,\sigma)}$ in momentum space, and then taking
Fourier transforms. While this procedure is well-known in the above two examples and in few others, the general case
(with arbitrary spin) has not been described explicitly in the mathematics literature, up to our knowledge.
It has been discussed in the physics literature (see for instance \cite{bk}), but the underlying geometric picture has not
been made apparent.\\

The first goal of the paper is to fill this gap. The mechanism to generate
equivariant differential operators turns out to have a simple description in
general terms. Given $m$, let $\mathcal{O}_m\subset V^*$ be the orbit for $H$
given by $\|p\|^2=m^2$ and lying in the forward timelike cone. (When $m=0$, we have to distinguish $p=0$ from the two branches of the open cone.) Choose a preferred point in $\mathcal{O}_m$, and let $K$ be the stabilizer of that point. The representation $W$ of $H$ that
defines the spin-tensor fields can be restricted to $K$, which allows the construction of
an $H$-equivariant vector bundle over $\mathcal{O}_m$. We will show in section \ref{spin-tensor fields}
that this bundle has a natural equivariant trivialization, which can be used to associate to every
$K$-equivariant linear map from $W$ to another $H$-module $E$, an $H$-equivariant symbol
$\zeta:\mathcal{O}_m\longrightarrow W^*\otimes E$. These symbols, in turn, give rise to
$G$-equivariant differential operators on spacetime.\\

From Mackey's little group method, we know that if $F_{\sigma}$ is the space of an irreducible
unitary representation of $K$, then $\mathcal{H}_{(m,\sigma)}:=\hbox{ind}_K^H F_{\sigma}$
carries an irreducible unitary representation of $G$. By a double application of Frobenius theorem,
we show (Theorem \ref{dF consequence}) that in order to realize $\mathcal{H}_{(m,\sigma)}$ in the space of spin-tensor fields
$\Gamma(\mathring{M},\mathbb{W})$, one has to choose $W$ in such a way that its decomposition
under the stabilizer $K$ contains $F_{\sigma}$. Then, an equivariant symbol can be obtained by
propagating along the orbit $\mathcal{O}_m$ a $K$-equivariant linear map on $W$ whose kernel
is $F_{\sigma}$. Here, it is worth pointing out that in order to obtain a nontrivial differential
operator (that is, a symbol which is not constant with respect to $p$), the chosen $K$-equivariant
linear map should not be $H$-equivariant. This symmetry breaking appears as a necessary condition for dynamics.\\

The second goal of the paper is to introduce the supersymmetric generalization
of the above mechanism. Now $G$ is the super Poincar\'e group, corresponding to
the super Lie algebra $\mathfrak{g}=\mathfrak{spin}(V)\oplus V\oplus S^*$,
where $S^*$ is a real irreducible spinor representation of
$\hbox{Spin}(V)$. The classification of the irreducible unitary representations
of the super Poincar\'e group can be achieved by a suitable adaptation of
the Mackey little group method. This has been done fully at the group level in \cite{cctv}. In dimension 4 with signature $(1,3)$, these irreducible
representations are classified by a nonnegative real number $m$ (the mass), and a half-integer
$\sigma\in\{0,\frac{1}{2},1,\frac{3}{2},...\}$ (the superspin). \\

For simplicity, consider the massive superspin 0 case in dimension $4$. The irreducible unitary
representation
$\mathcal{H}_{(m,0)}$ (with $m>0$) of $G$ can be obtained by induction from an irreducible unitary representation of the stabilizer on $F:=\bigwedge^{\bullet} S_+^*$, where $S_+^*$ is the two-dimensional complex half-spinor representation ($S_{\mathbb{C}}=S_+^*\oplus S_-^*$). At the infinitesimal level, the stabilizer is $\mathfrak{k}\oplus S^*$, where $\mathfrak{k}$ is the Lie algebra of the little group $K$ that has appeared above in the non-super case. As a $K$-module, $F$ is reducible: $F=\mathbb{C}\oplus S_+^*\oplus\bigwedge^2 S_+^*$. In other words, the superparticle of mass $m$ and superspin 0 is made of two particles of spin 0 (corresponding to $\mathbb{C}$ and $\bigwedge^2 S_+^*\simeq\mathbb{C}$) and one particle of spin $\frac{1}{2}$ (corresponding to $S_+^*$), having all the same mass $m$.\\

On the other hand, one can consider ``superfields" on Minkowski superspacetime.
The latter is defined as the linear cs supermanifold $M_{cs}$ associated to the super vector space $V_{\mathbb{C}}\oplus S_{\mathbb{C}}$. Thus, $M_{cs}=(\mathring{M},\mathcal{O}_{M_{cs}})$, where
$\mathcal{O}_{M_{cs}}(U)=\mathcal{C}^{\infty}(U,\bigwedge^{\bullet}S_{\mathbb{C}}^*)$ for every open set $U\subset \mathring{M}$ (``cs" refers to this structure, as opposed to a complex analytic supermanifold structure, {\it cf.} \cite{dm} for instance). Here, we consider a ``superfield" as being a superfunction on $M_{cs}$, possibly spin-tensor valued. (At first sight, this seems to be at odds with the notion of superfield that is found in the physics literature; however, as we explain in section \ref{superfunctions vs superfields} using the functor of points, it makes perfectly sense here to consider ``superfields" as being just superfunctions.) Minkowski superspacetime $M_{cs}$ carries of course a transitive action of the super Poincar\'e group $G$, and we have a representation of $G$ on the super vector space $\mathcal{O}_{M_{cs}}(\mathring{M})=\mathcal{C}^{\infty}(\mathring{M},\bigwedge^{\bullet}S_{\mathbb{C}}^*)$.\\

The question now is to realize the superparticle $\mathcal{H}_{(m,0)}$ as a
$G$-invariant sub-superspace of (a suitable super-Hilbert space completion of a quotient of) $\mathcal{C}^{\infty}(\mathring{M},\bigwedge^{\bullet}S_{\mathbb{C}}^*)$. This
should certainly be possible: the obvious choice for $W$ is
$\bigwedge^{\bullet}S_{\mathbb{C}}^*\simeq \bigwedge^{\bullet}S_{+}^*\otimes
\bigwedge^{\bullet}S_{-}^*$, whose decomposition under the stabilizer contains
a subrepresentation isomorphic to $F=\bigwedge^2 S_+^*$. To single out the
superparticle, it remains to find appropriate $(\mathfrak{k}\oplus
S^*)$-equivariant linear maps on $W$, propagate them into
$(\mathfrak{spin}(V)\oplus S^*)$-equivariant symbols, and then use some super
version of the Fourier transform to obtain super Poincar\'e equivariant
differential operators. We achieve this in sections \ref{supersymbols} and
\ref{super FT}. Inspired by \cite{deb}, we use the standard supermetric on
$M_{cs}$ induced by the inner product on $V$ and the natural
$\hbox{Spin}(V)$-invariant symplectic structures $\varepsilon_{\pm}$ on
$S_{\pm}$ to define the super Fourier transform of a (compactly supported)
superfunction $f\in
\mathcal{C}^{\infty}_c(\mathring{M},\bigwedge^{\bullet}S_{\mathbb{C}}^*)\simeq
\mathcal{C}^{\infty}_c(\mathring{M})[\theta^a,\bar{\theta}^{\dot{a}}]$ as
follows: it is the element\\ $\star\widehat{f}\in
\mathcal{C}^{\infty}(V^*)[\tau^a,\bar{\tau}^{\dot{a}}]$ defined by:
$$\star\widehat{f}:=\int_{M_{cs}} e^{-i(\langle p,x\rangle+\varepsilon_+(\tau,\theta)+\varepsilon_-(\bar{\tau},\bar{\theta}))}\;f\;dx\;d\theta\;d\bar{\theta}$$
This super Fourier transform has desirable properties such as exchanging $\displaystyle\frac{\partial}{\partial \theta^a}$ with exterior multiplication by $\varepsilon_+(\tau^a,\cdot)$, and multiplication by $\theta^a$ with contraction by $(\varepsilon_{+})^{-1}(\tau^a,\cdot)$. This already gives a hint of how the symbols of the fermionic differential operators should be constructed.\\

We also notice that restriction of a superfunction to the body ({\it i.e.} setting $\theta=\bar{\theta}=0$) corresponds to taking the Berezin integral of its super Fourier transform.\\

In fact, if we define the bosonic Fourier transform of $f$ to be:
$$\widehat{f}:=\int_{\mathring{M}} e^{-i\langle p,x\rangle}\;f\;dx$$
then $$\star\widehat{f}:=\int e^{-i(\varepsilon_+(\tau,\theta)+\varepsilon_-(\bar{\tau},\bar{\theta}))}\;\widehat{f}\;d\theta\;d\bar{\theta}$$
We show (Theorem \ref{odd super Fourier equals Hodge star}) that this purely odd super Fourier transform is nothing but the Hodge star with respect to the symplectic structure defined by $\varepsilon_{\pm}$, which explains our choice of notation.\\

Finally, using appropriate equivariant symbols on the orbit $\mathcal{O}_m$, we construct the super Poincar\'e equivariant differential operators whose kernel corresponds to the irreducible unitary representation of mass $m$ and superspin 0, realized as a linear sub-supermanifold of superfunctions. In this way, we obtain in Theorem \ref{super equations} supersymmetric differential equations in terms of {\it ordinary} ({\it i.e.} non-Grassmannian) fields. In particular, those equations reduce to a Klein-Gordon equation and a Dirac equation, the latter involving ordinary spinor fields with (commuting) complex-valued components. This seems to be at odds with the physics literature, where the spinor fields have typically anticommuting Grassmann-valued components. In fact, as we show via Theorem \ref{representability}, the two points of view are naturally related via the functor of points, which establishes the link between our supersymmetric equations, and the Wess-Zumino supersymmetric field equations for massive chiral superfields in dimension $(4|4)$. It is clear that our results allow generalizations to arbitrary superspin, and then to other spacetime dimensions. \\

The article is organized as follows. In section 2, we recall briefly the main
steps leading to the classification of the irreducible unitary representations
of the Poincar\'e group, while introducing some of the notations and
terminology that we will use later on. In section 3, we discuss the
construction of equivariant differential operators whose kernels correspond to
those irreducible representations. The content of section 3 is illustrated in
section 4, where we present two examples that also serve as toy models for the
analogous constructions that we perform later on in the super case. Section 5
is the super-analog of section 2: we recall briefly the main features from the
classification of super Poincar\'e group's irreducible unitary representations
that will play a role in the remaining part
of the paper. In section 6,
we recall some of the structure of Minkowski superspacetime in dimension $(4|4)$, and start discussing the relation to superparticles. In section 7, we construct the supersymmetric symbols that select the chiral representation and the irreducible massive representation of superspin 0. In section 8, we introduce the super Fourier transform on Minkowski superspacetime and use it to obtain the supersymmetric differential operators corresponding to our previously constructed symbols. Finally, we clarify in section 9 the link between superfunctions (in the sense of Berezin, Kostant, Leites...) and superfields (in the sense of the physicists).\\

\section{Unitary dual of the Poincar\'e group} \label{unitary dual of the Poincare group}

In all this paper, $V$ denotes a real vector space of dimension $d$ equipped
with an inner product $\langle,\rangle$ of signature $(1,d-1)$, and
$\mathring{M}$ an affine space directed by $V$ ({\it Minkowski spacetime}). We
denote by $C_+$ one of the two connected components of the timelike cone
$C=\{v\in V\;|\;\langle v,v\rangle>0\}$, and by $\hbox{Spin}(V)$ the spinorial double cover of the connected Lorentz group of $V$ (preserving space and time orientation).\\

We denote by $\Pi(V)$ the {\it Poincar\'e group} of $V$, defined as the semidirect product\\
$V\rtimes\hbox{Spin}(V)$. In sections \ref{unitary dual of the Poincare group} to \ref{examples}, we will often abbreviate the notations by using the letter $G$ for the Poincar\'e group $\Pi(V)$, and the letter $H$ for the group $\hbox{Spin}(V)$.\\

The {\it unitary dual} of $G$ is the set $\widehat{G}$ of all isomorphism classes of irreducible unitary representations of $G$. We recall briefly in this section the main steps leading to the description of $\widehat{G}$, that is, to the classification of irreducible unitary representations of $G$. For more details, see for example \cite{vdb}, \cite{var1} or \cite{var2}.\\

Let $\rho:G\longrightarrow\hbox{U}(\mathcal{H})$ be an irreducible unitary representation of $G$ on a Hilbert space $\mathcal{H}$. We start by looking at the action of the translation subgroup $V\subset G$: to the restricted representation $\rho_{|V}:V\longrightarrow \hbox{U}(\mathcal{H})$, there corresponds a unique projection-valued measure $P$ on the dual space $V^*$ such that for every $f\in L^1(V)$,
$$\int_V f(v)\;\rho_{|V}(-v)\;dv=\int_{V^*}\widehat{f}\;dP$$
where $\widehat{f}$ is the Fourier transform of $f$ ({\it cf.} \cite{vdb}). This is the spectral measure associated with the family of commuting unitary operators $\{\rho_{|V}(v)\;;\;v\in V\}$. The support $\mathcal{O}$ of this measure is by definition the {\it spectrum of the representation $\rho$}. \\

The map $\rho_{|V}:V\longrightarrow \hbox{U}(\mathcal{H})$ is $H$-equivariant: $\rho_{|V}(hv)=\rho_{|H}(h)\circ\rho_{|V}(v)\circ\rho_{|H}(h)^{-1}$ f
or all $h\in H$ and $v\in V$. As a result, the spectral measure $P$ is $H$-equivariant, and its support (the spectrum $\mathcal{O}$ of $\rho$) is $H$-invariant. In fact, it is not difficult to show that by irreducibility of $\rho$, the action of $H$ on $\mathcal{O}$ is transitive. Thus, $\mathcal{O}$ is an orbit for the action of $H$ on $V^*$. In summary, we have a map $\hbox{spec}:\widehat{G}\longrightarrow V^*/H$ that associates to every (isomorphism class of) irreducible unitary representation of $G$ its spectrum in $V^*/H$.\\

Next, choose a preferred point $q\in\mathcal{O}$ and let $K$ be the stabilizer of $q$ under the action of $H$. Also,
let $F:=\bigcap_{v\in V}\hbox{Ker}(\rho_{|V}(v)-e^{iq(v)}\hbox{\upshape{Id}}_{\mathcal{H}})$. It is not difficult to check that $F$
is invariant under $K$, and that the representation $\rho^F_{|K}:K\longrightarrow \hbox{U}(F)$ is irreducible (by irreducibility of $\rho$).\\

Thus, to every $[\rho]\in\widehat{G}$, one can associate a pair $(\mathcal{O},[\rho^F_{|K}])$ where $\mathcal{O}\in V^*/H$
and $[\rho^F_{|K}]\in \widehat{K}$.\\

Conversely, given a pair $(\mathcal{O},\lambda)$ where $\mathcal{O}\in V^*/H$ and $\lambda:K\longrightarrow\hbox{U}(F)$
is an irreducible unitary representation, one can define an irreducible unitary representation\\ $\rho:G\longrightarrow\hbox{U}(\mathcal{H})$
as follows. First, one induces a unitary representation $\mathcal{H}$ of $H$ from the unitary representation $F$ of $K$.
Concretely, $\mathcal{H}:=\hbox{ind}_{K}^H F$ may be defined as follows: let $\mathbb{H}$ be the $H$-equivariant
Hermitian vector bundle over $\mathcal{O}$ associated to the principal $K$-bundle $H\longrightarrow \mathcal{O}$
by the representation $\lambda$ of $K$ on $F$. Then let $\mathcal{H}$ be the space of $L^2$ sections of $\mathbb{H}$:
$$\mathcal{H}:=\Gamma_{L^2}(\mathcal{O},\mathbb{H})$$ (remark that $\mathcal{O}$ has an $H$-invariant measure, but otherwise one could have used half-densities).
We have a unitary representation of $H$ on $\mathcal{H}$ defined by $(h\cdot\Psi)_p:=h\cdot\Psi_{h^{-1}p}$,
and if we make $v\in V$ act by $(v\cdot\Psi)_p:=e^{ip(v)}\Psi_p$, we obtain an irreducible unitary representation
$\rho$ of $G$ on $\mathcal{H}$.\\

In conclusion, irreducible unitary representations of $G$ are classified by pairs $(\mathcal{O},\lambda)$ where $\mathcal{O}\in V^*/H$ and $\lambda$ is an irreducible unitary representation of the little group $K$.\\

\begin{rem} Alternatively, given an orbit $\mathcal{O}\in V^*/H$, the data of a unitary representation $\rho$ of
$G$ with spectrum $\mathcal{O}$ is equivalent to the data of a pair $(\gamma,P)$ where $\gamma$ is a unitary representation of $H$ and $P$ is an $H$-equivariant projection-valued measure on $\mathcal{O}$. Such a pair $(\gamma,P)$ (called ``system of imprimitivity") is in turn equivalent to a unitary representation of $K$, by the imprimitivity theorem (cf. \cite{vdb}). In particular, the following are equivalent: irreducible unitary representations of $G$ with spectrum $\mathcal{O}$, irreducible systems of imprimitivity on $\mathcal{O}$, and irreducible unitary representations of $K$.\\ In terms of the map $\;\hbox{\upshape{spec}}:\widehat{G}\longrightarrow V^*/H$, $$\hbox{\upshape{spec}}^{-1}(\{\mathcal{O}\})\simeq \widehat{K}$$
\end{rem}

The above method can of course be used, without significant change, to classify irreducible unitary representations of arbitrary
semidirect products $G=A\rtimes H$ where $A$ is abelian (cf. \cite{vdb} or \cite{var1} for instance). But here, we concentrate
on the Poincar\'e group. The orbits of $H=\hbox{\upshape{Spin}}(V)$ on $V^*$ are well-known. They are of several types.
We will focus on the case where the orbit is a sheet of hyperboloid $\;\mathcal{O}_m^+:=\{p\in V^*\;|\;\langle p,p\rangle =m^2\}\cap C^{\vee}_+$\\
for some $m>0$, called the {\it mass} of the representation $\rho$. Here, $C^{\vee}=\{p\in V^*\;|\;\langle p,p\rangle>0\}$ denotes the
dual timelike cone, and $C^{\vee}_+=\{p\in C^{\vee}\;|\;p_0>0\}$, where we have written \\$p=p_0e^0+p_1e^1+\dots+p_{d-1}e^{d-1}$ in
an orthonormal basis $(e^0,e^1,...,e^{d-1})$ of $V^*$. Thinking of $p_0$ as the energy, the
irreducible unitary representations of the Poincar\'e group corresponding to the orbit $\mathcal{O}_m^+$ are said to
be {\it massive positive energy representations}.\\

In this case, we choose $me^0$ as preferred point on the orbit $\mathcal{O}_m^+$, and we denote by
$K$ the stabilizer of $me^0$ under
the action of $H$. It is not difficult to check that $K\simeq \hbox{Spin}(d-1)$. In particular, $K$
is compact, and the elements of $\widehat{K}$ are labeled by the highest weights.\\

In particular, when $d=4$, we have $K=\hbox{Spin}(3)\simeq \hbox{SU}(2)$, and $\widehat{K}\simeq\{0,\frac{1}{2},1,\frac{3}{2},...\}$. For $\sigma\in \{0,\frac{1}{2},1,\frac{3}{2},...\}$, we denote by $F_{\sigma}$ be the ($(2\sigma+1)$-dimensional) space of the irreducible unitary representation of $K$ of {\it spin} $\sigma$, and by $\mathcal{H}_{(m,\sigma)}$ the corresponding irreducible unitary representation of $G$.\\

An important situation that we will not consider in the present paper is that of the massless irreducible unitary representations of the Poincar\'e group. These correspond to the case where the orbit is the one-sided cone $\;\mathcal{O}_0^+:=\{p\in V^*\;|\;\langle p,p\rangle =0\}\cap C^{\vee}_+$. These are also positive energy representations. When $d=4$, the stabilizer $K$ is a semidirect product $\mathbb{C}\rtimes \hbox{U}(1)$. The irreducible unitary representations of $K$ that are finite-dimensional correspond to a trivial action of $\mathbb{C}$ and are classified by $\widehat{\hbox{U}(1)}\simeq\mathbb{Z}$ ({\it helicity}). But there are also representations of $K$ on which $\mathbb{C}$ acts nontrivially, and these are infinite-dimensional ({\it cf.} \cite{var1} for more details).\\

Finally, let us note that the theory, at this level, allows perfectly for representations which are not of positive energy. For example, one has also the negative energy representations, corresponding to the sheet of hyperboloid $\;\mathcal{O}_m^-:=\{p\in V^*\;|\;\langle p,p\rangle =m^2\}\cap C^{\vee}_-$ for some $m>0$, where $C^{\vee}_-=\{p\in C^{\vee}\;|\;p_0<0\}$. There are also representations corresponding to orbits of the type $\{p\in V^*\;|\;\langle p,p\rangle =(i\mu)^2\}$ for some $\mu>0$: these representations of imaginary mass $i\mu$ are not of positive energy (their energy is not even of definite sign).\\

\section{Realization in terms of spin-tensor fields} \label{spin-tensor fields}

We know that the Poincar\'e group $G:=V\rtimes H$ (where $H:=\hbox{Spin}(V)$) acts transitively on Minkowski spacetime $\mathring{M}$, and if we choose an origin $o\in\mathring{M}$, we obtain a splitting $i_{o}:H\longrightarrow G$ of the short exact sequence
$$0\longrightarrow V \longrightarrow G \longrightarrow H \longrightarrow 1$$
which allows for a right action of $H$ on $G$, by setting $g\cdot h=g\;i_o(h)$ for every $g\in G$ and $h\in H$. Then, we can view $G$ as a $G$-equivariant principal bundle over $\mathring{M}$ with structural group $H$. Denoting by $L:G\longrightarrow H$ the group morphism in the above exact sequence, we see that any representation of $H$ lifts trivially to a representation of $G$, by making any Poincar\'e transformation $g$ act through its linear part $L(g)$.\\

Let $W$ be an irreducible representation of $H$, and $\mathbb{W}:=G\times_H W$, so that $\Gamma(\mathring{M},\mathbb{W})$ carries a representation of $G$. Recall that $G$ acts on $\mathbb{W}$ as follows: $a\cdot [g,w]:=[ag,w]$ (so $\mathbb{W}$ is a $G$-equivariant vector bundle). Next, if $\Phi\in \Gamma(\mathring{M},\mathbb{W})$, then $(g\cdot\Phi)_x:=g\cdot\Phi_{g^{-1}x}$.\\

We start by the following lemma.\\

\begin{lem} \label{useful lemma} The $G$-equivariant vector bundle $\mathbb{W}$ has a natural, $G$-equivariant trivialization, inducing a $G$-isomorphism between $\Gamma(\mathring{M},\mathbb{W})$ and $\mathcal{C}^{\infty}(\mathring{M},W)$, the latter space carrying the action of $G$ given by $\;(g\cdot\phi)(x)=L(g)(\phi(g^{-1}x))\;$ for every $g\in G$, $\phi\in \mathcal{C}^{\infty}(\mathring{M},W)$ and $x\in\mathring{M}$.
\end{lem}

\noindent {\bf Proof~:}
Let $\Theta:\mathbb{W}\longrightarrow \mathring{M}\times W$ be defined by $\Theta([g,w])=(\pi(g),L(g)(w))$. Note that $\Theta$ is well-defined: $(\pi(g\cdot h),L(g\cdot h)(h^{-1}\cdot w))=(\pi(g),L(g\;i_o(h))(h^{-1}\cdot w))=(\pi(g),L(g)(L(i_o(h))(h^{-1}\cdot w)))=(\pi(g),L(g)(h\cdot(h^{-1}\cdot w)))=(\pi(g),L(g)(w))$. Moreover, $\Theta$ is $G$-equivariant: $\Theta(a\cdot[g,w])=\Theta([ag,w])=(\pi(ag),L(ag)(w))=(a\pi(g),L(a)(L(g)(w)))$\\
$=a\cdot\Theta([g,w])$. The induced isomorphism $\Gamma(\mathring{M},\mathbb{W})\longrightarrow\mathcal{C}^{\infty}(\mathring{M},W)$ sends the section $\Phi$ of $\mathbb{W}$ to the map $\phi:\mathring{M}\longrightarrow W$ defined by $\phi(x)=\hbox{pr}_W(\Theta(\Phi_x))$. It is clearly equivariant since $\hbox{pr}_W(\Theta((g\cdot\Phi)_x))=\hbox{pr}_W(\Theta(g\cdot\Phi_{g^{-1}x}))=\hbox{pr}_W(g\cdot\Theta(\Phi_{g^{-1}x}))=L(g)(\hbox{pr}_W(\Theta(\Phi_{g^{-1}x})))=L(g)(\phi(g^{-1}x))=(g\cdot\phi)(x)$. $\hfill\square$\\

\begin{rem} \label{important remark} It is important to note here that the above lemma is more generally valid for any equivariant vector bundle over any homogeneous space, provided that the action of the little group on the typical fiber happens to extend into an action of the full group on that typical fiber. If this condition is satisfied, then the equivariant vector bundle admits an equivariant trivialization.\\
\end{rem}

Let $\mathcal{H}_{(m,\sigma)}$ be the space of an irreducible unitary representation of $G$ of mass $m>0$ and spin $\sigma$. Let $F_{\sigma}$ be the space of the irreducible unitary representation of the stabilizer $K$ of the preferred point $me^0\in\mathcal{O}_m$.\\

Choose the spin $\sigma$ such that $F_{\sigma}$ appears in $W_{|K}$. We are interested in determining a
subspace of (a suitable Hilbert space completion of a quotient of) $\Gamma(\mathring{M},\mathbb{W})$ isomorphic to $\mathcal{H}_{(m,\sigma)}$. Since we are working at mass
$m$, it is natural to start by considering the vector bundle $\mathbb{W}_{(m)}:=H\times_{K} W$, associated to the principal bundle $H\longrightarrow\mathcal{O}_m$ by the representation of $K$ on $W$ obtained by restricting that of $H$. Note that $\mathbb{W}_{(m)}$ is an $H$-equivariant vector bundle over $\mathcal{O}_m$ (the action is given by $b\cdot [h,w]:=[bh,w]$), and therefore we have a representation of $H$ on $\Gamma(\mathcal{O}_m,\mathbb{W}_{(m)})$ (given by $(b\cdot\Psi)_p:=b\cdot\Psi_{b^{-1}p}$).\\

\begin{prop} Suppose there exists an $H$-module $E$, and a $K$-equivariant morphism $u:W\longrightarrow E$ such
that $\hbox{\upshape{Ker}}\;u=F_{\sigma}$. Suppose also that there exists a $K$-invariant sesquilinear form $\langle,\rangle_0$
on $W$ whose restriction to $F_{\sigma}$ is Hermitian positive definite. Then $u$ determines an $H$-equivariant Hermitian
subbundle $\mathbb{D}_{(m,\sigma)}$ of $\mathbb{W}_{(m)}$ isomorphic to $\mathbb{H}_{(m,\sigma)}$. Moreover,
$\Gamma_{L^2}(\mathcal{O}_m,\mathbb{D}_{(m,\sigma)})$ is unitarily equivalent to $\Gamma_{L^2}(\mathcal{O}_m,\mathbb{H}_{(m,\sigma)})$
as representations of $H$ (and then of $G$).
\end{prop}

\noindent {\bf Proof~:}
Let $\mathbb{E}_{(m)}=H\times_{K} E$. Then $\mathbb{E}_{(m)}$ is an $H$-equivariant vector bundle over $\mathcal{O}_m$.
The $K$-morphism $u$ determines a morphism of vector bundles $\tilde{u}: \mathbb{W}_{(m)}\longrightarrow \mathbb{E}_{(m)}$,
defined by $\tilde{u}([h,w])=[h,u(w)]$. This is well-defined by $K$-equivariance of $u$. Also, $\tilde{u}$ is $H$-equivariant:
$\tilde{u}(b\cdot[h,w])=\tilde{u}([bh,w])=[bh,u(w)]=b\cdot[h,u(w)]=b\cdot\tilde{u}([h,w])$. Let $\mathbb{D}_{(m,\sigma)}:=\hbox{Ker}\;\tilde{u}$.
Then $\mathbb{D}_{(m,\sigma)}$ is an $H$-equivariant vector subbundle of $\mathbb{W}_{(m)}$, with typical fiber
$F_{\sigma}\subset W$. It is not difficult to see that $\mathbb{D}_{(m,\sigma)}$ is isomorphic to $\mathbb{H}_{(m,\sigma)}$
as $H$-equivariant bundles, and that $\Gamma(\mathcal{O}_m,\mathbb{D}_{(m,\sigma)})$ is isomorphic to
$\Gamma(\mathcal{O}_m,\mathbb{H}_{(m,\sigma)})$ as representations of $H$ (and then of $G$).
Now we need to take care of the inner products. For $p\in\mathcal{O}_m$, define\\
$h_p:(\mathbb{W}_{(m)})_p\times (\mathbb{W}_{(m)})_p\longrightarrow \mathbb{C}$ by
$g_p([h,w],[h,w'])=\langle w,w'\rangle_0$. This is well-defined by $K$-invariance of
$\langle,\rangle_0$. Thus, we have a section $g\in\Gamma(\mathcal{O}_m,\mathbb{W}_{(m)}^*\otimes\overline{\mathbb{W}_{(m)}^*})$.
This section is $H$-equivariant: $g_{bp}(b\cdot[h,w],b\cdot[h,w'])=g_{bp}([bh,w],[bh,w'])=\langle w,w'\rangle_0=g_p([h,w],[h,w'])$.
Suppose $[h,w]\in\mathbb{D}_{(m,\sigma)}$, so that $\tilde{u}([h,w])=0$. Then $[h,u(w)]=0$, so $u(w)=0$, and so $w\in F_{\sigma}$.
Then, if $[h,w]$ is not on the zero section of $\mathbb{D}_{(m,\sigma)}$, we have $w\neq 0$ and
$g_p([h,w],[h,w])=\langle w,w\rangle_0 >0$ since $w\in F_{\sigma}-\{0\}$ and the restriction of $\langle,\rangle_0$
to $F_{\sigma}$ is Hermitian positive definite. Thus, $\mathbb{D}_{(m,\sigma)}$ is a Hermitian vector bundle.
If $\Psi,\Psi'\in\Gamma_{L^2}(\mathcal{O}_m,\mathbb{D}_{(m,\sigma)})$, set
$\displaystyle\langle\Psi,\Psi'\rangle:=\int_{\mathcal{O}_m} g_p(\Psi_p,\Psi'_p)\;d\beta_m(p)$.
It is not difficult to see that $\mathbb{D}_{(m,\sigma)}$ is isomorphic to $\mathbb{H}_{(m,\sigma)}$
as $H$-equivariant Hermitian bundles, and that the Hilbert space $\Gamma_{L^2}(\mathcal{O}_m,\mathbb{D}_{(m,\sigma)})$
is equivalent to $\Gamma_{L^2}(\mathcal{O}_m,\mathbb{H}_{(m,\sigma)})$ as unitary representations of $H$ (and then of $G$).
$\hfill\square$\\

\begin{rem} In fact, the morphism of vector bundles $\tilde{u}$ induces in turn an $H$-equivariant linear map
$\tilde{\tilde{u}}:\Gamma_{L^2}(\mathcal{O}_m,\mathbb{W}_{(m)})\longrightarrow\Gamma_{L^2}(\mathcal{O}_m,\mathbb{E}_{(m)})$
given by $\tilde{\tilde{u}}(\Psi)_p=\tilde{u}(\Psi_p)$, and\\ $\Gamma_{L^2}(\mathcal{O}_m,\mathbb{D}_{(m,\sigma)})=\hbox{\upshape{Ker}}\;\tilde{\tilde{u}}$,
which gives another way to see that $\Gamma_{L^2}(\mathcal{O}_m,\mathbb{D}_{(m,\sigma)})$ carries a representation of $H$.\\
\end{rem}

\begin{rem} We could have taken $E$ to be just a $K$-module in the above proposition. The action of $H$ on $E$ (as well as on $W$) was not used anywhere. Now it is going to be used.\\
\end{rem}

By Remark \ref{important remark}, we can apply Lemma \ref{useful lemma} to each of the $H$-equivariant vector
bundles $\mathbb{W}_{(m)}$ and $\mathbb{E}_{(m)}$. Thus, $\mathbb{W}_{(m)}$ has a natural, $H$-equivariant
trivialization $\Theta_{\mathbb{W}_{(m)}}:\mathbb{W}_{(m)}\longrightarrow\mathcal{O}_m\times W$,
inducing an $H$-isomorphism between $\Gamma(\mathcal{O}_m,\mathbb{W}_{(m)})$ and $\mathcal{C}^{\infty}(\mathcal{O}_m,W)$,
the latter space carrying the action of $H$ given by $\;(h\cdot\psi)(p)=h\cdot\psi(h^{-1}p)\;$ for every $h\in H$,
$\psi\in \mathcal{C}^{\infty}(\mathcal{O}_m,W)$ and $p\in\mathcal{O}_m$. The same applies for $\mathbb{E}_{(m)}$.
We use these trivializations to associate a symbol to every $K$-equivariant linear map $u:W\longrightarrow E$.\\

Given $u:W\longrightarrow E$, define $\zeta_u:\mathcal{O}_m\longrightarrow W^*\otimes E$ as follows:
$$\zeta_u(p)(w):=\hbox{pr}_E(\Theta_{\mathbb{E}_{(m)}}(\tilde{u}(\Theta_{\mathbb{W}_{(m)}}^{-1}(p,w))))$$
In other words, $\zeta_u$ is the vector bundle morphism $\tilde{u}:\mathbb{W}_{(m)}\longrightarrow\mathbb{E}_{(m)}$ ``read in the trivializations".\\

\begin{prop} The map $\zeta_u:\mathcal{O}_m\longrightarrow W^*\otimes E$ is $H$-equivariant.
\end{prop}

\noindent {\bf Proof~:} If $T_u:\mathcal{C}^{\infty}(\mathcal{O}_m,W)\longrightarrow\mathcal{C}^{\infty}(\mathcal{O}_m,E)$ is defined by $T_u(\psi)(p):=\hbox{\upshape{pr}}_E(\Theta_{\mathbb{E}_{(m)}}(\tilde{\tilde{u}}(\Psi)_p))$, where $\Psi_p:=\Theta_{\mathbb{W}_{(m)}}^{-1}(p,\psi(p))$, then $T_u(\psi)(p)=\hbox{\upshape{pr}}_E(\Theta_{\mathbb{E}_{(m)}}(\tilde{u}(\Theta_{\mathbb{W}_{(m)}}^{-1}(p,\psi(p)))))$, and we clearly have
$$T_u(\psi)(p)=\zeta_u(p)(\psi(p))$$
The $H$-equivariance of $\tilde{\tilde{u}}$ and that of the trivializations imply that \\
$T_u(h\cdot\psi)=(\hbox{\upshape{pr}}_E\circ\Theta_{\mathbb{E}_{(m)}}\circ\tilde{\tilde{u}})
(h\cdot\Psi)=h\cdot(\hbox{\upshape{pr}}_E\circ\Theta_{\mathbb{E}_{(m)}}\circ\tilde{\tilde{u}})(\Psi)=h\cdot T_u(\psi)$,\\
so $T_u$ is $H$-equivariant. This, in turn, can be used in the above expression
of $T_u$:\\

$T_u(h\cdot \psi)(p)=\zeta_u(p)((h\cdot\psi)(p))$ implies that $(h\cdot T_u(\psi))(p)=\zeta_u(p)((h\cdot\psi)(p))$, so\\
$h\cdot T_u(\psi)(h^{-1}p)=\zeta_u(p)(h\cdot \psi(h^{-1}p))$ and so $T_u(\psi)(h^{-1}p)=h^{-1}\cdot\zeta_u(p)(h\cdot \psi(h^{-1}p))$.\\
Thus, $\zeta_u(h^{-1}p)(\psi(h^{-1}p))=h^{-1}\cdot\zeta_u(p)(h\cdot \psi(h^{-1}p))$, which implies that $$\zeta_u(hp)=\rho(h)\circ\zeta_u(p)\circ\rho(h)^{-1}$$
for every $h\in H$ and $p\in\mathcal{O}_m$. In other words,
the map $\;\zeta_u:\mathcal{O}_m\longrightarrow W^*\otimes E\;$ is\\ $H$-equivariant. $\hfill\square$\\

The above proof implies in particular that $\zeta_u(k\;me^0)=\rho(k)\circ\zeta_u(me^0)\circ\rho(k)^{-1}$ for every $k\in K$. Since $k\;me^0=me^0$, we deduce that $\zeta_u(me^0)\in (W^*\otimes E)^{K}$, {\it i.e.} $\zeta_u(me^0):W\longrightarrow E$ is $K$-equivariant. In fact, since $H$ acts transitively on $\mathcal{O}_m$, the map $\;\zeta_u:\mathcal{O}_m\longrightarrow W^*\otimes E\;$ is entirely determined by $\zeta_u(me^0)$. Indeed, any $p\in\mathcal{O}_m$ can be written as $p=h_p(me^0)$ for some $h_p\in H$. Now set $\zeta_u(p):=\rho(h_p)\circ\zeta_u(me^0)\circ\rho(h_p)^{-1}$. By $K$-equivariance of $\zeta_u(me^0)$, this does not depend on the choice of $h_p$. Actually, from the definition of $\zeta_u$, one can check that $\;\zeta_u(me^0)=u$, so that
$$\zeta_u(p):=\rho(h_p)\circ u\circ\rho(h_p)^{-1}$$

Note that $\Gamma(\mathcal{O}_m,\mathbb{D}_{(m,\sigma)})\simeq \{\psi:\mathcal{O}_m\longrightarrow W\;|\;\zeta_u(p)(\psi(p))=0\;\;\;\forall p\in\mathcal{O}_m\}$.\\

The Hermitian bundle metric can also be trivialized equivariantly. To every $p\in\mathcal{O}_m$, we can associate a sesquilinear form $\langle,\rangle_p$ on $W$ in such a way that $\Theta_{|(\mathbb{W}_{(m)})_p}:(\mathbb{W}_{(m)})_p\longrightarrow\{p\}\times W$ is an isometry. Since $\Theta_{|(\mathbb{W}_{(m)})_p}([h,w])=(p,hw)$, we want to have $g_p([h,w],[h,w'])=\langle hw,hw'\rangle_p$, that is $\langle w,w'\rangle_0=\langle hw,hw'\rangle_p$. Now any $p\in\mathcal{O}_m$ can be written as $p=h_p(me^0)$ for some $h_p\in H$. Now set
 $$\langle w,w'\rangle_p:=\langle h_p^{-1}\cdot w,h_p^{-1}\cdot w'\rangle_0$$
 By $K$-invariance of $\langle,\rangle_0$, this does not depend on the choice of $h_p$. Note that the map $\mathcal{O}_m\longrightarrow W^*\otimes\overline{W^*}$ which takes $p$ to $\langle,\rangle_p$ is $H$-equivariant. Note that if $\psi,\psi':\mathcal{O}_m\longrightarrow W$ are square-integrable maps such that for all $p\in\mathcal{O}_m$,\\
 $\zeta_u(p)(\psi(p))=0=\zeta_u(p)(\psi'(p))$, then
$$\langle\psi,\psi'\rangle=\int_{\mathcal{O}_m} \langle\psi(p),\psi'(p)\rangle_p\;d\beta_m(p)=\int_{\mathcal{O}_m} \langle h_p^{-1}\cdot\psi(p),h_p^{-1}\cdot\psi'(p)\rangle_0\;d\beta_m(p)$$

It remains to define the equivariant differential operator corresponding to a map $\zeta_u:\mathcal{O}_m\longrightarrow W^*\otimes E$. Of course, the notions of symbols and differential operators that we discuss here are special cases of the corresponding standard notions, formulated in an equivariant setting. See for instance \cite{cs}.\\

\begin{defi}
We say that $\;\zeta_u:\mathcal{O}_m\longrightarrow W^*\otimes E\;$ is the {\bf symbol of a linear differential operator of order $r$ at most} if there exists a linear map $\;\displaystyle\Xi:\bigoplus_{k=0}^r\hbox{\upshape{Sym}}^k V_{\mathbb{C}}^*\otimes W\longrightarrow E\;$ such that for every $p\in\mathcal{O}_m$ and $w\in W$,
$$\zeta_u(p)(w)=\Xi(w\;,\;p\otimes w\;,\;p\otimes p\otimes w\;,\;\dots\;,\;p\otimes ... \otimes p\otimes w)$$
\end{defi}

We write $\Xi=\bigoplus_{k=0}^r\Xi^{(k)}$, where for each $k$, $\;\Xi^{(k)}:\hbox{\upshape{Sym}}^k V_{\mathbb{C}}^*\otimes W\longrightarrow E$.\\
Then, the above expression becomes:
$$\zeta_u(p)(w)=\Xi^{(0)}(w)+\Xi^{(1)}(p\otimes w)+\Xi^{(2)}(p\otimes p\otimes w)+\dots+\Xi^{(r)}(p\otimes ... \otimes p\otimes w)$$

\begin{defi}
If $\phi:\mathring{M}\longrightarrow W$ is a smooth map, and $x\in\mathring{M}$, we define the {\bf $r^{th}$-jet of $\phi$ at $x$} (or the {$r^{th}$-order Taylor polynomial generated by $\phi$ at $x$}) to be the element $\;j^{(r)}_x\phi\in\displaystyle\bigoplus_{k=0}^r\hbox{\upshape{Sym}}^k V_{\mathbb{C}}^*\otimes W$ given by :
$$j^{(r)}_x\phi=(\phi(x)\;,\;-id\phi(x)\;,\;-d^{(2)}\phi(x)\;,\;\dots\;,\;(-i)^rd^{(r)}\phi(x))$$
\end{defi}

\begin{defi} Suppose $\;\zeta_u:\mathcal{O}_m\longrightarrow W^*\otimes E\;$ is the symbol of a linear differential operator of order $r$. The {\bf linear differential operator of order $r$ at most} associated to $\zeta_u$ is the linear map $D:\mathcal{C}^{\infty}(\mathring{M},W)\longrightarrow\mathcal{C}^{\infty}(\mathring{M},E)$ defined as follows: for every $\phi\in\mathcal{C}^{\infty}(\mathring{M},W)$ and $x\in\mathring{M}$,
$$(D\phi)(x):=\Xi(j^{(r)}_x\phi)=\Xi^{(0)}(\phi(x))+\Xi^{(1)}(-id\phi(x))+\Xi^{(2)}(-d^{(2)}\phi(x))+\dots+\Xi^{(r)}((-i)^rd^{(r)}\phi(x))$$
\end{defi}

\begin{prop}
If $$\phi(x)=\int_{\mathcal{O}_m}e^{ip(x)}\;\widehat{\phi}(p)\;d\beta_m(p)$$
for some $\widehat{\phi}:\mathcal{O}_m\longrightarrow W$, then $$(D\phi)(x)=\int_{\mathcal{O}_m}e^{ip(x)}\;\zeta_u(p)(\widehat{\phi}(p))\;d\beta_m(p)$$
\end{prop}

\noindent {\bf Proof~:}
 $$d^{(k)}\phi(x)=\int_{\mathcal{O}_m}i^k e^{ip(x)}\;p\otimes...\otimes p\otimes\widehat{\phi}(p)\;d\beta_m(p)$$
and so $$(-i)^k d^{(k)}\phi(x)=\int_{\mathcal{O}_m}e^{ip(x)}\;p\otimes...\otimes p\otimes\widehat{\phi}(p)\;d\beta_m(p)$$
This implies that
$$\Xi^{(k)}((-i)^k d^{(k)}\phi(x))=\int_{\mathcal{O}_m}e^{ip(x)}\;\Xi^{(k)}(p\otimes...\otimes p\otimes\widehat{\phi}(p))\;d\beta_m(p)$$
As a result,
$$(D\phi)(x)=\sum_{k=0}^r\Xi^{(k)}((-i)^k d^{(k)}\phi(x))=\int_{\mathcal{O}_m}e^{ip(x)}\;\zeta_u(p)(\widehat{\phi}(p))\;d\beta_m(p)$$
$\hfill\square$\\

Thus, $\;\zeta_u(p)(\widehat{\phi}(p))=0$ for all $p\in\mathcal{O}_m\;$ is equivalent to partial differential equation $$D\phi=0$$

Finally, we want to determine the multiplicity of a given irreducible unitary representation $\mathcal{H}_{(m,\sigma)}$ of $G$ in $\Gamma(\mathring{M},\mathbb{W})$.\\

\begin{lem} \label{double Frobenius}
$$\hbox{\upshape{Hom}}_G(\mathcal{H}_{(m,\sigma)},\Gamma(\mathring{M},\mathbb{W}))\simeq\hbox{\upshape{Hom}}_{K}(F,\hbox{\upshape{res}}^H_{K} W)$$
\end{lem}

\noindent {\bf Proof~:} By Frobenius reciprocity, we have
$\;\hbox{\upshape{Hom}}_G(\mathcal{H}_{(m,\sigma)},\hbox{ind}_H^G W)\simeq \hbox{\upshape{Hom}}_H(\hbox{res}^G_H\mathcal{H}_{(m,\sigma)},W)$.\\
Also by Frobenius reciprocity, we have
$\;\hbox{\upshape{Hom}}_H(W,\hbox{ind}_{K}^H F)\simeq\hbox{\upshape{Hom}}_{K}(\hbox{res}^H_{K} W,F)$.\\
But $\hbox{res}^G_H\mathcal{H}_{(m,\sigma)}=\hbox{ind}_{K}^H F$ (and $\Gamma(\mathring{M},\mathbb{W})=\hbox{ind}_H^G W$). $\hfill\square$\\

We are ready to state the following important consequence of the above lemma.\\

\begin{thm} \label{dF consequence} Let $\mathcal{H}_{(m,\sigma)}$ be the positive energy irreducible unitary representation of the Poincar\'e group of mass $m>0$ and spin $\sigma$ (obtained by induction from an irreducible unitary representation $F$ of the little group $K$). Also, let $\mathbb{W}$ be the vector bundle on Minkowski spacetime $\mathring{M}$ associated to a representation $W$ of the group $\hbox{\upshape{Spin}}(V)$. The multiplicity of $\mathcal{H}_{(m,\sigma)}$ in  $\Gamma(\mathring{M},\mathbb{W})$ (as representations of the Poincar\'e group) is equal to the multiplicity of $F$ in the decomposition of $W$ under the subgroup $K$ of $\hbox{\upshape{Spin}}(V)$. \\
\end{thm}

For example, in dimension 4 with signature $(1,3)$, the tensor field representations in which the particle $\mathcal{H}_{m,\sigma}$
will appear are the $\mathcal{C}^{\infty}(\mathring{M},\hbox{Sym}^{2\alpha}S_+^*\otimes  \hbox{Sym}^{2\beta}S_-^*)$
where $\alpha+\beta\geq \sigma$.\\

\section{Examples in dimension 4: arbitrary spin} \label{examples}

We illustrate the content of the preceding section with a couple of examples in four-dimensional Minkowski spacetime ($d=4$).\\

{\bf Massive particle of spin $\frac{1}{2}$}\\

The Clifford action is defined by the map $\gamma:V^*\longrightarrow \hbox{End}(S_{\mathbb{C}})$, where $S_{\mathbb{C}}\simeq \mathbb{C}^4$ is the space of Dirac spinors. We will also use the associated map $\tilde{\gamma}:V^*\otimes S_{\mathbb{C}}\longrightarrow  S_{\mathbb{C}}$. We know that $\mathcal{H}_{(m,\frac{1}{2})}$ is obtained by induction from the irreducible unitary representation of $K\simeq \hbox{Spin}(3)\simeq\hbox{SU}(2)$ of spin $\frac{1}{2}$. The space of this representation is $F_{\frac{1}{2}}\simeq\mathbb{C}^2$.\\

We want to realize $\mathcal{H}_{(m,\frac{1}{2})}$ in the space of Dirac spinor fields $\mathcal{C}^{\infty}(\mathring{M},S_{\mathbb{C}})$. Thus, in the notations of the preceding section, we have $W=S_{\mathbb{C}}$, and we need an $\hbox{SU}(2)$-equivariant map $u$ on the $\hbox{Spin}(V)$-module $S_{\mathbb{C}}$ whose kernel is $F_{\frac{1}{2}}$. We take $u=\gamma^0-\hbox{\upshape{Id}}_{S_{\mathbb{C}}}$ where $\gamma^0:=\gamma(e^0)$. Note that $e^0:V\longrightarrow\mathbb{R}$ is $\hbox{SU}(2)$-invariant, since the point $me^0$ is stabilized by $\hbox{SU}(2)$. Therefore, $u:S_{\mathbb{C}}\longrightarrow S_{\mathbb{C}}$ is $\hbox{SU}(2)$-equivariant, and since $\gamma^0 \circ\gamma^0=\hbox{\upshape{Id}}_{S_{\mathbb{C}}}$, the subspace $\hbox{Ker}\;u\subset S_{\mathbb{C}}$ is two-dimensional (and $\hbox{SU}(2)$-invariant).\\

The corresponding symbol $\;\zeta_u:\mathcal{O}_m\longrightarrow S_{\mathbb{C}}^*\otimes S_{\mathbb{C}}\;$ is given by:
$$\zeta_u(p)=\frac{1}{m}\gamma(p)-\hbox{\upshape{Id}}_{S_{\mathbb{C}}}$$
(since if $h_p\in \hbox{Spin}(V)$ is such that $h_p\cdot (me^0)=p$, we have $\zeta_u(p)=h_p\cdot (\gamma^0-\hbox{\upshape{Id}}_{S_{\mathbb{C}}})\cdot h_p^{-1}=h_p\cdot \gamma(e^0)\cdot h_p^{-1} - \hbox{\upshape{Id}}_{S_{\mathbb{C}}}=\gamma(h_p\cdot e^0)-\hbox{\upshape{Id}}_{S_{\mathbb{C}}}=\gamma(\frac{1}{m}p)-\hbox{\upshape{Id}}_{S_{\mathbb{C}}}=\frac{1}{m}\gamma(p)-\hbox{\upshape{Id}}_{S_{\mathbb{C}}}$).\\

Let $\;\Xi:S_{\mathbb{C}}\oplus (V^*\otimes S_{\mathbb{C}})\longrightarrow S_{\mathbb{C}}\;$ be defined by $$\Xi(s\;,\;p\otimes s)=-s+\frac{1}{m}\gamma(p)(s)=-s+\frac{1}{m}\tilde{\gamma}(p\otimes s)$$
Then $\;\zeta_u(p)(s)=\Xi(s\;,\;p\otimes s)\;$ for every $p\in\mathcal{O}_m$ and $s\in S_{\mathbb{C}}$. Thus, $\zeta_u$ is the symbol of a first-order differential operator $D_u:\mathcal{C}^{\infty}(\mathring{M},S_{\mathbb{C}})\longrightarrow \mathcal{C}^{\infty}(\mathring{M},S_{\mathbb{C}})$. For $\psi\in\mathcal{C}^{\infty}(\mathring{M},S_{\mathbb{C}})$,
$$D_u\psi(x)=\Xi(\psi(x)\;,\;-id\psi(x))=-\psi(x)+\frac{1}{m}\tilde{\gamma}(-id\psi(x))=-\frac{i}{m}(-im\psi(x)+\tilde{\gamma}(d\psi(x)))$$
and so
$$D_u=-\frac{i}{m}(\slashed D-im)$$
where $\slashed D:=\tilde{\gamma}\circ d\psi$ is the Dirac operator.\\

Thus, the irreducible unitary representation $\mathcal{H}_{(m,\frac{1}{2})}$ of the Poincar\'e group is selected by $\;\gamma(p)(\widehat{\psi}(p))=m\widehat{\psi}(p)\;$ in momentum space, and by the Dirac equation $\;\slashed D\psi-im\psi=0\;$ in spacetime. (In principle, one should also impose the Klein-Gordon equation as well, but here this is not necessary as the Klein-Gordon equation is already implied by the Dirac equation).\\

{\bf Massive particle of spin $\sigma\geq 1$}\\

We know that $\mathcal{H}_{(m,\sigma)}$ is obtained by induction from the irreducible unitary representation of $K\simeq\hbox{Spin}(3)\simeq\hbox{SU}(2)$ of spin $\sigma$. The space of this representation is $F_{\sigma}\simeq\hbox{Sym}^{2\sigma}\mathbb{C}^2$.

We want to realize $\mathcal{H}_{(m,\sigma)}$ in a space of spin-tensor fields $\mathcal{C}^{\infty}(\mathring{M},W)$. From Theorem \ref{dF consequence}, we know that we can take $W=\hbox{Sym}^{2\alpha}S_+\otimes\hbox{Sym}^{2\beta}S_-$ with $\alpha+\beta\geq\sigma$. We consider the minimal choice $\alpha+\beta=\sigma$. Of course, $\alpha,\beta\in\{0,\frac{1}{2},1,\frac{3}{2},2,...\}$, and we consider in what follows the interesting case $\alpha>0$ and $\beta>0$. We need an $\hbox{SU}(2)$-equivariant map $u$ on the $\hbox{Spin}(V)$-module $W$ whose kernel is $F_{\sigma}$. At this point, we need to know how to decompose $\hbox{Sym}^{2\alpha}S_+^*\otimes\hbox{Sym}^{2\beta}S_-^*$ into irreducible representations of $\hbox{SU}(2)$. \\

  Recall that the irreducible complex representations of $\hbox{SU}(2)$ are classified by $\{0,\frac{1}{2},1,\frac{3}{2},2,...\}$. More precisely, for each $\sigma\in \{0,\frac{1}{2},1,\frac{3}{2},2,...\}$, the vector space $\hbox{Sym}^{2\sigma}S_+^*$ carries the irreducible representation of $\hbox{SU}(2)$ of highest weight $\sigma$. We say that $\hbox{Sym}^{2\sigma}S_+^*$ is the irreducible representation of spin $\sigma$ of $\hbox{SU}(2)$; its dimension is $2\sigma+1$, and its internal structure can be described as follows.\\

A canonical choice of Cartan subalgebra of $\mathfrak{su}(2)$ is $\mathfrak{t}:=\{ \left( \begin{array}{cc} i\theta & 0 \\ 0 & -i\theta \end{array} \right)\;;\;\theta\in\mathbb{R} \}$. Under $\mathfrak{t}$, the representation $\hbox{Sym}^{2\sigma}S_+^*$ decomposes into one-dimensional weight spaces:
 $$\hbox{Sym}^{2\sigma}S_+^*=L_{-\sigma}\oplus L_{-\sigma+1}\oplus L_{-\sigma+2}\oplus\dots\oplus L_{\sigma-2}\oplus L_{\sigma-1}\oplus L_{\sigma}$$
where $\left( \begin{array}{cc} i\theta & 0 \\ 0 & -i\theta \end{array} \right)$ acts on $L_j$ by $s\mapsto 2j(i\theta)s$.\\

For instance, the spin 0 representation is the trivial representation on $\mathbb{C}$, the spin $\frac{1}{2}$ representation is $S_+^*=L_{-\frac{1}{2}}\oplus L_{\frac{1}{2}}$, and the spin 1 representation is $\hbox{Sym}^{2}S_+^*=L_{-1}\oplus L_0\oplus L_1$.\\

\begin{lem} \label{decomposition lemma}
Assume in addition that $\alpha\geq\beta$. Then, as representation of $\hbox{\upshape{SU}}(2)$,
$$\hbox{\upshape{Sym}}^{2\alpha}S_+^*\otimes\hbox{\upshape{Sym}}^{2\beta}S_-^*\;\simeq\; \hbox{\upshape{Sym}}^{2(\alpha+\beta)}S_+^*\;\oplus\; \hbox{\upshape{Sym}}^{2(\alpha+\beta-1)}S_+^* \;\oplus\; \dots \;\oplus\; \hbox{\upshape{Sym}}^{2(\alpha-\beta)}S_+^*$$
\end{lem}

\noindent {\bf Proof~:} As representations of $\hbox{SU}(2)$, $S_+^*$ and $S_-^*$ become equivalent. Thus, we need to decompose $\hbox{Sym}^{2\alpha}S_+^*\otimes\hbox{Sym}^{2\beta}S_+^*$. We start by writing the weight-space decomposition of each factor: we have
$$\hbox{Sym}^{2\alpha}S_+^*=L_{-\alpha}\oplus L_{-\alpha+1}\oplus L_{-\alpha+2}\oplus\dots\oplus L_{\alpha-2}\oplus L_{\alpha-1}\oplus L_{\alpha}$$
and
$$\hbox{Sym}^{2\beta}S_+^*=L_{-\beta}\oplus L_{-\beta+1}\oplus L_{-\beta+2}\oplus\dots\oplus L_{\beta-2}\oplus L_{\beta-1}\oplus L_{\beta}$$
Taking the tensor product, and using the fact that $L_j\otimes L_k=L_{j+k}$, we obtain
$$\hbox{Sym}^{2\alpha}S_+^*\otimes\hbox{Sym}^{2\beta}S_+^*=L_{-\alpha-\beta}\oplus 2L_{-\alpha-\beta+1}\oplus 3L_{-\alpha-\beta+2}\oplus\dots\oplus 3L_{\alpha+\beta-2}\oplus 2L_{\alpha+\beta-1}\oplus L_{\alpha+\beta}$$
which implies easily the result. $\hfill\square$\\

Notice that by the above lemma, $$\hbox{\upshape{Sym}}^{2\alpha-1}S_+^*\otimes\hbox{\upshape{Sym}}^{2\beta-1}S_-^*\;\simeq\; \hbox{\upshape{Sym}}^{2(\alpha+\beta-1)}S_+^*\;\oplus\; \hbox{\upshape{Sym}}^{2(\alpha+\beta-2)}S_+^* \;\oplus\; \dots \;\oplus\; \hbox{\upshape{Sym}}^{2(\alpha-\beta)}S_+^*$$
and therefore we have (also by the above lemma):
$$\hbox{\upshape{Sym}}^{2\alpha}S_+^*\otimes\hbox{\upshape{Sym}}^{2\beta}S_-^*\;\simeq\; \hbox{\upshape{Sym}}^{2\sigma}S_+^*\;\oplus\;(\hbox{\upshape{Sym}}^{2\alpha-1}S_+^*\otimes\hbox{\upshape{Sym}}^{2\beta-1}S_-^*)$$

Consequently, we define
 $$u:\hbox{\upshape{Sym}}^{2\alpha}S_+^*\otimes\hbox{\upshape{Sym}}^{2\beta}S_-^*\longrightarrow \hbox{\upshape{Sym}}^{2\alpha-1}S_+^*\otimes\hbox{\upshape{Sym}}^{2\beta-1}S_-^*$$
proceeding as follows. First, extend $e^0\in V^*$ by $\mathbb{C}$-linearity to obtain an element $e^0\in V_{\mathbb{C}}^*$. Then compose with $\Gamma_{\mathbb{C}}:S_+^*\otimes S_-^*\longrightarrow V_{\mathbb{C}}$. This gives a map $e^0\circ\Gamma_{\mathbb{C}}:S_+^*\otimes S_-^*\longrightarrow \mathbb{C}$.\\ Now let $\iota:\hbox{\upshape{Sym}}^{2\alpha}S_+^*\otimes\hbox{\upshape{Sym}}^{2\beta}S_-^*\hookrightarrow S_+^*\otimes S_-^*\otimes  \hbox{\upshape{Sym}}^{2\alpha-1}S_+^*\otimes\hbox{\upshape{Sym}}^{2\beta-1}S_-^*$ be the canonical inclusion. Finally, set $u:=((e^0\circ\Gamma_{\mathbb{C}})\otimes\hbox{\upshape{Id}})\circ \iota$. Then $u$ is $\hbox{SU}(2)$-equivariant (since $e^0$ is $\hbox{SU}(2)$-equivariant).\\

 In fact, we have the following exact sequence of $\hbox{SU}(2)$-modules:
 $$0\longrightarrow \hbox{Sym}^{2\sigma} S_+^*\longrightarrow \hbox{Sym}^{2\alpha}S_+^*\otimes \hbox{Sym}^{2\beta}S_-^* \longrightarrow \hbox{\upshape{Sym}}^{2\alpha-1}S_+^*\otimes\hbox{\upshape{Sym}}^{2\beta-1}S_-^*\longrightarrow 0$$
the second nontrivial map being $u$, and the first nontrivial map being $(\hbox{\upshape{Id}}\otimes \tilde{c}) \circ \iota_{\alpha,\beta}$, where $\iota_{\alpha,\beta}: \hbox{Sym}^{2\sigma} S_+^*\hookrightarrow \hbox{Sym}^{2\alpha}S_+^*\otimes \hbox{Sym}^{2\beta}S_+^*$ is the canonical inclusion and $c:S_+^*\longrightarrow S_-^*$ is an $\hbox{SU}(2)$-equivariant isomorphism.\\

It is easy to check that the corresponding symbol $\;\zeta_u:\mathcal{O}_m\longrightarrow W^*\otimes E\;$ (where $E:=\hbox{\upshape{Sym}}^{2\alpha-1}S_+^*\otimes\hbox{\upshape{Sym}}^{2\beta-1}S_-^*$) is given by
$$\zeta_u(p)=((p\circ\Gamma_{\mathbb{C}})\otimes\hbox{\upshape{Id}}_E)\circ \iota$$
Let $\;\Xi:W\oplus (V^*\otimes W)\longrightarrow E\;$ be defined by $\;\Xi(w\;,\;p\otimes w)=\Xi^{(1)}(p\otimes w)\;$, where $\;\Xi^{(1)}:V^*\otimes W\longrightarrow E\;$ is given by
$$\Xi^{(1)}=(\hbox{tr}\otimes\hbox{\upshape{Id}}_E)\circ (\hbox{\upshape{Id}}_{V_{\mathbb{C}}^*}\otimes\Gamma_{\mathbb{C}}\otimes\hbox{\upshape{Id}}_E) \circ (j\otimes \iota)$$
where $j:V^*\hookrightarrow V_{\mathbb{C}}^*$ is the canonical inclusion.\\

Then $\;\Xi(w\;,\;p\otimes w)=\Xi^{(1)}(p\otimes w)=((\hbox{tr}\otimes\hbox{\upshape{Id}}_E)\circ (\hbox{\upshape{Id}}_{V_{\mathbb{C}}^*}\otimes\Gamma_{\mathbb{C}}\otimes\hbox{\upshape{Id}}_E))\;(p\otimes \iota(w))$\\
$=(\hbox{tr}\otimes\hbox{\upshape{Id}}_E)(p\otimes (\Gamma_{\mathbb{C}}\otimes\hbox{\upshape{Id}}_E)(\iota(w)))=((p\circ\Gamma_{\mathbb{C}})\otimes\hbox{\upshape{Id}}_E)(\iota(w))=\zeta_u(p)(w)$\\
for every $p\in\mathcal{O}_m$ and $w\in W$. Thus, $\zeta_u$ is the symbol of a first-order differential operator $D_u:\mathcal{C}^{\infty}(\mathring{M},W)\longrightarrow \mathcal{C}^{\infty}(\mathring{M},E)$. For $\phi\in\mathcal{C}^{\infty}(\mathring{M},W)$,
$$D_u\phi(x)=\Xi(\phi(x)\;,\;-id\phi(x))=\Xi^{(1)}(-id\phi(x))$$
We denote this ``divergence-type" differential operator $D_u$ by $\delta_{\alpha,\beta}$.\\

In conclusion, we have the following theorem:\\

\begin{thm} \label{equations} Let $\mathcal{H}_{(m,\sigma)}$ be the positive energy irreducible unitary representation of the Poincar\'e group of mass $m>0$ and spin $\sigma$ (obtained by induction from an irreducible unitary representation $F$ of the little group $K$). Also, let $\mathbb{W}$ be the vector bundle on Minkowski spacetime $\mathring{M}$ associated to the representation $W=\hbox{\upshape{Sym}}^{2\alpha}S_+^*\otimes\hbox{\upshape{Sym}}^{2\beta}S_-^*$ of the group $\hbox{\upshape{Spin}}(V)$. Assume that $F$ appears in the decomposition of $W$ under $K$ (which is equivalent to $\alpha+\beta\geq\sigma$). Then $\mathcal{H}_{(m,\sigma)}$ is selected by the condition $\;\zeta_u(p)(\widehat{\phi}(p))=0\;$ in momentum space, and by the following equations in spacetime:
$$\left\{\begin{array}{rcl} (\square+m^2)\phi & = & 0\\
\delta_{\alpha,\beta}\phi & = & 0  \end{array}\right.$$
where $\;\delta_{\alpha,\beta}:\mathcal{C}^{\infty}(\mathring{M},\hbox{\upshape{Sym}}^{2\alpha}S_+^*\otimes\hbox{\upshape{Sym}}^{2\beta}S_-^*)\longrightarrow \mathcal{C}^{\infty}(\mathring{M},\hbox{\upshape{Sym}}^{2\alpha-1}S_+^*\otimes\hbox{\upshape{Sym}}^{2\beta-1}S_-^*)\,.$\\
\end{thm}

\section{Unitary dual of the super-Poincar\'e group}

In this section, we recall briefly some of the main aspects of the irreducible unitary representations of
the super Poincar\'e group. We start by recalling quickly the notions of super-Hilbert space, super-adjoint,
and the notion of unitary representation of super Lie groups (viewed as super Harish-Chandra pairs).
Here we follow closely \cite{cctv}, to which we refer the reader for more details. Then, we discuss
the main ingredients of the classification of superparticles, restricting our attention to the massive
case in even spacetime dimension, in particular in $d=4$ (which is the case of interest for us in the
coming sections).\\

\begin{defi}
A {\bf super Hilbert space} is a $\mathbb{Z}_2$-graded complex vector space\\ $\mathcal{H}=\mathcal{H}_0\oplus\mathcal{H}_1$ equipped with a scalar product $\langle,\rangle:\mathcal{H}\times\mathcal{H}\longrightarrow\mathbb{C}$ satisfying the following conditions:
\begin{enumerate}
\item $\langle,\rangle$ is an even map (and so $\langle\mathcal{H}_0,\mathcal{H}_1\rangle=\langle\mathcal{H}_1,\mathcal{H}_0\rangle=0$).
\item $\langle,\rangle$ is sesquilinear (we adopt the convention where sesquilinear forms are linear in the first argument and conjugate-linear in the second).
\item $\langle,\rangle$ has graded-Hermitian symmetry: $\langle \Psi',\Psi\rangle=(-1)^{|\Psi|\cdot |\Psi'|}\langle \Psi,\Psi'\rangle$ for all homogeneous $\Psi,\Psi'\in\mathcal{H}$.
\item $\langle,\rangle$ is positive-definite in the following sense: $\langle\Psi_0,\Psi_0\rangle >0$ for every $\Psi_0\in\mathcal{H}_0-\{0\}$ and $-i\langle\Psi_1,\Psi_1\rangle >0$ for every $\Psi_1\in\mathcal{H}_1-\{0\}$.\\
\end{enumerate}
\end{defi}

\begin{rem}
One can associate to $\langle,\rangle$ another scalar product $\langle,\rangle_0$ defined by: $\langle\Psi,\Psi'\rangle_0:=\langle\Psi,\Psi'\rangle$
if $\Psi,\Psi'\in\mathcal{H}_0$, $\langle\Psi,\Psi'\rangle_0:=-i\langle\Psi,\Psi'\rangle$ if $\Psi,\Psi'\in\mathcal{H}_1$ and
$\langle\Psi,\Psi'\rangle_0:=0$ if $\Psi$ and $\Psi'$ are of different parity. Then $(\mathcal{H},\langle,\rangle_0)$ is
an ordinary Hilbert space, in which the subspaces $\mathcal{H}_0$ and $\mathcal{H}_1$ are orthogonal. \\
\end{rem}

Recall that a densely defined operator $T:D\longrightarrow\mathcal{H}$ is said to be {\it even} (resp. {\it odd}) if $D=(D\cap\mathcal{H}_0)\oplus (D\cap\mathcal{H}_1)$ and for every $\Psi\in D\cap\mathcal{H}_k$ ($k\in\{0,1\}$), $T(\Psi)\in\mathcal{H}_k$ (resp. $T(\Psi)\in\mathcal{H}_{1-k}$).\\

\begin{defi}
Let $T:D\longrightarrow\mathcal{H}$ be a densely defined operator. The {\bf super-adjoint} of $T$ is the linear operator $T^{\dagger}:D^*\longrightarrow\mathcal{H}$ defined by $T^{\dagger}=T^*$ if $T$ is even and $T^{\dagger}=-iT^*$ if $T$ is odd (here, $T^*:D^*\longrightarrow\mathcal{H}$ is the adjoint of $T$ in $(\mathcal{H},\langle,\rangle_0)$).\\
\end{defi}

\begin{rem} It is not difficult to check that for $\Psi\in D$ and $\Psi'\in D^*$, we have $\langle T(\Psi),\Psi'\rangle=(-1)^{|T|\cdot |\Psi|}\langle\Psi,T^{\dagger}(\Psi')\rangle$.\\
\end{rem}

\begin{defi}
A {\bf super Harish-Chandra pair} is a pair $(G_0,\mathfrak{g})$ where $\mathfrak{g}=\mathfrak{g}_0\oplus\mathfrak{g}_1$ is
a super Lie algebra, and $G_0$ is a Lie group with Lie algebra
$\mathfrak{g}_0$, such that there is a linear action of $G_0$ on $\mathfrak{g}$ restricting to the adjoint action on
$\mathfrak{g}_0$ and whose differential is the adjoint action of $\mathfrak{g}_0$ on $\mathfrak{g}$. \\
\end{defi}

\begin{rem}\
\begin{enumerate}
\item We shall always assume in the present paper that the group $G_0$ is connected.
\item It is well-known ({\it cf.} \cite{dm} for instance) that the category of super Lie groups is equivalent to the category of super
Harish-Chandra pairs. In what follows, we will refer to super Harish-Chandra pairs as super Lie groups.\\
\end{enumerate}
\end{rem}

A finite-dimensional unitary representation of a super Lie group $(G_0,\mathfrak{g})$ is a pair $(\rho,\eta)$ where $\rho:G_0\longrightarrow \hbox{\upshape{U}}(F)$ is an even unitary representation of $G_0$ on a finite-dimensional super Hilbert space $F$, and $\eta:\mathfrak{g}\longrightarrow\mathfrak{gl}(F)$ is a morphism of super Lie algebras such that:
\begin{itemize}
\item $\eta_{|\mathfrak{g}_0}=\rho_*$
\item $\eta(gX)=\rho(g)\circ\eta(X)\circ\rho(g)^{-1}$ for all $g\in G_0$ and $X\in\mathfrak{g}$
\item $\eta(X)^{\dagger}=-\eta(X)$ for all $X\in\mathfrak{g}$
\end{itemize}
Define $\alpha:\mathfrak{g}_1\longrightarrow\mathfrak{gl}(F)_1$ by $\alpha(X):=e^{-i\frac{\pi}{4}}\eta(X)$. It is easy to check that for $X\in\mathfrak{g}_1$, the condition $\eta(X)^{\dagger}=-\eta(X)$ is equivalent to $\alpha(X)^*=\alpha(X)$. Moreover, since $\eta$ is a morphism of super Lie algebras, we have in particular $\eta([X,Y])=\eta(X)\circ\eta(Y)+\eta(Y)\circ\eta(X)$ for all $X,Y\in\mathfrak{g}_1$. Taking into account the fact that $\eta_{|\mathfrak{g}_0}=\rho_*$, this becomes $-i\rho_*([X,Y])=\alpha(X)\circ\alpha(Y)+\alpha(Y)\circ\alpha(X)$ for all $X,Y\in\mathfrak{g}_1$.\\

On the other hand, if $\rho:G\longrightarrow\hbox{U}(\mathcal{H})$ is a unitary representation of a Lie group $G$ on a Hilbert space $\mathcal{H}$, an element $\Psi\in\mathcal{H}$ is called a {\it smooth vector} for the representation $\rho$ if the map $g\mapsto \rho(g)(\Psi)$ from $G$ to $\mathcal{H}$ is smooth. We denote by $\mathcal{H}^{\infty}$ the subspace of smooth vectors of $\mathcal{H}$. It is clear that $\mathcal{H}^{\infty}$ is $G$-invariant, and one can define a representation of $\mathfrak{g}$ on $\mathcal{H}^{\infty}$ by setting $X\Psi:=\frac{d}{dt}[\pi(e^{tX})(\Psi)]_{|t=0}$ for every $X\in\mathfrak{g}$ and $\Psi\in\mathcal{H}^{\infty}$. In addition, $\mathcal{H}^{\infty}$ is dense in $\mathcal{H}$.\\

We are ready to define unitary representations of super Lie groups on general, possibly infinite-dimensional, super Hilbert spaces.\\

\begin{defi}
A {\bf unitary representation} of a super Lie group $(G_0,\mathfrak{g})$ is a pair $(\rho,\alpha)$ where $\rho:G_0\longrightarrow \hbox{\upshape{U}}(\mathcal{H})$ is an even unitary representation of $G_0$ on a super Hilbert space $\mathcal{H}$, and $\alpha:\mathfrak{g}_1\longrightarrow\mathfrak{gl}(\mathcal{H}^{\infty})_1$ is a linear map such that:
\begin{itemize}
\item $-i\rho_*([X,Y])=\alpha(X)\circ\alpha(Y)+\alpha(Y)\circ\alpha(X)$ for all $X,Y\in\mathfrak{g}_1$
\item $\alpha(gX)=\rho(g)\circ\alpha(X)\circ\rho(g)^{-1}$ for all $g\in G_0$ and $X\in\mathfrak{g}_1$
\item $\alpha(X):\mathcal{H}^{\infty}\longrightarrow \mathcal{H}^{\infty}$ is a symmetric odd operator for all $X\in\mathfrak{g}_1$\\
\end{itemize}
\end{defi}

\begin{rem}
If $(\rho,\alpha)$ is a unitary representation of $(G_0,\mathfrak{g})$ on $\mathcal{H}$, then the linear map $\eta:\mathfrak{g}\longrightarrow\mathfrak{gl}(\mathcal{H}^{\infty})$ defined by $\eta(X):=\rho_*(X_0)+e^{i\frac{\pi}{4}}\alpha(X_1)$ for all $X=X_0+X_1\in\mathfrak{g}$ is a morphism of super Lie algebras. \\
\end{rem}

Let $S$ be an irreducible real Clifford module for $V$. The action of the Clifford algebra $\hbox{C}\ell(V)$ on $S$ induces the spin representation $\hbox{Spin}(V)\longrightarrow\hbox{GL}(S)$, and there is a $\hbox{Spin}(V)$-equivariant symmetric morphism $\Gamma:S^*\otimes S^*\longrightarrow V$ which is positive in the sense that $\Gamma(s\otimes s)\in\bar{C}_+$ for all $s\in S^*$, and definite ($\Gamma(s\otimes s)=0\Longleftrightarrow s=0$).\\

The {\it super-Poincar\'e algebra} of $V$ is the super Lie algebra $\mathfrak{s}\pi(V):=(\mathfrak{spin}(V)\oplus V)\oplus S^*$, the super-bracket being defined as follows:\\
$[\mathfrak{spin}(V),\mathfrak{spin}V]\subset\mathfrak{spin}V$ is given by the ordinary Lie bracket of the Lie algebra $\mathfrak{spin}(V)$,\\
$[\mathfrak{spin}(V),V]\subset V$ is given by the standard representation of $\mathfrak{spin}(V)$ on $V$,\\
$[\mathfrak{spin}(V),S^*]\subset S^*$ is given by the spin representation $\mathfrak{spin}(V)\longrightarrow\mathfrak{gl}(S^*)$,\\
$[V,V]=[V,S^*]=0$,\\
$[S^*,S^*]\subset V$ is given by $\Gamma$: for $s_1,s_2\in S^*$, $[s_1,s_2]:=-2\Gamma(s_1\otimes s_2)$.\\

The {\it super Poincar\'e group} of $V$ is the super Lie group $\hbox{S}\Pi(V)=(\Pi(V),\mathfrak{s}\pi(V))$. \\

We recall now the main steps leading to the classification of irreducible unitary representations of the super Poincar\'e group $\hbox{S}\Pi(V)$. In sections 5 and 6, we will often abbreviate the notations by using the letter $G$ for the super Poincar\'e group $\hbox{S}\Pi(V)$, and the letter $H$ for the super Lie group $(\hbox{Spin}(V),\mathfrak{spin}(V)\oplus S^*)$. \\

Let $(\rho,\alpha)$ be an irreducible unitary representation of $\hbox{S}\Pi(V)$ on a super Hilbert space $\mathcal{H}$, so that $\rho:\Pi(V)\longrightarrow \hbox{\upshape{U}}(\mathcal{H})$ is an even unitary representation of $\Pi(V)$ on $\mathcal{H}$, and $\alpha:S^*\longrightarrow\mathfrak{gl}(\mathcal{H}^{\infty})_1$ is a linear map such that:
\begin{itemize}
\item $-i\rho_*([s_1,s_2])=\alpha(s_1)\circ\alpha(s_2)+\alpha(s_2)\circ\alpha(s_1)$ for all $s_1,s_2\in S^*$
\item $\alpha(gs)=\rho(g)\circ\alpha(s)\circ\rho(g)^{-1}$ for all $g\in \Pi(V)$ and $s\in S^*$
\item $\alpha(s):\mathcal{H}^{\infty}\longrightarrow \mathcal{H}^{\infty}$ is a symmetric odd operator for all $s\in S^*$\\
\end{itemize}

Similarly to the case of the Poincar\'e group, we start by looking at the action of the translation subgroup $V\subset G$. The corresponding spectral measure $P$ on $V^*$ is $H$-equivariant ($H$ acts on $V^*$ through $\hbox{Spin}(V)$, the action of $S^*$ being trivial), and its support $\mathcal{O}$ (the spectrum of $(\rho,\alpha)$) is $H$-invariant. In fact, $\mathcal{O}$ is an orbit for the action of $\hbox{Spin}(V)$ on $V^*$ (by irreducibility of $(\rho,\alpha)$).\\

Next, choose a preferred point $q\in\mathcal{O}$ and consider the sub super Lie group $\tilde{K}:=(K,\mathfrak{k}\oplus S^*)$ of $H$, where $K$ is the stabilizer of $q$ under the action of $\hbox{Spin}(V)$. Then $\tilde{K}$ is the stabilizer of $q$ under the action of $H$. Also, let $F:=\bigcap_{v\in V}\hbox{Ker}(\rho_{|V}(v)-e^{iq(v)}\hbox{\upshape{Id}}_{\mathcal{H}})$. It is not difficult to check that $F$ is invariant under $K$, and since $[V,S^*]=0$, $F$ is invariant under $S^*$ as well. Thus, $F$ is invariant under $\tilde{K}$, and the irreducibility of $(\rho,\alpha)$ implies that the unitary representation $(\rho^F_{|K},\alpha^F)$ of $\tilde{K}$ on $F$ is irreducible.\\

Thus, to every $[(\rho,\alpha)]\in\widehat{G}$, one can associate a pair $(\mathcal{O},[(\rho^F_{|K},\alpha^F)])$ where $\mathcal{O}\in V^*/\hbox{Spin}(V)$ and $[(\rho^F_{|K},\alpha^F)]\in\widehat{\tilde{K}}$. \\

In fact, if $p\in\mathcal{O}$, and $\mathcal{H}_p:=\bigcap_{v\in V}\hbox{Ker}(\rho_{|V}(v)-e^{ip(v)}\hbox{\upshape{Id}}_{\mathcal{H}})=\bigcap_{v\in V}\hbox{Ker}(\rho_*(v)-ip(v)\hbox{\upshape{Id}}_{\mathcal{H}})$, then $\eta(s_1)\circ\eta(s_2)+ \eta(s_2)\circ\eta(s_1)=\rho_*([s_1,s_2])$ implies $\eta(s)\circ\eta(s)=-\rho_*(\Gamma(s\otimes s))$. On $\mathcal{H}_p$, this becomes $\eta(s)\circ\eta(s)=-ip(\Gamma(s\otimes s))\;\hbox{\upshape{Id}}_{\mathcal{H}_p}$. Now for each $s\in S^*$, the operator $\eta(s)$ is odd super antihermitian, and therefore has its spectrum on the second bisector $\mathbb{R}e^{-i\frac{\pi}{4}}$. It follows that $\eta(s)\circ\eta(s)$ is even super Hermitian, and therefore has its spectrum on the half-line $\mathbb{R}_+(-i)$. Consequently, $p(\Gamma(s\otimes s))\geq 0$ for all $s\in S^*$, and therefore $p\in C_+^{\vee}$ (since $\Gamma$ is positive). This forces the orbit $\mathcal{O}$ to be contained in the forward timelike cone $C_+^{\vee}$. Such an orbit will be called {\it admissible}. We see that contrary to the case of the Poincar\'e group, there is already a restriction on the type of orbit that can arise. In other words, supersymmetry already implies the positivity of the energy.\\

Conversely, given a pair $(\mathcal{O},(\lambda,\beta))$ where $\mathcal{O}$ is an admissible orbit and $(\lambda,\beta)$ is an irreducible unitary representation of $\tilde{K}$ on a finite-dimensional super Hilbert space $F$, one can define an irreducible unitary representation $(\rho,\alpha)$ of $G$ on a super Hilbert space as follows. First, one induces a unitary representation $\mathcal{H}$ of $H$ from the unitary representation of $\tilde{K}$ on $F$. Concretely, $\mathcal{H}:=\hbox{ind}_{\tilde{K}}^H F$ may be defined as follows. Let $\mathbb{H}$ be the $H$-equivariant Hermitian super vector bundle over $\mathcal{O}$ associated to the principal $\tilde{K}$-bundle $H\longrightarrow\mathcal{O}$ by the representation $(\lambda,\beta)$ of $\tilde{K}$ on $F$. Then let $\mathcal{H}$ be the space of $L^2$ sections of $\mathbb{H}$: $$\mathcal{H}:=\Gamma_{L^2}(\mathcal{O},\mathbb{H})$$ (remark that $\mathcal{O}$ has an $H$-invariant measure, but otherwise one could have used half-densities). Then we get a unitary representation of $H$ on $\mathcal{H}$, and if we make $v\in V$ act by $(v\cdot\Psi)_p:=e^{ip(v)}\Psi_p$, we obtain an irreducible unitary representation $(\rho,\alpha)$ of $G$ on $\mathcal{H}$.\\

\begin{rem} Alternatively, given an admissible orbit $\mathcal{O}$, the data of a unitary representation $(\rho,\alpha)$ of $G$ with spectrum $\mathcal{O}$ is equivalent to the data of a pair $(\gamma,P)$ where $\gamma$ is a unitary representation of $H$ and $P$ is an $H$-equivariant projection-valued measure on $\mathcal{O}$. Such a pair $(\gamma,P)$ (``super system of imprimitivity") is in turn equivalent to a unitary representation of $\tilde{K}$, by a super version of the imprimitivity theorem ({\it cf.} \cite{cctv}). In particular, the following are equivalent: irreducible unitary representations of $G$ with spectrum $\mathcal{O}$, irreducible systems of imprimitivity on $\mathcal{O}$, and irreducible unitary representations of $\tilde{K}$.\\
\end{rem}

Now we need to investigate the structure of $F=\mathcal{H}_q$. We will do this only in the massive case, that is when the orbit $\mathcal{O}$ is a sheet of hyperboloid $\;\mathcal{O}_m:=\{p\in V^*\;|\;\langle p,p\rangle =m^2\}\cap C^{\vee}_+$\\ for some $m>0$. We will also restrict ourselves very soon to the case where $d$ is even, and then to $d=4$, since this the case where we will focus our study in sections 5 and 6.\\

Let $\langle,\rangle_q:=-^t\Gamma(q)\in S\otimes S$ (so that $\langle s_1,s_2\rangle_q=-q(\Gamma(s_1\otimes s_2))$ for all $s_1,s_2\in S^*$). \\

\begin{prop}\
\begin{enumerate}
\item The pairing $\langle,\rangle_q$ is a negative-definite inner product on $S^*$.
\item The subspace $F$ is a Clifford module for $(S^*,\langle,\rangle_q)$.
\end{enumerate}
\end{prop}

\noindent {\bf Proof~:} 1. The symmetry of $\langle,\rangle_q$ is a consequence of that of $\Gamma$. For all $s\in S^*$, we have $\langle s,s\rangle_q=-q(\Gamma(s\otimes s))\leq 0$ since $q\in C^{\vee}_+$ and $\Gamma(s\otimes s)\in \bar{C}_+$ (positivity of $\Gamma$). On the other hand, if $\langle s,s\rangle_q=0$, then $q(\Gamma(s\otimes s))=0$, and so $\Gamma(s\otimes s)\in \hbox{Ker}\;q\cap \bar{C}_+$. But $\hbox{Ker}\;q\cap \bar{C}_+=\{0\}$ since $q\in C^{\vee}_+$. Thus, $\Gamma(s\otimes s)=0$, and so $s=0$ by definiteness of $\Gamma$.\\ 2. For all $s_1,s_2\in S^*$ and $\Psi\in F$, we have $(\eta(s_1)\circ\eta(s_2)+\eta(s_2)\circ\eta(s_1))(\Psi)=[\eta(s_1),\eta(s_2)](\Psi)=\eta([s_1,s_2])(\Psi)=\eta(-2\Gamma(s_1\otimes s_2))(\Psi)=iq(-2\Gamma(s_1\otimes s_2))\Psi=2i\langle s_1,s_2\rangle_q\Psi$. Thus,
$$\eta(s_1)\circ\eta(s_2)+\eta(s_2)\circ\eta(s_1)=2i \langle s_1,s_2\rangle_q\;\hbox{\upshape{Id}}_{F}$$ $\hfill\square$\\

From now on, we assume that the dimension $d$ of $V$ is even.

Since $S^*$ is even dimensional, it has a unique irreducible Clifford module, constructed as
follows. First, we complexify $S^*$: let $S_{\mathbb{C}}^*:=S^*\otimes_{\mathbb{R}}\mathbb{C}$.
Since $d$ is even, then we know that $S_{\mathbb{C}}^*$ decomposes into two inequivalent irreducible representations of $\hbox{Spin}(V)$:
$$S_{\mathbb{C}}^*=S_{+}^*\oplus S_{-}^*$$
If $\Gamma_{\mathbb{C}}:S_{\mathbb{C}}^*\otimes S_{\mathbb{C}}^*\longrightarrow V_{\mathbb{C}}$ is the
complexification of $\Gamma:S^*\otimes S^*\longrightarrow V$, then
$(\Gamma_{\mathbb{C}})_{|S_+^*\otimes S_+^*}=(\Gamma_{\mathbb{C}})_{|S_-^*\otimes S_-^*}=0$, so we are left
with $\Gamma_{\mathbb{C}}:S_+^*\otimes S_-^*\longrightarrow V_{\mathbb{C}}$.
As a result, $\langle s_+,s_+'\rangle_q=\langle s_-,s_-'\rangle_q=0$ for all $s_+,s_+'\in S_+^*$ and $s_-,s_-'\in S_-^*$,
and $\langle,\rangle_q:S_+^*\times S_-^*\longrightarrow \mathbb{C}$ is nondegenerate, since
$\langle,\rangle_q:S_{\mathbb{C}}^*\times S_{\mathbb{C}}^*\longrightarrow \mathbb{C}$ is nondegenerate.
This shows that $S_{\mathbb{C}}^*=S_{+}^*\oplus S_{-}^*$ is a decomposition of $S_{\mathbb{C}}^*$
into two maximal isotropic subspaces.\\

$\bigwedge^{\bullet}S_{+}^*$ is the irreducible Clifford module we were looking for. We could have
chosen $\bigwedge^{\bullet}S_{-}^*$, which is equivalent. An element $s=s_++s_-\in S_{\mathbb{C}}^*=S_{+}^*\oplus S_{-}^*$
acts by sending $a\in\bigwedge^{\bullet}S_{+}^*$ to $s_+\wedge a+i_{s_-}a$. In particular,
we see that the elements of $S_+^*$ act on $\bigwedge^{\bullet}S_{+}^*$ as creation operators
(sending $\bigwedge^{k}S_{+}^*$ to $\bigwedge^{k+1}S_{+}^*$), whereas the elements of $S_-^*$ act
on $\bigwedge^{\bullet}S_{+}^*$ as annihilation operators (sending $\bigwedge^{k}S_{+}^*$ to $\bigwedge^{k-1}S_{+}^*$).\\

Going back to the Clifford module $F$, we deduce that it is necessarily a direct sum of copies of the
unique irreducible Clifford module $\bigwedge^{\bullet}S_{+}^*$. Thus, we have\\

 $\hspace{60mm}F\simeq \bigwedge^{\bullet}S_{+}^*\otimes E$\\

for some finite-dimensional vector space $E$ on which $S^*$ does not act, and whose dimension is
the multiplicity of $\bigwedge^{\bullet}S_{+}^*$ in $F$.\\

Note that we have $F_0\simeq \bigwedge^{even}S_{+}^*\otimes E$ and $F_1\simeq \bigwedge^{odd}S_{+}^*\otimes E$. Moreover, we see immediately (from its explicit description given above) that the action of $S^*$ exchanges $F_0$ and $F_1$.\\

Now we need to determine the possibilities for the vector space $E$. To this end, we notice that if $K$ is the stabilizer of $q$ under the action of $\hbox{Spin}(V)$ on $\mathcal{O}_m$, we have a natural unitary representation of $K$ on $F$ (since $\rho(k)(\mathcal{H}_{q})=\mathcal{H}_{k\cdot q}=\mathcal{H}_{q}$ for all $k\in K$). It is not difficult to check that since $m>0$, we have $K\simeq \hbox{Spin}(d-1)$.\\

\begin{prop}
The vector space $E$ carries an irreducible representation of $K$.
\end{prop}

\noindent {\bf Proof~:} Suppose $K$ acted reducibly on $E$. Then $\mathfrak{k}\oplus S^*$ would act reducibly on $F$. But then $(\rho,\alpha)$ would itself be reducible, contrary to our assumption. $\hfill\square$\\

Consequently, we have\\

 $\hspace{60mm}F\simeq \bigwedge^{\bullet}S_{+}^*\otimes E^{(\sigma)}$\\

for some highest weight $\sigma$, called the {\it superspin} of the representation $(\rho,\alpha)$.\\

In conclusion, the irreducible unitary representations of the super-Poincar\'e group are classified by the mass and the superspin.\\

Of course, $F$ is reducible under $K$: $$F=\bigoplus_{\omega\in \Delta_{K}}E^{(\omega)}$$ for some spectrum $\Delta_{K}$ that has to be determined, where $E^{(\omega)}$ is the irreducible representation of $K$ of highest weight $\omega$. For each $\omega\in\Delta_{K}$, the irreducible unitary representation $E^{(\omega)}$ of $K$ induces an irreducible unitary representation $\mathcal{H}^+_{(m,\omega)}$ of the Poincar\'e group $\Pi(V)$. It is the representation of mass $m$ and spin $\omega$, and it is a direct factor of $\mathcal{H}$. Namely,
$$\mathcal{H}^{(m,\sigma)}=\bigoplus_{\omega\in\Delta_{K}}\mathcal{H}_{(m,\omega)}^+$$

Let us illustrate the above in the case $d=4$.\\

 Then $\hbox{dim}_{\mathbb{R}}S^*=4$ and $\hbox{dim}_{\mathbb{C}}S_{+}^*=\hbox{dim}_{\mathbb{C}}S_{-}^*=2$. Morover, $K\simeq \hbox{Spin}(3)\simeq\hbox{SU}(2)$.\\

Consider an irreducible unitary representation $(\rho,\alpha)$ of $\hbox{S}\Pi(V)$ of mass $m$ and superspin $\sigma$. Then\\

 $\hspace{60mm} F\simeq \bigwedge^{\bullet}S_{+}^*\otimes\hbox{Sym}^{2\sigma}S_+^*$\\

For simplicity, let us consider first the special case of superspin 0, so that $F$ is an irreducible Clifford module for $S^*$:\\

 $\hspace{60mm} F\simeq \bigwedge^{\bullet}S_{+}^*$\\

We can describe explicitly the Clifford action on $\bigwedge^{\bullet}S_{+}^*=\mathbb{C}\oplus S_+^*\oplus\bigwedge^2 S_+^*$. For instance, the element $s_+\in S_+^*$ acts as a creation operator, sending the vacuum $1\in\mathbb{C}$ to $s_+\in S_+^*$, while the element $s'_+\cdot s_+$ of the Clifford algebra sends $1$ to $s'_+\wedge s_+\in \bigwedge^2 S_+^*$. On the other hand, the element $s_-\in S_-^*$ acts as an annihilation operator, sending the vacuum $1\in\mathbb{C}$ to $0$, the vector $s_+\in S_+^*$ to $\langle s_+,s_-\rangle_p\in\mathbb{C}$, and the bivector $s_+\wedge s'_+\in \bigwedge^2 S_+^*$ to $\langle s'_+,s_-\rangle_p\;s_+\in S_+^*$.\\

The decomposition of $F$ under $K$ is $$F=(E^{(0)}\otimes\mathbb{C}^2)\oplus E^{(\frac{1}{2})}$$
Indeed, this is just the decomposition $\bigwedge^{\bullet}S_{+}^*=\mathbb{C}\oplus S_+^*\oplus\bigwedge^2 S_+^*$, where each of $\mathbb{C}$ and $\bigwedge^2 S_+^*$ carries the trivial one-dimensional representation of $\hbox{SU}(2)$ (of spin 0), and $S_+^*$ carries the standard (two-dimensional) representation of $\hbox{SU}(2)$ (of spin $\frac{1}{2}$). \\

Thus, a superparticle of superspin 0 contains two particles of spin 0 and one particle of spin $\frac{1}{2}$. Note that $\hbox{dim}\;F_0=\hbox{dim}(E^{(0)}\otimes\mathbb{C}^2)=2$ and $\hbox{dim}\;F_1=\hbox{dim}(E^{(\frac{1}{2})})=2$, so we have indeed equality between the bosonic and fermonic degrees of freedom.\\

Now we turn to the case of a superspin $\sigma>0$. Then \\

 $F\simeq \bigwedge^{\bullet}S_{+}^*\otimes\hbox{Sym}^{2\sigma}S_+^*=(\mathbb{C}\otimes\hbox{Sym}^{2\sigma}S_+^*)\oplus(S_+^*\otimes\hbox{Sym}^{2\sigma}S_+^*)\oplus(\bigwedge^2 S_+^*\otimes\hbox{Sym}^{2\sigma}S_+^*)$\\

The first and the third term are clearly equivalent to $\hbox{Sym}^{2\sigma}S_+^*$; we just need to
decompose the second term into irreducible representations of $\hbox{SU}(2)$. To this end, recall that\\ $S_+^*=L_{-\frac{1}{2}}\oplus L_{\frac{1}{2}}$ and $\hbox{Sym}^{2\sigma}S_+^*=L_{-\sigma}\oplus L_{-\sigma+1}\oplus L_{-\sigma+2}\oplus\dots\oplus L_{\sigma-2}\oplus L_{\sigma-1}\oplus L_{\sigma}$.\\ Taking the tensor product, and using the fact that $L_j\otimes L_k=L_{j+k}$, we obtain that
$$S_+^*\otimes\hbox{Sym}^{2\sigma}S_+^*=L_{-\sigma-\frac{1}{2}}\oplus 2L_{-\sigma+\frac{1}{2}}\oplus 2L_{-\sigma+\frac{3}{2}}\oplus\dots\oplus 2L_{\sigma-\frac{3}{2}}\oplus 2L_{\sigma-\frac{1}{2}}\oplus L_{\sigma+\frac{1}{2}}$$
This shows that $$S_+^*\otimes\hbox{Sym}^{2\sigma}S_+^*=(\hbox{Sym}^{2\sigma-1}S_+^*)\oplus (\hbox{Sym}^{2\sigma+1}S_+^*)$$

The decomposition of $F$ under $K$ becomes $$F=(E^{(\sigma)}\otimes\mathbb{C}^2)\oplus E^{(\sigma-\frac{1}{2})}\oplus E^{(\sigma+\frac{1}{2})}$$

Thus, a superparticle of superspin $\sigma>0$ contains two particles of spin $\sigma$, one
particle of spin $\sigma-\frac{1}{2}$ and one particle of spin $\sigma+\frac{1}{2}$.
Note that $\hbox{dim}(E^{(\sigma)}\otimes\mathbb{C}^2)=2(2\sigma+1)=4\sigma+2$ and $\hbox{dim}(E^{(\sigma-\frac{1}{2})}\oplus E^{(\sigma+\frac{1}{2})})=2\sigma+(2\sigma+2)=4\sigma+2$, so we have indeed equality between the bosonic and fermonic degrees of freedom.\\

\section{Minkowski superspacetime in dimension $(4|4)$} \label{superspacetime}

In this section, we recall the definition and some properties of Minkowski
superspacetime in dimension $(4|4)$, focusing on the construction of the
fundamental (resp. invariant) vector fields associated to supertranslations. At the end of
this section, we discuss $W$-valued superfunctions (playing the role of
``spin-tensor superfields"),
and make the link with the irreducible unitary representations of the super Poincar\'e group.\\

We think of $S_+^*$ as being isomorphic to $\mathbb{C}^2$ with the standard action of
$\hbox{SL}_{2}(\mathbb{C})_{\mathbb{R}}$ (call this representation $\rho_+$), and of $S_-^*$
as being isomorphic to $\mathbb{C}^2$ on which $A\in\hbox{SL}_{2}(\mathbb{C})_{\mathbb{R}}$ acts
by left multiplication with $-A^{\dagger}$ (call this representation $\rho_-$). The representation
$S_-^*$ is conjugate to $S_+^*$: the map $\zeta:(\overline{\mathbb{C}^2},\rho_+)\longrightarrow (\mathbb{C}^2,\rho_-)$
defined by $\zeta(z_1,z_2)=(-i\bar{z}_2,i\bar{z}_1)$ is a $\mathbb{C}$-linear, $\hbox{SL}_{2}(\mathbb{C})_{\mathbb{R}}$-equivariant
isomorphism. Equivalently, we may view $\zeta$ as a $\mathbb{C}$-antilinear, $\hbox{SL}_{2}(\mathbb{C})_{\mathbb{R}}$-equivariant
isomorphism from $(\mathbb{C}^2,\rho_+)$ to $(\mathbb{C}^2,\rho_-)$.
Note that we have $\zeta^2=\hbox{\upshape{Id}}_{\mathbb{C}^2}$. Consider $S_{\mathbb{C}}^*=S_+^*\oplus S_-^*$
with the direct sum representation, and let $c_1:S_{\mathbb{C}}^*\longrightarrow S_{\mathbb{C}}^*$
be defined as follows: $c_1(z_1,z_2,z_3,z_4)=(\zeta(z_3,z_4),\zeta(z_1,z_2))$. Then $c_1$ is a $\mathbb{C}$-antilinear,
$\hbox{SL}_{2}(\mathbb{C})_{\mathbb{R}}$-equivariant automorphism of $S_{\mathbb{C}}^*$ which satisfies
$(c_1)^2=\hbox{\upshape{Id}}_{S_{\mathbb{C}}^*}$. In other words, $c_1$ is a conjugation of the representation $S_{\mathbb{C}}^*$.
We obtain a real irreducible representation of $\hbox{SL}_{2}(\mathbb{C})_{\mathbb{R}}$ by taking
$S^*=\hbox{Ker}(c_1-\hbox{\upshape{Id}}_{S_{\mathbb{C}}})$, so that $S^*=\{(z_1,z_2,\zeta(z_1,z_2))\;;\;(z_1,z_2)\in\mathbb{C}^2\}$.\\

Choose a basis $\{f^1,f^2\}$ of $S_+^*$ (for e.g. $f^1:=(1,0,0,0)$ and $f^2:=(0,1,0,0)$). Let $\bar{f}^{1}=c_1(f^1)$ ($=(0,0,0,i)$) and $\bar{f}^{2}=c_1(f^2)$ ($=(0,0,-i,0)$). Then $\{\bar{f}^{1},\bar{f}^{2}\}$ is a basis of $S_-^*$.
Also, $\{f^1,f^2,\bar{f}^{1},\bar{f}^{2}\}$ is a basis of $S_{\mathbb{C}}^*$, while the elements $f^1+\bar{f}^{1}$ and $f^2+\bar{f}^{2}$ belong to $S^*$.\\

We define {\bf complex superspacetime} as being the complex supermanifold $M_{cs}=(\mathring{M},\mathcal{O}_{M_{cs}})$,
where $\mathcal{O}_{M_{cs}}(U):=\mathcal{C}^{\infty}(U,\mathbb{C})\otimes \bigwedge S_{\mathbb{C}}^*$ for every open
set $U\subset\mathring{M}$. A superfunction on $M_{cs}$ can be written:

$f=\varphi+\psi_a\theta^a+\eta_{a}\bar{\theta}^{a}+F\theta^1\theta^2+G\bar{\theta}^{1}\bar{\theta}^{2}+iA_{\mu}\Gamma^{\mu}_{ab}\theta^a\bar{\theta}^{b}+\lambda_{a}\theta^1\theta^2\bar{\theta}^{a}+\mu_{a}\bar{\theta}^{1}\bar{\theta}^{2}\theta^a+H\theta^1\theta^2\bar{\theta}^{1}\bar{\theta}^{2}$\\

where $\varphi\;,\psi_a,\;\eta_{a},\;F,\;G,\;A_{\mu},\;\lambda_{a},\;\mu_{a},\;H\in \mathcal{C}^{\infty}(U,\mathbb{C})$ (we have $\hbox{dim}\bigwedge^{\bullet}S_{\mathbb{C}}^*=16$ complex-valued functions).\\

There is a canonical conjugation $c=(\check{c},c^{\sharp})$ on $M_{cs}$, where $\check{c}=\hbox{\upshape{Id}}_{\mathring{M}}$, and for every open set $U\subset\mathring{M}$, $c^{\sharp}_U:\mathcal{C}^{\infty}(U,\mathbb{C})\otimes \bigwedge S_{\mathbb{C}}^*\longrightarrow \mathcal{C}^{\infty}(U,\mathbb{C})\otimes \bigwedge S_{\mathbb{C}}^*$ is defined by:\\

$c^{\sharp}_U(f)=\bar{\varphi}+\bar{\psi}_a\bar{\theta}^a+\bar{\eta}_{a}\theta^{a}+\bar{F}\bar{\theta}^1\bar{\theta}^2+\bar{G}\theta^{1}\theta^{2}-i\bar{A}_{\mu}\overline{\Gamma^{\mu}_{ab}}\bar{\theta}^a\theta^{b}+\bar{\lambda}_{a}\bar{\theta}^1\bar{\theta}^2\theta^{a}+\bar{\mu}_{a}\theta^{1}\theta^{2}\bar{\theta}^a+\bar{H}\bar{\theta}^1\bar{\theta}^2\theta^{1}\theta^{2}$\\

Now $\overline{\Gamma^{\mu}_{ab}}=\Gamma^{\mu}_{ba}$, $\bar{\theta}^a\theta^b=-\theta^b\bar{\theta}^a$ and $\bar{\theta}^1\bar{\theta}^2\theta^{1}\theta^{2}=\theta^{1}\theta^{2}\bar{\theta}^1\bar{\theta}^2$. Setting $\bar{f}:=c^{\sharp}_U(f)$, we get:\\

$\bar{f}=\bar{\varphi}+\bar{\psi}_a\bar{\theta}^a+\bar{\eta}_{a}\theta^{a}+\bar{F}\bar{\theta}^1\bar{\theta}^2+\bar{G}\theta^{1}\theta^{2}+i\bar{A}_{\mu}\Gamma^{\mu}_{ab}\theta^{a}\bar{\theta}^b+\bar{\lambda}_{a}\bar{\theta}^1\bar{\theta}^2\theta^{a}+\bar{\mu}_{a}\theta^{1}\theta^{2}\bar{\theta}^a+\bar{H}\theta^{1}\theta^{2}\bar{\theta}^1\bar{\theta}^2$\\

We obtain {\bf real superfunctions} by imposing $\bar{f}=f$, which gives:\\

$\bar{\varphi}=\varphi$, $\eta_a=\bar{\psi}_a$, $G=\bar{F}$, $\bar{A}_{\mu}=A_{\mu}$, $\mu_a=\bar{\lambda}_a$ and $\bar{H}=H$. Thus,\\

$f=\varphi+\psi_a\theta^a+\bar{\psi}_a\bar{\theta}^{a}+F\theta^1\theta^2+\bar{F}\bar{\theta}^{1}\bar{\theta}^{2}+iA_{\mu}\Gamma^{\mu}_{ab}\theta^a\bar{\theta}^{b}+\lambda_{a}\theta^1\theta^2\bar{\theta}^{a}+\bar{\lambda}_{a}\bar{\theta}^{1}\bar{\theta}^{2}\theta^a+H\theta^1\theta^2\bar{\theta}^{1}\bar{\theta}^{2}$\\

where $\psi_a,\;F,\;\lambda_a\in\mathcal{C}^{\infty}(U,\mathbb{C})$ and $\varphi,\;A_{\mu},\;H\in\mathcal{C}^{\infty}(U,\mathbb{R})$. We have 5 complex-valued functions and 6 real-valued functions; this corresponds to $\hbox{dim}\bigwedge^{\bullet}S^*=16$ real degrees of freedom.\\

\begin{rem}
Note that $M_{cs}$ is the linear complex supermanifold associated with the super-vector space $V_{\mathbb{C}}\oplus S_{\mathbb{C}}$. We write $M_{cs}=\hbox{\upshape{L}}((V_{\mathbb{C}}\oplus S_{\mathbb{C}})$. It represents the functor $\mathcal{L}(V_{\mathbb{C}}\oplus S_{\mathbb{C}}):\mathbf{sMan}_{fd}^{cs}\longrightarrow \mathbf{Set}$ defined by $\mathcal{L}(V_{\mathbb{C}}\oplus S_{\mathbb{C}})(B)=(\mathcal{O}_B(|B|)_0\otimes V)\oplus (\mathcal{O}_B(|B|)_1\otimes S_{\mathbb{C}})$\\
$=(\mathcal{O}_B(|B|)_0\otimes V)\oplus (\mathcal{O}_B(|B|)_1\otimes S_+)\oplus (\mathcal{O}_B(|B|)_1\otimes S_-)$ for every complex supermanifold $B$.\\
\end{rem}

\begin{prop}
$M_{cs}$ has a natural structure of complex super Lie group, whose associated super Lie algebra is the
complex super Lie algebra $V_{\mathbb{C}}\oplus S_{\mathbb{C}}$. 
\end{prop}

\noindent {\bf Proof~:} We want to show that $M_{cs}$ is a group object in the
category of complex supermanifolds. This is equivalent to showing that the
image of the functor $\mathcal{L}(V_{\mathbb{C}}\oplus S_{\mathbb{C}})$ is contained in
the category of groups. Let $B$ be a complex supermanifold. The group structure
on $\mathcal{L}(V_{\mathbb{C}}\oplus S_{\mathbb{C}})(B)$ comes from the super
Lie algebra structure on $V_{\mathbb{C}}\oplus S_{\mathbb{C}}$, via the
Campbell-Baker-Hausdorff formula (the exponential map being the identity, and
the triple brackets vanishing). Namely, if $u=v^{\mu}\otimes e_{\mu}+s^a\otimes
f_a+t^b\otimes \bar{f}_b$ and $u'=v'^{\mu}\otimes e_{\mu}+s'^a\otimes
f_a+t'^b\otimes \bar{f}_b$ are elements of $\mathcal{L}(V_{\mathbb{C}}\oplus
S_{\mathbb{C}})(B)$, then
$$u*u'=u+u'+\frac{1}{2}[u,u']=(v^{\mu}+v'^{\mu}+i\Gamma_{ab}^{\mu} (s^a t'^b - s'^a t^b))\otimes e_{\mu} + (s^a+s'^a)\otimes f_a + (t^b+t'^b)\otimes \bar{f}_b$$ $\hfill\square$\\

Consider the (free and transitive) action of $M_{cs}$ on itself from the left:
we denote by $P_{\mu}$, $Q_a$ and $\overline{Q}_b$ the fundamental vector fields associated by this action to $e_{\mu}$, $f_a$ and $\bar{f}_b$
respectively. Thus, we have an infinitesimal action
$V_{\mathbb{C}}\oplus S_{\mathbb{C}}\longrightarrow\mathcal{T}_{M_{cs}}(\mathring{M})$,
{\it i.e.} a morphism of super Lie algebras $V_{\mathbb{C}}\oplus S_{\mathbb{C}}\longrightarrow \mathcal{T}_{M_{cs}}(\mathring{M})_0\oplus \mathcal{T}_{M_{cs}}(\mathring{M})_1$. It sends $e_{\mu}\in V$ to $P_{\mu}\in \mathcal{T}_{M_{cs}}(\mathring{M})_0$, $f_a\in S_+$ to $Q_a\in\mathcal{T}_{M_{cs}}(\mathring{M})_1$ and $\bar{f}_b\in S_-$ to $\overline{Q}_b\in\mathcal{T}_{M_{cs}}(\mathring{M})_1$.\\

In order to derive expressions for $P_{\mu}$, $Q_a$ and $\overline{Q}_b$, and for later use as well,
we introduce the following convenient terminology. \\

\begin{defi}
Let $N$ be a supermanifold. For any supermanifold $B$, we often write $N(B)$ for $\hbox{Hom}(B,N)$.
\begin{enumerate}
\item If $f\in\mathcal{O}_N(|N|)$ is a super function
 on $N$, then for every supermanifold $B$, the {\bf $B$-function} associated to $f$ is the map $f_B:N(B)\longrightarrow \mathcal{O}_B(|B|)$ defined by $f_B(\beta)=\beta^{\#}(f)$ for all $\beta\in N(B)$.
\item If $X\in\mathcal{T}_N(|N|)$ is a vector field on $N$, then for every supermanifold $B$, the {\bf $B$-vector field} associated to $X$ is the map $X_B:\hbox{Hom}(N(B),\mathcal{O}_B(|B|))\longrightarrow\hbox{Hom}(N(B),\mathcal{O}_B(|B|))$ defined by $X_B f_B=(Xf)_B$ for all $f\in\mathcal{O}_N(|N|)$.\\
\end{enumerate}
\end{defi}

 Denote by $\displaystyle\frac{\partial}{\partial x^{\mu}}\in\mathcal{T}_{M_{cs}}(\mathring{M})_0$, $\displaystyle\frac{\partial}{\partial \theta^a}\in\mathcal{T}_{M_{cs}}(\mathring{M})_1$ and $\displaystyle\frac{\partial}{\partial \bar{\theta}^a}\in\mathcal{T}_{M_{cs}}(\mathring{M})_1$ the vector fields associated with the standard coordinates $x^{\mu},\theta^a,\bar{\theta}^a$ on $M_{cs}$.

Let $B$ be an auxiliary complex supermanifold, and $M_{cs}(B):=\hbox{Hom}(B,M_{cs})$ the set of $B$-points of $M_{cs}$. Given $\beta\in M_{cs}(B)$, we denote here by $y^{\mu}\in\mathcal{O}_B(|B|)_0$, $\xi^a\in\mathcal{O}_B(|B|)_1$ and $\bar{\xi}^b\in\mathcal{O}_B(|B|)_1$ the images of the standard coordinates of $M_{cs}$ by $\beta^{\sharp}_{\mathring{M}}:\mathcal{C}^{\infty}(\mathring{M})\otimes\bigwedge^{\bullet}S_{\mathbb{C}}^*\longrightarrow \mathcal{O}_B(|B|)$. Then we write $\beta=(y^{\mu},\xi^a,\bar{\xi}^b)$, thinking of $y^{\mu},\xi^a,\bar{\xi}^b$ as the ``coordinates" of the $B$-point $\beta$. Finally, we denote by $\displaystyle\frac{\partial}{\partial y^{\mu}}$, $\displaystyle\frac{\partial}{\partial \xi^a}$ and $\displaystyle\frac{\partial}{\partial \bar{\xi}^a}$ the corresponding $B$-vector fields:

$\displaystyle\frac{\partial}{\partial y^{\mu}}:=\left(\frac{\partial}{\partial x^{\mu}}\right)_B:\hbox{Hom}(M_{cs}(B),\mathcal{O}_B(|B|))\longrightarrow\hbox{Hom}(M_{cs}(B),\mathcal{O}_B(|B|))$\\

$\displaystyle\frac{\partial}{\partial \xi^a}:=\left(\frac{\partial}{\partial \theta^a}\right)_B:\hbox{Hom}(M_{cs}(B),\mathcal{O}_B(|B|))\longrightarrow\hbox{Hom}(M_{cs}(B),\mathcal{O}_B(|B|))$\\

$\displaystyle\frac{\partial}{\partial \bar{\xi}^a}:=\left(\frac{\partial}{\partial \bar{\theta}^a}\right)_B:\hbox{Hom}(M_{cs}(B),\mathcal{O}_B(|B|))\longrightarrow\hbox{Hom}(M_{cs}(B),\mathcal{O}_B(|B|))$\\

\begin{prop}
$P_{\mu}=\displaystyle\frac{\partial}{\partial x^{\mu}}$, $\;\;Q_a=\displaystyle\frac{\partial}{\partial \theta^a} +i\Gamma_{ab}^{\mu}\bar{\theta}^b\displaystyle\frac{\partial}{\partial x^{\mu}}\;\;$ and $\;\;\overline{Q}_b=\displaystyle\frac{\partial}{\partial \bar{\theta}^b} +i\Gamma_{ab}^{\mu}\theta^a\displaystyle\frac{\partial}{\partial x^{\mu}}$.
\end{prop}

\noindent {\bf Proof~:} Let $f\in \mathcal{O}_{M_{cs}}(\mathring{M})=\mathcal{C}^{\infty}(U,\mathbb{C})\otimes \bigwedge S_{\mathbb{C}}^*$. For any supermanifold $B$, let $f_B:M_{cs}(B)\longrightarrow \mathcal{O}_B(|B|)$ be the $B$-function associated with $f$. If $\gamma=(v^{\mu},\varepsilon^a,\bar{\varepsilon}^b)\in M_{cs}(B)$, we calculate $\gamma\cdot f_B:M_{cs}(B)\longrightarrow\mathcal{O}_B(|B|)$:\\

$(\gamma\cdot f_B)(\beta)=f_B(\gamma^{-1}*\beta)=f_B((-v^{\mu},-\varepsilon^a,-\bar{\varepsilon}^b)*(y^{\mu},\xi^a,\bar{\xi}^b))$\\
$=f_B(-v^{\mu}+y^{\mu}+i\Gamma_{ab}^{\mu}(-\varepsilon^a\bar{\xi}^b+\xi^a\bar{\varepsilon}^b),-\varepsilon^a+\xi^a,-\bar{\varepsilon}^b+\bar{\xi}^b)$\\
$=f_B(y^{\mu}-(v^{\mu}+i\Gamma_{ab}^{\mu}(\varepsilon^a\bar{\xi}^b+\bar{\varepsilon}^b\xi^a)),\xi^a-\varepsilon^a,\bar{\xi}^b-\bar{\varepsilon}^b)$\\
$\displaystyle=f_B(y^{\mu},\xi^a,\bar{\xi}^b)-(v^{\mu}+i\Gamma_{ab}^{\mu}(\varepsilon^a\bar{\xi}^b+\bar{\varepsilon}^b\xi^a))\frac{\partial f_B}{\partial y^{\mu}}(y^{\mu},\xi^a,\bar{\xi}^b)-\varepsilon^a\frac{\partial f_B}{\partial \xi^a}(y^{\mu},\xi^a,\bar{\xi}^b)-\bar{\varepsilon}^b\frac{\partial f_B}{\partial \bar{\xi}^b}(y^{\mu},\xi^a,\bar{\xi}^b)$\\
$+\dots$\\
$\displaystyle=f_B(\beta)-v^{\mu}\frac{\partial f_B}{\partial y^{\mu}}(\beta)-\varepsilon^a\left(\frac{\partial f_B}{\partial \xi^a}(\beta)+i\Gamma_{ab}^{\mu}\bar{\xi}^b\frac{\partial f_B}{\partial y^{\mu}}(\beta)\right)-\bar{\varepsilon}^b\left(\frac{\partial f_B}{\partial \bar{\xi}^b}(\beta)+i\Gamma_{ab}^{\mu}\xi^a\frac{\partial f_B}{\partial y^{\mu}}(\beta)\right)+\dots$\\
$\displaystyle=f_B(\beta)-v^{\mu}\left(\displaystyle\frac{\partial f}{\partial x^{\mu}}\right)_B(\beta)-\varepsilon^a\left(\displaystyle\frac{\partial f}{\partial \theta^a} +i\Gamma_{ab}^{\mu}\bar{\xi}^b\displaystyle\frac{\partial f}{\partial x^{\mu}}\right)_B(\beta)-\bar{\varepsilon}^b\left(\displaystyle\frac{\partial f}{\partial \bar{\theta}^b} +i\Gamma_{ab}^{\mu}\xi^a\displaystyle\frac{\partial f}{\partial x^{\mu}}\right)_B(\beta)+\dots$ $\hfill\square$\\

Note that the vector fields $Q_a$ and $\overline{Q}_b$, being left-fundamental, are also right-invariant
(whereas the $P_{\mu}$ are bi-invariant). Then we have: $[P_{\mu},P_{\nu}]=0$, $[P_{\mu},Q_a]=[P_{\mu},\overline{Q}_b]=0$,
$[Q_a,Q_b]=[\overline{Q}_a,\overline{Q}_b]=0$ and $[Q_a,\overline{Q}_b]=-2\Gamma_{ab}^{\mu}P_{\mu}$.\\

We will also need the left-invariant vector fields $D_a$ and $\overline{D}_b$ on $M_{cs}$ associated to $f_a$ and $\bar{f}_b$ respectively. A similar calculation shows that\\

 $D_a=\displaystyle\frac{\partial}{\partial \theta^a} -i\Gamma_{ab}^{\mu}\bar{\theta}^b\displaystyle\frac{\partial}{\partial x^{\mu}}\;\;$ and $\;\;\overline{D}_b=\displaystyle\frac{\partial}{\partial \bar{\theta}^b} -i\Gamma_{ab}^{\mu}\theta^a\displaystyle\frac{\partial}{\partial x^{\mu}}$\\

and we have: $[P_{\mu},D_a]=[P_{\mu},\overline{D}_b]=0$, $[D_a,D_b]=[\overline{D}_a,\overline{D}_b]=0$ and $[D_a,\overline{D}_b]=2\Gamma_{ab}^{\mu}P_{\mu}$.\\

The left and right actions are different (since $M_{cs}$ is not abelian), but of course they commute. This is expressed infinitesimally by: $[Q_a,D_b]=[\overline{Q}_a,\overline{D}_b]=[\overline{Q}_a,D_b]=[Q_a,\overline{D}_b]=0$.\\

Since the super-Poincar\'e group $\hbox{S}\Pi(V)$ acts (transitively) on the supermanifold $M_{cs}$, there is a natural representation of $\hbox{S}\Pi(V)$ on the super-vector space $\mathcal{O}_{M_{cs}}(\mathring{M})$. Viewing $\hbox{S}\Pi(V)$ as a Harish-Chandra pair $(\Pi(V),\mathfrak{s}\pi(V))$, this representation is equivalent to a pair $(\rho,\eta)$ where $\rho:\Pi(V)\longrightarrow\hbox{Aut}(\mathcal{O}_{M_{cs}}(\mathring{M}))$ is a morphism of Lie groups, and $\eta:\mathfrak{s}\pi(V)\longrightarrow\mathfrak{gl}(\mathcal{O}_{M_{cs}}(\mathring{M}))$ is a morphism of super Lie algebras, and we have seen that $\eta(e_{\mu})=P_{\mu}$, $\eta(f_a)=Q_a$ and $\eta(\bar{f}_b)=\overline{Q}_b$.\\

Recall that $\;\mathcal{O}_{M_{cs}}(\mathring{M})\;=\;\mathcal{C}^{\infty}(\mathring{M},\mathbb{C})\otimes\bigwedge^{\bullet}S_{\mathbb{C}}^*\;
\simeq\;\mathcal{C}^{\infty}(\mathring{M},\mathbb{C})\otimes\bigwedge^{\bullet}S_{+}^*\otimes\bigwedge^{\bullet}S_{-}^*$.\\

Let $W$ be a representation of $\hbox{Spin}(V)$. In the next two sections, we will be interested in realizing the
irreducible unitary representations of $\hbox{S}\Pi(V)$ in the super-vector space of ``spin-tensor superfields"
$\mathcal{O}_{M_{cs}}(\mathring{M})\otimes W=\mathcal{C}^{\infty}(\mathring{M},\mathbb{C})\otimes\bigwedge^{\bullet}S_{\mathbb{C}}^*\otimes W\;\simeq\;\mathcal{C}^{\infty}(\mathring{M},\mathbb{C})\otimes\bigwedge^{\bullet}S_{+}^*\otimes\bigwedge^{\bullet}S_{-}^*\otimes W$ (which carries of course a representation of $\hbox{S}\Pi(V)$). \\

First, we have the following supersymmetric generalization of Theorem \ref{dF consequence}.\\ 

\begin{thm} \label{super dF consequence} Let $\mathcal{H}_{(m,\sigma)}$ be the irreducible unitary representation of the super-Poincar\'e group of mass $m>0$ and superspin $\sigma$ (obtained by induction from an irreducible unitary representation $F$ of the little group $\tilde{K}$). Also, let $W$ be a representation of the super Lie group $H=(\hbox{\upshape{Spin}}(V),\mathfrak{spin}(V)\oplus S^*)$. The multiplicity of $\mathcal{H}_{(m,\sigma)}$ in  $\mathcal{O}_{M_{cs}}(\mathring{M})\otimes W$ (as representations of the super-Poincar\'e group) is equal to the multiplicity of $F$ in the decomposition of $W$ under the sub-super Lie group $\tilde{K}$ of $H$. \\
\end{thm}

For example, the ``spin-tensor superfield" representations in which the superparticle $\mathcal{H}_{m,\sigma}$
will appear are the $\mathcal{O}_{M_{cs}}(\mathring{M})\otimes\hbox{Sym}^{2\alpha}S_+^*\otimes  \hbox{Sym}^{2\beta}S_-^*$ where $\alpha+\beta\geq \sigma$.\\

Let $\mathcal{H}_{(m,\sigma)}$ be an irreducible unitary representation of $\hbox{S}\Pi(V)$. One can take
$W:=\hbox{Sym}^{2\alpha}S_+^*\otimes\hbox{Sym}^{2\beta}S_-^*$ with
$\alpha+\beta=\sigma$ (the minimal choice satisfying the above condition). \emph{A priori}, we have $\alpha,\beta\in\{0,\frac{1}{2},1,\frac{3}{2},2,...\}$, but we assume in what
follows that $\alpha>0$ and $\beta>0$. Let $f:\mathring{M}\longrightarrow W$ be
a $W$-valued superfunction.
Imposing only the Klein-Gordon equation on $f$ would imply that the Fourier transform of $f$ is a $(\bigwedge^{\bullet}S_+^*\otimes\bigwedge^{\bullet}S_-^*\otimes W)$-valued measure supported on $\mathcal{O}_m$, of the form $\Psi d\beta_m$ for some function $\Psi:\mathcal{O}_m\longrightarrow \bigwedge^{\bullet}S_+^*\otimes\bigwedge^{\bullet}S_-^*\otimes W$. This is clearly not enough. Roughly speaking, we should constraint $f$ sufficiently so that $\Psi$ ``becomes valued in the $(\bigwedge^{\bullet}S_+^*\otimes\bigwedge^{\bullet}S_-^*\otimes\hbox{Sym}^{2\sigma} S_+^*)$-component of $W$". We know how to achieve this: by imposing the condition $\delta_{\alpha,\beta}f=0$, where $\delta_{\alpha,\beta}$ is the differential operator defined at the end of section \ref{examples}.  \\

This is still not enough however: we know that an irreducible unitary representation of the super-Poincar\'e group should be constructed out of a Hermitian bundle with typical fiber $\bigwedge^{\bullet}S_+^*\otimes\hbox{Sym}^{2\sigma} S_+^*$, whereas here, the typical fiber is $\bigwedge^{\bullet}S_+^*\otimes\bigwedge^{\bullet}S_-^*\otimes\hbox{Sym}^{2\sigma} S_+^*$. This means that the unitary representation of $\hbox{S}\Pi(V)$ that we obtain by imposing only $\delta_{\alpha,\beta}f=0$ (in addition the the Klein-Gordon equation) is actually reducible. To determine its irreducible components, consider first the case of a scalar superfunction, so that the typical fiber is $\bigwedge^{\bullet}S_+^*\otimes\bigwedge^{\bullet}S_-^*$ (in this case, $\sigma=0$, and therefore $W=\mathbb{C}$). \\

Observe that $\bigwedge^{\bullet}S_+^*\otimes\bigwedge^{\bullet}S_-^*$ is a Clifford module for $S^*$. Since $\hbox{C}\ell(S^*)$ has a unique irreducible Clifford module, say $\bigwedge^{\bullet}S_+^*$, we are sure that $\bigwedge^{\bullet}S_-^*$ is just a multiplicity: it does not carry an action of $S^*$. This allows us to decompose: $\bigwedge^{\bullet}S_+^*\otimes\bigwedge^{\bullet}S_-^*=\bigwedge^{\bullet}S_+^*\otimes(\mathbb{C}\oplus S_-^*\oplus \bigwedge^{2}S_-^*)$, and so\\

$\hspace{30mm} \bigwedge^{\bullet}S_+^*\otimes\bigwedge^{\bullet}S_-^*\;\simeq\;(\bigwedge^{\bullet}S_+^*)\;\oplus\; (\bigwedge^{\bullet}S_+^*\otimes S_-^*)\;\oplus\; \bigwedge^{\bullet}S_+^*$\\

Thus, a scalar superfunction $f$ subjected to $\delta_{\alpha,\beta}f=0$ in addition to the Klein-Gordon equation will give rise to a reducible unitary representation of the super-Poincar\'e group, containing two superparticles of superspin 0, and one superparticle of superspin $\frac{1}{2}$. \\

The discussion is similar for a $W$-valued superfunction; we can decompose:\\

 $\bigwedge^{\bullet}S_+^*\otimes\bigwedge^{\bullet}S_-^*\otimes\hbox{Sym}^{2\sigma} S_+^*=\bigwedge^{\bullet}S_+^*\otimes(\mathbb{C}\oplus S_-^*\oplus \bigwedge^{2}S_-^*)\otimes\hbox{Sym}^{2\sigma} S_+^*$\\

$\simeq\bigwedge^{\bullet}S_+^*\otimes(\hbox{Sym}^{2\sigma} S_+^*\oplus (S_-^*\otimes\hbox{Sym}^{2\sigma} S_+^*)\oplus \hbox{Sym}^{2\sigma} S_+^*)$\\

$\simeq\bigwedge^{\bullet}S_+^*\otimes(\hbox{Sym}^{2\sigma} S_+^*\oplus \hbox{Sym}^{2\sigma-1} S_+^*\oplus\hbox{Sym}^{2\sigma+1} S_+^*\oplus \hbox{Sym}^{2\sigma} S_+^*)$\\

and so\\

 $\bigwedge^{\bullet}S_+^*\otimes\bigwedge^{\bullet}S_-^*\otimes\hbox{Sym}^{2\sigma} S_+^*\;\simeq\; (\bigwedge^{\bullet}S_+^*\otimes\hbox{Sym}^{2\sigma} S_+^*)\;\oplus\; (\bigwedge^{\bullet}S_+^*\otimes\hbox{Sym}^{2\sigma-1} S_+^*)$\\

 $\oplus\;(\bigwedge^{\bullet}S_+^*\otimes\hbox{Sym}^{2\sigma+1} S_+^*)\;\oplus\;(\bigwedge^{\bullet}S_+^*\otimes \hbox{Sym}^{2\sigma} S_+^*)$\\

Thus, a $W$-valued superfunction $f$ of type $(\alpha,\beta)$ (with $\alpha+\beta=\sigma$) subjected to $\delta_{\alpha,\beta}f=0$ in addition to the Klein-Gordon equation will give rise to a reducible unitary representation of the super Poincar\'e group, containing two superparticles of superspin $\sigma$, one superparticle of superspin $\sigma-\frac{1}{2}$, and one superparticle of superspin $\sigma+\frac{1}{2}$. \\

If we want to fall on one of these irreducible components, it is necessary to subject the superfield $f$ to a further differential equation in superspacetime. This leads to {\it chirality}.\\

\section{Supersymmetric symbols} \label{supersymbols}

Our notations from now on are as follows:\\

$G=\hbox{S}\Pi(V)=(\Pi(V),\mathfrak{spin}(V)\oplus V\oplus S^*)$\\

$H=(\hbox{Spin}(V),\mathfrak{spin}(V)\oplus S^*)$ \\

$W=\bigwedge^{\bullet} S_+^*\otimes \bigwedge^{\bullet} S_-^*$\\

$\tilde{K}=(K,\mathfrak{k}\oplus S^*)$ \\

$F\simeq\bigwedge^{\bullet} S_+^*\;$ (it appears in the decomposition of $W$ under $\tilde{K}$)\\

We first give an overview of the contents of this section, then we present the details.\\

In the spirit of section \ref{spin-tensor fields}, we proceed first at the algebraic level,
looking for $\tilde{K}$-equivariant linear maps on $W$. On $\bigwedge^{\bullet}S^*_{\pm}$, the exterior multiplication by elements of $S^*_{\pm}$ is defined. Using the $K$-invariant pairing $\Gamma^0:S_+^*\times S_-^*\longrightarrow\mathbb{C}$
(whose definition is recalled below), we obtain also an interior multiplication on $\bigwedge^{\bullet}S^*_{\pm}$. Using these two operations, we define the following endomorphisms of $W$:\\

$\bar{d}_{\bar{\tau}^a}:=(\hbox{\upshape{Id}}\otimes e_{\bar{\tau}^a})+(i_{\bar{\tau}^a}\otimes\hbox{\upshape{Id}})\;\;$ and $\;\;d_{\tau^a}:=(e_{\tau^a}\otimes \hbox{\upshape{Id}} )+(\hbox{\upshape{Id}}\otimes i_{\tau^a})$\\

Each of these endomorphisms is $S^*$-equivariant (but not $\tilde{K}$-equivariant, if taken individually). However, the {\it chiral subspace}
$$W_{chiral}:=\hbox{Ker}\;\bar{d}_{\bar{\tau}^1}\cap\hbox{Ker}\;\bar{d}_{\bar{\tau}^2}$$ is $\tilde{K}$-invariant, and we give a characterization of its elements. Then, we use the natural $\hbox{Spin}(V)$-invariant symplectic structure $\varepsilon_{\pm}$ on $S_{\pm}^*$ to define the endomorphism
$$d^2:=\varepsilon_{ab}\;d_{\tau^a}\circ d_{\tau^b}$$
of $W$, which turns out to be not only $S^*$-invariant, but also $K$-invariant (and thus, $\tilde{K}$-invariant). This is not surprising since $d^2$ is defined via the invariant symplectic structure.\\

Second, we leave the purely algebraic level and propagate our endomorphisms along the orbit $\mathcal{O}_m$, to obtain equivariant symbols $\;\zeta_{\bar{d}_{\bar{\tau}^a}},\zeta_{d^2}:\mathcal{O}_m\longrightarrow\hbox{End}(W)$. There is a characterization of the maps $f:\mathcal{O}_m\longrightarrow W$ satisfying $\;f(p)\in \hbox{Ker}\;\zeta_{\bar{d}_{\bar{\tau}^1}}(p)\cap\hbox{Ker}\;\zeta_{\bar{d}_{\bar{\tau}^2}}(p)\;$ for every $p\in\mathcal{O}_m$. These maps may be called {\it chiral maps}. Finally, the irreducible unitary representation of mass $m$ and superspin 0 is selected by the subspace of chiral maps $f:\mathcal{O}_m\longrightarrow W$ satisfying the condition:
$$\zeta_{d^2}(p)(f(p))=m\bar{f}(p)$$
for all $p\in\mathcal{O}_m$.\\

Here are the details.\\

{\underline{\bf 1.The algebraic level}}

There is a representation of $(\rho,\alpha)$ of $H$ on $W$, namely:\\
$\rho:\hbox{Spin}(V)\longrightarrow \hbox{Aut}(\bigwedge^{\bullet} S_+^*\otimes \bigwedge^{\bullet} S_-^*)\quad$ and $\quad \alpha:S^*\longrightarrow \mathfrak{gl}(\bigwedge^{\bullet} S_+^*\otimes \bigwedge^{\bullet} S_-^*)_1$ \\

Set $\;\langle s_1,s_2\rangle_{e^0}:=e^0(\Gamma(s_1,s_2))\;$ for all $s_1,s_2\in S^*$.\\
 Then $\langle,\rangle_{e^0}$ is a positive definite inner product on $S^*$, which is in addition $K$-invariant. The map $\alpha$ is the restriction of the Clifford algebra representation:\\
$\hbox{C}\ell(S^*,\langle,\rangle_{e^0})\longrightarrow\hbox{\underline{End}}(\bigwedge^{\bullet} S_+^*\otimes \bigwedge^{\bullet} S_-^*)$\\

This representation is reducible: we can think of $\bigwedge^{\bullet} S_-^*$ as a multiplicity space. Then we have the following decomposition under $\hbox{C}\ell(S^*,\langle,\rangle_{e^0})$, but also under $H$ ({\it i.e.} under $\mathfrak{spin}(V)\oplus S^*$):\\

$\bigwedge^{\bullet} S_+^*\otimes \bigwedge^{\bullet} S_-^*=(\bigwedge^{\bullet} S_+^*\otimes\mathbb{C})\;\oplus\; (\bigwedge^{\bullet} S_+^*\otimes S_-^*)\;\oplus\; (\bigwedge^{\bullet} S_+^*\otimes \bigwedge^2 S_-^*)$\\

In fact, we always consider the action of the complexification of $S^*$, that is, the action of $S_{\mathbb{C}}^*=S_+^*\oplus S_-^*$. Also, we will need the nondegenerate bilinear form $\Gamma^0:S_+^*\times S_-^*\longrightarrow\mathbb{C}$ given by $\Gamma^0:=e^0\circ(\Gamma_{\mathbb{C}})_{|S_+^*\times S_-^*}$.\\

For $s_+\in S_+^*$, we denote by $e_{s_+}:\bigwedge^{\bullet} S_+^*\longrightarrow \bigwedge^{\bullet} S_+^*$ the exterior multiplication by $s_+$: $\;e_{s_+}(a)=s_+\wedge a\;$, and we denote by $i_{s_+}:\bigwedge^{\bullet} S_-^*\longrightarrow \bigwedge^{\bullet} S_-^*$ the interior multiplication by $s_+$: $\;i_{s_+}(\lambda+t+r\wedge r')=\Gamma^0(s_+,t)+\Gamma^0(s_+,r)r'-\Gamma^0(s_+,r')r$.\\

For $s_-\in S_-^*$, we denote by $i_{s_-}:\bigwedge^{\bullet} S_+^*\longrightarrow \bigwedge^{\bullet} S_+^*$ the interior multiplication by $s_-$: $\;i_{s_-}(\lambda+t+r\wedge r')=\Gamma^0(t,s_-)+\Gamma^0(r,s_-)r'-\Gamma^0(r',s_-)r\;$, and we denote by\\ $e_{s_-}:\bigwedge^{\bullet} S_-^*\longrightarrow \bigwedge^{\bullet} S_-^*$ the exterior multiplication by $s_-$: $\;e_{s_-}(b)=s_-\wedge b$.\\

\begin{prop}
We have the following (anti)commutation relations:\\
$i_{\tau^a}\circ e_{\bar{\tau}^b}\;+\;e_{\bar{\tau}^b}\circ i_{\tau^a}\;=\;\Gamma^0(\tau^a,\bar{\tau}^b)\;\hbox{\upshape{Id}}$\\
$i_{\tau^a}\circ i_{\tau^b}\;+\;i_{\tau^b}\circ i_{\tau^a}\;=\;0$\\
$e_{\bar{\tau}^a}\circ e_{\bar{\tau}^b}\;+\;e_{\bar{\tau}^b}\circ e_{\bar{\tau}^a}\;=\;0$\\
\end{prop}

\begin{rem}
The first commutation relation is equivalent to the four following equations:\\
$i_{\tau^1}\circ e_{\bar{\tau}^2}\;+\;e_{\bar{\tau}^2}\circ i_{\tau^1}\;=\;0$\\
$i_{\tau^2}\circ e_{\bar{\tau}^1}\;+\;e_{\bar{\tau}^1}\circ i_{\tau^2}\;=\;0$\\
$i_{\tau^1}\circ e_{\bar{\tau}^1}\;+\;e_{\bar{\tau}^1}\circ i_{\tau^1}\;=\;\hbox{\upshape{Id}}$\\
$i_{\tau^2}\circ e_{\bar{\tau}^2}\;+\;e_{\bar{\tau}^2}\circ i_{\tau^2}\;=\;\hbox{\upshape{Id}}$\\
and can be thought of as the odd counterpart of Heisenberg uncertainty relation.\\
\end{rem}

Define the following endomorphisms of $W$:\\

$\bar{d}_{\bar{\tau}^a}:=(\hbox{\upshape{Id}}\otimes e_{\bar{\tau}^a})+(i_{\bar{\tau}^a}\otimes\hbox{\upshape{Id}})\;\;$ and $\;\;d_{\tau^a}:=(e_{\tau^a}\otimes \hbox{\upshape{Id}} )+(\hbox{\upshape{Id}}\otimes i_{\tau^a})$,\\

$\bar{q}_{\bar{\tau}^a}:=(\hbox{\upshape{Id}}\otimes e_{\bar{\tau}^a})-(i_{\bar{\tau}^a}\otimes\hbox{\upshape{Id}})\;\;$ and $\;\;q_{\tau^a}:=(e_{\tau^a}\otimes \hbox{\upshape{Id}} )-(\hbox{\upshape{Id}}\otimes i_{\tau^a})$.\\

\begin{prop} \label{susy invariance} We have: $\;\;[q_{\tau^a},d_{\tau^b}]=[\bar{q}_{\bar{\tau}^a},d_{\tau^b}]=0\;\;$ and $\;\;[q_{\tau^a},\bar{d}_{\bar{\tau}^b}]=[\bar{q}_{\bar{\tau}^a},\bar{d}_{\bar{\tau}^a}]=0$.\\
\end{prop}

\begin{defi} The {\bf chiral subspace} of $W$ is $$W_{chiral}:=\hbox{Ker}\;\bar{d}_{\bar{\tau}^1}\cap\hbox{Ker}\;\bar{d}_{\bar{\tau}^2}$$
\end{defi}

Write $\;\;f\;\;=\;\;H\;\;+\;\;\mu_1\;\tau^1\;\;+\;\;\mu_2\;\tau^2\;\;+\;\;\lambda_1\;\bar{\tau}^1\;\;+\;\;\lambda_2\;\bar{\tau}^2\;\;+\;\;G\;\tau^1\wedge\tau^2\;\;+\;\;F\;\bar{\tau}^1\wedge\bar{\tau}^2$\\

$+\;\;A_{11}\;\tau^1\otimes\bar{\tau}^1\;\;+\;\;A_{12}\;\tau^1\otimes\bar{\tau}^2\;\;+\;\;A_{21}\;\tau^2\otimes\bar{\tau}^1\;\;+\;\;A_{22}\;\tau^2\otimes\bar{\tau}^2$\\

$+\;\;\eta_1\;(\tau^{1}\wedge\tau^{2})\otimes\bar{\tau}^1\;\;+\;\;\eta_2\;(\tau^{1}\wedge\tau^{2})\otimes\bar{\tau}^2$\\

$+\;\;\psi_1\;\tau^1\otimes(\bar{\tau}^1\wedge\bar{\tau}^2)\;\;+\;\;\psi_2\;\tau^2\otimes(\bar{\tau}^1\wedge\bar{\tau}^2)\;\;+\;\;\varphi\;(\tau^1\wedge\tau^2)\otimes(\bar{\tau}^1\wedge\bar{\tau}^2)$\\

for a generic element of $W$ (here, $\;H,\mu_a,\lambda_a,G,F,A_{ab},\eta_a,\psi_a,\varphi\;$ are complex numbers).\\

\begin{prop} An element $f\in W$ belongs to the chiral subspace $W_{chiral}$ if and only if \\

$f\;\;=\;\;(\Gamma^0(\tau^1,\bar{\tau}^1)\;\Gamma^0(\tau^2,\bar{\tau}^2)-\Gamma^0(\tau^2,\bar{\tau}^1)\;\Gamma^0(\tau^1,\bar{\tau}^2))\;\varphi$\\

$+\;\;(\Gamma^0(\tau^1,\bar{\tau}^2)\;\psi_1+\Gamma^0(\tau^2,\bar{\tau}^2)\;\psi_2)\;\bar{\tau}^1\;\;+\;\;(-\Gamma^0(\tau^1,\bar{\tau}^1)\;\psi_1-\Gamma^0(\tau^2,\bar{\tau}^1)\;\psi_2)\;\bar{\tau}^2$\\

$+\;\;F\;\bar{\tau}^1\wedge\bar{\tau}^2\;\;+\;\;(-\Gamma^0(\tau^2,\bar{\tau}^2)\;\varphi)\;\tau^1\otimes\bar{\tau}^1\;\;+\;\;(\Gamma^0(\tau^2,\bar{\tau}^1)\;\varphi)\;\tau^1\otimes\bar{\tau}^2$\\

$+\;\;(\Gamma^0(\tau^1,\bar{\tau}^2)\;\varphi)\;\tau^2\otimes\bar{\tau}^1\;\;+\;\;(-\Gamma^0(\tau^1,\bar{\tau}^1)\;\varphi)\;\tau^2\otimes\bar{\tau}^2$\\

$+\;\;\psi_1\;\tau^1\otimes(\bar{\tau}^1\wedge\bar{\tau}^2)\;\;+\;\;\psi_2\;\tau^2\otimes(\bar{\tau}^1\wedge\bar{\tau}^2)\;\;+\;\;\varphi\;(\tau^1\wedge\tau^2)\otimes(\bar{\tau}^1\wedge\bar{\tau}^2)$\\

for some complex numbers $\;\varphi,\psi,F$.
\end{prop}

\noindent {\bf Proof~:}
We calculate $\bar{d}_{\bar{\tau}^1}(f)$.\\

$\bar{d}_{\bar{\tau}^1}(f)\;\;=\;\;H\;\bar{\tau}^1\;\;+\;\;\mu_1\;\tau^1\otimes\bar{\tau}^1\;\;+\;\;\Gamma^0(\tau^1,\bar{\tau}^1)\;\mu_1\;\;+\;\;\mu_2\;\tau^2\otimes\bar{\tau}^1\;\;+\;\;\Gamma^0(\tau^2,\bar{\tau}^1)\;\mu_2$\\

$+\;\;\lambda_1\;\bar{\tau}^1\wedge\bar{\tau}^1\;\;+\;\;\lambda_2\;\bar{\tau}^1\wedge\bar{\tau}^2\;\;+\;\;G\;(\tau^1\wedge\tau^2)\otimes \bar{\tau}^1\;\;+\;\;\Gamma^0(\tau^1,\bar{\tau}^1)\;G\;\tau^2\;\;-\;\;\Gamma^0(\tau^2,\bar{\tau}^1)\;G\;\tau^1$\\

$+\;\;A_{11}\;\tau^1\otimes(\bar{\tau}^1\wedge\bar{\tau}^1)\;\;+\;\;\Gamma^0(\tau^1,\bar{\tau}^1)\;A_{11}\;\bar{\tau}^1\;\;+\;\;A_{12}\;\tau^1\otimes(\bar{\tau}^1\wedge\bar{\tau}^2)\;\;+\;\;\Gamma^0(\tau^1,\bar{\tau}^1)\;A_{12}\;\bar{\tau}^2$\\

$+\;\;A_{21}\;\tau^2\otimes(\bar{\tau}^1\wedge\bar{\tau}^1)\;\;+\;\;\Gamma^0(\tau^2,\bar{\tau}^1)\;A_{21}\;\bar{\tau}^1\;\;+\;\;A_{22}\;\tau^2\otimes(\bar{\tau}^1\wedge\bar{\tau}^2)\;\;+\;\;\Gamma^0(\tau^2,\bar{\tau}^1)\;A_{22}\;\bar{\tau}^2$\\

$+\;\;\eta_1\;(\tau^{1}\wedge\tau^{2})\otimes(\bar{\tau}^1\wedge\bar{\tau}^1)\;\;+\;\;\Gamma^0(\tau^1,\bar{\tau}^1)\;\eta_1\;\tau^2\otimes\bar{\tau}^1\;\;-\;\;\Gamma^0(\tau^2,\bar{\tau}^1)\;\eta_1\;\tau^1\otimes\bar{\tau}^1$\\

$+\;\;\eta_2\;(\tau^{1}\wedge\tau^{2})\otimes(\bar{\tau}^1\wedge\bar{\tau}^2)\;\;+\;\;\Gamma^0(\tau^1,\bar{\tau}^1)\;\eta_2\;\tau^2\otimes\bar{\tau}^2\;\;-\;\;\Gamma^0(\tau^2,\bar{\tau}^1)\;\eta_2\;\tau^1\otimes\bar{\tau}^2$\\

$+\;\;\Gamma^0(\tau^1,\bar{\tau}^1)\;\psi_1\;\bar{\tau}^1\wedge\bar{\tau}^2\;\;+\;\;\Gamma^0(\tau^2,\bar{\tau}^1)\;\psi_2\;\bar{\tau}^1\wedge\bar{\tau}^2\;\;+\;\;\Gamma^0(\tau^1,\bar{\tau}^1)\;\varphi\;\tau^2\otimes(\bar{\tau}^1\wedge\bar{\tau}^2)$\\

$-\;\;\Gamma^0(\tau^2,\bar{\tau}^1)\;\varphi\;\tau^1\otimes(\bar{\tau}^1\wedge\bar{\tau}^2)$\\

We calculate $\bar{d}_{\bar{\tau}^2}(f)$.\\

$\bar{d}_{\bar{\tau}^2}(f)\;\;=\;\;H\;\bar{\tau}^2\;\;+\;\;\mu_1\;\tau^1\otimes\bar{\tau}^2\;\;+\;\;\Gamma^0(\tau^1,\bar{\tau}^2)\;\mu_1\;\;+\;\;\mu_2\;\tau^2\otimes\bar{\tau}^2\;\;+\;\;\Gamma^0(\tau^2,\bar{\tau}^2)\;\mu_2$\\

$+\;\;\lambda_1\;\bar{\tau}^2\wedge\bar{\tau}^1\;\;+\;\;\lambda_2\;\bar{\tau}^2\wedge\bar{\tau}^2\;\;+\;\;G\;(\tau^1\wedge\tau^2)\otimes \bar{\tau}^2\;\;+\;\;\Gamma^0(\tau^1,\bar{\tau}^2)\;G\;\tau^2\;\;-\;\;\Gamma^0(\tau^2,\bar{\tau}^2)\;G\;\tau^1$\\

$+\;\;A_{11}\;\tau^1\otimes(\bar{\tau}^2\wedge\bar{\tau}^1)\;\;+\;\;\Gamma^0(\tau^1,\bar{\tau}^2)\;A_{11}\;\bar{\tau}^1\;\;+\;\;A_{12}\;\tau^1\otimes(\bar{\tau}^2\wedge\bar{\tau}^2)\;\;+\;\;\Gamma^0(\tau^1,\bar{\tau}^2)\;A_{12}\;\bar{\tau}^2$\\

$+\;\;A_{21}\;\tau^2\otimes(\bar{\tau}^2\wedge\bar{\tau}^1)\;\;+\;\;\Gamma^0(\tau^2,\bar{\tau}^2)\;A_{21}\;\bar{\tau}^1\;\;+\;\;A_{22}\;\tau^2\otimes(\bar{\tau}^2\wedge\bar{\tau}^2)\;\;+\;\;\Gamma^0(\tau^2,\bar{\tau}^2)\;A_{22}\;\bar{\tau}^2$\\

$+\;\;\eta_1\;(\tau^{1}\wedge\tau^{2})\otimes(\bar{\tau}^2\wedge\bar{\tau}^1)\;\;+\;\;\Gamma^0(\tau^1,\bar{\tau}^2)\;\eta_1\;\tau^2\otimes\bar{\tau}^1\;\;-\;\;\Gamma^0(\tau^2,\bar{\tau}^2)\;\eta_1\;\tau^1\otimes\bar{\tau}^1$\\

$+\;\;\eta_2\;(\tau^{1}\wedge\tau^{2})\otimes(\bar{\tau}^2\wedge\bar{\tau}^2)\;\;+\;\;\Gamma^0(\tau^1,\bar{\tau}^2)\;\eta_2\;\tau^2\otimes\bar{\tau}^2\;\;-\;\;\Gamma^0(\tau^2,\bar{\tau}^2)\;\eta_2\;\tau^1\otimes\bar{\tau}^2$\\

$+\;\;\Gamma^0(\tau^1,\bar{\tau}^2)\;\psi_1\;\bar{\tau}^1\wedge\bar{\tau}^2\;\;+\;\;\Gamma^0(\tau^2,\bar{\tau}^2)\;\psi_2\;\bar{\tau}^1\wedge\bar{\tau}^2\;\;+\;\;\Gamma^0(\tau^1,\bar{\tau}^2)\;\varphi\;\tau^2\otimes(\bar{\tau}^1\wedge\bar{\tau}^2)$\\

$-\;\;\Gamma^0(\tau^2,\bar{\tau}^2)\;\varphi\;\tau^1\otimes(\bar{\tau}^1\wedge\bar{\tau}^2)$\\

We see that $\;\bar{d}_{\bar{\tau}^1}(f)=0\;$ and $\;\bar{d}_{\bar{\tau}^2}(f)=0\;$ is equivalent to:\\

$G=\mu_1=\mu_2=\eta_1=\eta_2=0$\\

$H=-\Gamma^0(\tau^1,\bar{\tau}^1)\;A_{11}-\Gamma^0(\tau^2,\bar{\tau}^1)\;A_{21}\quad$,$\quad\lambda_2=-\Gamma^0(\tau^1,\bar{\tau}^1)\;\psi_1-\Gamma^0(\tau^2,\bar{\tau}^1)\;\psi_2$\\

$A_{12}=\Gamma^0(\tau^2,\bar{\tau}^1)\;\varphi\quad$,$\quad A_{22}=-\Gamma^0(\tau^1,\bar{\tau}^1)\;\varphi$\\

$H=-\Gamma^0(\tau^1,\bar{\tau}^2)\;A_{12}-\Gamma^0(\tau^2,\bar{\tau}^2)\;A_{22}\quad$,$\quad\lambda_1=\Gamma^0(\tau^1,\bar{\tau}^2)\;\psi_1+\Gamma^0(\tau^2,\bar{\tau}^2)\;\psi_2$\\

$A_{11}=-\Gamma^0(\tau^2,\bar{\tau}^2)\;\varphi\quad$,$\quad A_{21}=\Gamma^0(\tau^1,\bar{\tau}^2)\;\varphi$\\

It follows that $\;H=(\Gamma^0(\tau^1,\bar{\tau}^1)\;\Gamma^0(\tau^2,\bar{\tau}^2)-\Gamma^0(\tau^2,\bar{\tau}^1)\;\Gamma^0(\tau^1,\bar{\tau}^2))\;\varphi$\\

and we obtain the result. $\hfill\square$\\

By Proposition \ref{susy invariance}, the subspace $W_{chiral}$ is $S^*$-invariant, and it is not difficult to check that it is $K$-invariant as well. Thus, it is $\tilde{K}$-invariant.\\

Now define the ``second-order" endomorphisms of $W$:\\

$\;\bar{d}^2:=\varepsilon_{ab}\;\bar{d}_{\bar{\tau}^a}\circ \bar{d}_{\bar{\tau}^b}\;$ and $\;d^2:=\varepsilon_{ab}\;d_{\tau^a}\circ d_{\tau^b}\;$.\\

Another expression for $\bar{d}^2$ and $d^2$ can be obtained as follows: set $\;i^2:=-\frac{1}{2}\varepsilon_{ab}\;i_{\tau^a}\circ i_{\tau^b}$.\\
Note that $\;i^2:\bigwedge^{\bullet} S_-^* \longrightarrow \bigwedge^{\bullet} S_-^* \;$, but we see it as $\;i^2:\bigwedge^2 S_-^* \longrightarrow \mathbb{C} \;$ since it kills any term of degree strictly lower than 2. So we calculate:\\
$i^2(r\wedge r')=-\frac{1}{2}\varepsilon_{ab}\;i_{\tau^a}(\Gamma^0(\tau^b,r)r'-\Gamma^0(\tau^b,r')r)=-\frac{1}{2}\varepsilon_{ab}\;(\Gamma^0(\tau^b,r)\Gamma^0(\tau^a,r')-\Gamma^0(\tau^b,r')\Gamma^0(\tau^a,r))$\\
Thus, $$i^2(r\wedge r')=\varepsilon_{ab}\;\Gamma^0(\tau^a,r)\;\Gamma^0(\tau^b,r')$$
In particular, $$i^2(\bar{\tau}^1\wedge\bar{\tau}^2)=\Gamma^0(\tau^1,\bar{\tau}^1)\;\Gamma^0(\tau^2,\bar{\tau}^2)-\Gamma^0(\tau^1,\bar{\tau}^2)\;\Gamma^0(\tau^2,\bar{\tau}^1)$$
Set $e^2=\varepsilon_{ab}\;e_{\tau^a}\circ e_{\tau^b}=\varepsilon_{ab}\;e_{\tau^a\wedge\tau^b}=e_{\varepsilon_{ab}\tau^a\wedge\tau^b}=e_{\varepsilon}$.\\
Note that $\;e^2:\bigwedge^{\bullet} S_+^* \longrightarrow \bigwedge^{\bullet} S_+^* \;$, but we see it as $\;e^2: \mathbb{C} \longrightarrow \bigwedge^2 S_+^* \;$ since it kills any term of degree strictly higher than 0. Thus, $\;e^2(\lambda)=\lambda\varepsilon\;$ for every $\lambda\in\mathbb{C}$. Similarly, one can also define $\;\bar{i}^2:\bigwedge^{\bullet} S_+^* \longrightarrow \bigwedge^{\bullet} S_+^* \;$ and $\;\bar{e}^2:\bigwedge^{\bullet} S_-^* \longrightarrow \bigwedge^{\bullet} S_-^*$.\\

It is easy to check that $\;\bar{d}^2=(\hbox{\upshape{Id}}\otimes \bar{e}^2)+(\bar{i}^2\otimes\hbox{\upshape{Id}})\;$ and $\;d^2=(e^2\otimes \hbox{\upshape{Id}} )+(\hbox{\upshape{Id}}\otimes i^2)$.\\

Consequently, $\;\;d^2(f)\;\;=\;\;(\Gamma^0(\tau^1,\bar{\tau}^1)\;\Gamma^0(\tau^2,\bar{\tau}^2)-\Gamma^0(\tau^2,\bar{\tau}^1)\;\Gamma^0(\tau^1,\bar{\tau}^2))\;\varphi\;\tau^1\wedge\tau^2$\\

$+\;\;(\Gamma^0(\tau^1,\bar{\tau}^2)\;\psi_1+\Gamma^0(\tau^2,\bar{\tau}^2)\;\psi_2)\;(\tau^1\wedge\tau^2)\otimes\bar{\tau}^1$\\

$+\;\;(-\Gamma^0(\tau^1,\bar{\tau}^1)\;\psi_1-\Gamma^0(\tau^2,\bar{\tau}^1)\;\psi_2)\;(\tau^1\wedge\tau^2)\otimes\bar{\tau}^2$\\

$+\;\;F\;(\tau^1\wedge\tau^2)\otimes(\bar{\tau}^1\wedge\bar{\tau}^2)\;\;+\;\;(\Gamma^0(\tau^1,\bar{\tau}^1)\;\Gamma^0(\tau^2,\bar{\tau}^2)-\Gamma^0(\tau^1,\bar{\tau}^2)\;\Gamma^0(\tau^2,\bar{\tau}^1))\;F$\\

$+\;\;(\Gamma^0(\tau^1,\bar{\tau}^1)\;\Gamma^0(\tau^2,\bar{\tau}^2)-\Gamma^0(\tau^1,\bar{\tau}^2)\;\Gamma^0(\tau^2,\bar{\tau}^1))\;\psi_1\;\tau^1$\\

$+\;\;(\Gamma^0(\tau^1,\bar{\tau}^1)\;\Gamma^0(\tau^2,\bar{\tau}^2)-\Gamma^0(\tau^1,\bar{\tau}^2)\;\Gamma^0(\tau^2,\bar{\tau}^1))\;\psi_2\;\tau^2$\\

$+\;\;(\Gamma^0(\tau^1,\bar{\tau}^1)\;\Gamma^0(\tau^2,\bar{\tau}^2)-\Gamma^0(\tau^1,\bar{\tau}^2)\;\Gamma^0(\tau^2,\bar{\tau}^1))\;\varphi\;\tau^1\wedge\tau^2$\\

{\underline{\bf 2. The supersymmetric symbols}}

\begin{prop}\
\begin{enumerate}
\item $i^2$ is $K$-equivariant.
\item Let $\zeta_{i^2}(p):=m^2\;i^2\cdot h_p^{-1}$. Then $$\zeta_{i^2}(p)(r\wedge r')=\varepsilon_{ab}\;p(\Gamma_{\mathbb{C}}(\tau^a,r))\;p(\Gamma_{\mathbb{C}}(\tau^b,r'))$$
\end{enumerate}
\end{prop}

\noindent {\bf Proof~:}
1. For every $k\in K$, we have:\\

$i^2(k^{-1}(r\wedge r'))=i^2(k^{-1}r\wedge k^{-1}r')=\varepsilon_{ab}\;\Gamma^0(\tau^a,k^{-1}r)\Gamma^0(\tau^b,k^{-1}r')$\\

 $= \varepsilon_{ab}\;e^0(\Gamma_{\mathbb{C}}(\tau^a,k^{-1}r))\;e^0(\Gamma_{\mathbb{C}}(\tau^b,k^{-1}r'))=\varepsilon_{ab}\;e^0(k^{-1}\Gamma_{\mathbb{C}}(k\tau^a,r))\;e^0(k^{-1}\Gamma_{\mathbb{C}}(k\tau^b,r'))$\\

$=\varepsilon_{ab}\;e^0(\Gamma_{\mathbb{C}}(k\tau^a,r))\;e^0(\Gamma_{\mathbb{C}}(k\tau^b,r'))=\varepsilon_{ab}\;e^0(\Gamma_{\mathbb{C}}(k^a_c\tau^c,r))\;e^0(\Gamma_{\mathbb{C}}(k^b_d\tau^d,r'))$\\

$= k^a_c k^b_d\varepsilon_{ab}\;e^0(\Gamma_{\mathbb{C}}(\tau^c,r))\;e^0(\Gamma_{\mathbb{C}}(\tau^d,r'))= \varepsilon_{cd}\;e^0(\Gamma_{\mathbb{C}}(\tau^c,r))\;e^0(\Gamma_{\mathbb{C}}(\tau^d,r'))$\\

$=\varepsilon_{ab}\;\Gamma^0(\tau^a,r)\Gamma^0(\tau^b,r')=i^2(r\wedge r')$.\\

2. $i^2(h_p^{-1}(r\wedge r'))=i^2(h_p^{-1}r\wedge h_p^{-1}r')=\varepsilon_{ab}\;\Gamma^0(\tau^a,h_p^{-1}r)\Gamma^0(\tau^b,h_p^{-1}r')$\\

 $= \varepsilon_{ab}\;e^0(\Gamma_{\mathbb{C}}(\tau^a,h_p^{-1}r))\;e^0(\Gamma_{\mathbb{C}}(\tau^b,h_p^{-1}r'))=\varepsilon_{ab}\;e^0(h_p^{-1}\Gamma_{\mathbb{C}}(h_p\tau^a,r))\;e^0(h_p^{-1}\Gamma_{\mathbb{C}}(h_p\tau^b,r'))$\\

$=\varepsilon_{ab}\;(h_p e^0)(\Gamma_{\mathbb{C}}(h_p\tau^a,r))\;(h_p  e^0)(\Gamma_{\mathbb{C}}(h_p\tau^b,r'))=\varepsilon_{ab}\;(h_p  e^0)(\Gamma_{\mathbb{C}}((h_p)^a_c\tau^c,r))\;(h_p e^0)(\Gamma_{\mathbb{C}}((h_p)^b_d\tau^d,r'))$\\

$= (h_p)^a_c (h_p)^b_d\varepsilon_{ab}\;(h_p e^0)(\Gamma_{\mathbb{C}}(\tau^c,r))\;(h_p e^0)(\Gamma_{\mathbb{C}}(\tau^d,r'))= \varepsilon_{cd}\;\frac{1}{m}p(\Gamma_{\mathbb{C}}(\tau^c,r))\;\frac{1}{m}p(\Gamma_{\mathbb{C}}(\tau^d,r'))$.\\

 $\hfill\square$\\

Equivalently, $$\zeta_{i^2}(p)=\varepsilon_{ab}\;\zeta_{i_{\tau^a}}(p)\circ\zeta_{i_{\tau^b}}(p)$$
where $$\zeta_{i_{s_+}}(p)(\lambda+t+r\wedge r'):=p(\Gamma_{\mathbb{C}}(s_+,t))+p(\Gamma_{\mathbb{C}}(s_+,r))r'-p(\Gamma_{\mathbb{C}}(s_+,r'))r$$
Similarly, one can define
$$\zeta_{i_{s_-}}(p)(\lambda+t+r\wedge r'):=p(\Gamma_{\mathbb{C}}(t,s_-))+p(\Gamma_{\mathbb{C}}(r,s_-))r'-p(\Gamma_{\mathbb{C}}(r',s_-))r$$ \\

\begin{prop} The symbols corresponding to the endomorphisms $\bar{d}_{\bar{\tau}^a}$ and $d_{\tau^a}$ of $W$ are given by: \\

$\zeta_{\bar{d}_{\bar{\tau}^a}}(p):=(\hbox{\upshape{Id}}\otimes e_{\bar{\tau}^a})+(\zeta_{i_{\bar{\tau}^a}}(p)\otimes\hbox{\upshape{Id}})\;\;$ and $\;\;\zeta_{d_{\tau^a}}(p):=(e_{\tau^a}\otimes \hbox{\upshape{Id}} )+(\hbox{\upshape{Id}}\otimes \zeta_{i_{\tau^a}}(p))$.\\
\end{prop}

Write $\;\;f(p)\;\;=\;\;H(p)\;\;+\;\;\mu_1(p)\;\tau^1\;\;+\;\;\mu_2(p)\;\tau^2\;\;+\;\;\lambda_1(p)\;\bar{\tau}^1\;\;+\;\;\lambda_2(p)\;\bar{\tau}^2$\\

$+\;\;G(p)\;\tau^1\wedge\tau^2\;\;+\;\;F(p)\;\bar{\tau}^1\wedge\bar{\tau}^2$\\

$+\;\;A_{11}(p)\;\tau^1\otimes\bar{\tau}^1\;\;+\;\;A_{12}(p)\;\tau^1\otimes\bar{\tau}^2\;\;+\;\;A_{21}(p)\;\tau^2\otimes\bar{\tau}^1\;\;+\;\;A_{22}(p)\;\tau^2\otimes\bar{\tau}^2$\\

$+\;\;\eta_1(p)\;(\tau^{1}\wedge\tau^{2})\otimes\bar{\tau}^1\;\;+\;\;\eta_2(p)\;(\tau^{1}\wedge\tau^{2})\otimes\bar{\tau}^2$\\

$+\;\;\psi_1(p)\;\tau^1\otimes(\bar{\tau}^1\wedge\bar{\tau}^2)\;\;+\;\;\psi_2(p)\;\tau^2\otimes(\bar{\tau}^1\wedge\bar{\tau}^2)\;\;+\;\;\varphi(p)\;(\tau^1\wedge\tau^2)\otimes(\bar{\tau}^1\wedge\bar{\tau}^2)$\\

for the expression of a generic map $f:\mathcal{O}_m\longrightarrow W$.\\

\begin{prop} We have $f(p)\in \hbox{Ker}\;\zeta_{\bar{d}_{\bar{\tau}^1}}(p)\cap\hbox{Ker}\;\zeta_{\bar{d}_{\bar{\tau}^2}}(p)$ if and only if \\

$f(p)\;\;=\;\;(p(\Gamma_{\mathbb{C}}(\tau^1,\bar{\tau}^1))\;p(\Gamma_{\mathbb{C}}(\tau^2,\bar{\tau}^2))-p(\Gamma_{\mathbb{C}}(\tau^2,\bar{\tau}^1))\;p(\Gamma_{\mathbb{C}}(\tau^1,\bar{\tau}^2)))\;\varphi(p)$\\

$+\;\;(p(\Gamma_{\mathbb{C}}(\tau^1,\bar{\tau}^2))\;\psi_1(p)+p(\Gamma_{\mathbb{C}}(\tau^2,\bar{\tau}^2))\;\psi_2(p))\;\bar{\tau}^1$\\

$+\;\;(-p(\Gamma_{\mathbb{C}}(\tau^1,\bar{\tau}^1))\;\psi_1(p)-p(\Gamma_{\mathbb{C}}(\tau^2,\bar{\tau}^1))\;\psi_2(p))\;\bar{\tau}^2$\\

$+\;\;F(p)\;\bar{\tau}^1\wedge\bar{\tau}^2\;\;+\;\;(-p(\Gamma_{\mathbb{C}}(\tau^2,\bar{\tau}^2))\;\varphi(p))\;\tau^1\otimes\bar{\tau}^1\;\;+\;\;(p(\Gamma_{\mathbb{C}}(\tau^2,\bar{\tau}^1))\;\varphi(p))\;\tau^1\otimes\bar{\tau}^2$\\

$+\;\;(p(\Gamma_{\mathbb{C}}(\tau^1,\bar{\tau}^2))\;\varphi(p))\;\tau^2\otimes\bar{\tau}^1\;\;+\;\;(-p(\Gamma_{\mathbb{C}}(\tau^1,\bar{\tau}^1)\;\varphi(p))\;\tau^2\otimes\bar{\tau}^2$\\

$+\;\;\psi_1(p)\;\tau^1\otimes(\bar{\tau}^1\wedge\bar{\tau}^2)\;\;+\;\;\psi_2(p)\;\tau^2\otimes(\bar{\tau}^1\wedge\bar{\tau}^2)\;\;+\;\;\varphi(p)\;(\tau^1\wedge\tau^2)\otimes(\bar{\tau}^1\wedge\bar{\tau}^2)$\\
\end{prop}

\begin{prop} The symbols corresponding to the ``second-order" endomorphisms $\bar{d}^2$ and $d^2$ of $W$ are given by: \\

$\zeta_{\bar{d}^2}(p)=(\hbox{\upshape{Id}}\otimes \bar{e}^2)+(\zeta_{\bar{i}^2}(p)\otimes\hbox{\upshape{Id}})\;\;$ and $\;\;\zeta_{d^2}(p)=(e^2\otimes \hbox{\upshape{Id}} )+(\hbox{\upshape{Id}}\otimes \zeta_{i^2}(p))$.\\
\end{prop}

Also, we have:\\

$\zeta_{d^2}(p)(f(p))\;=\;\;(p(\Gamma_{\mathbb{C}}(\tau^1,\bar{\tau}^1))\;p(\Gamma_{\mathbb{C}}(\tau^2,\bar{\tau}^2))-p(\Gamma_{\mathbb{C}}(\tau^2,\bar{\tau}^1))\;p(\Gamma_{\mathbb{C}}(\tau^1,\bar{\tau}^2)))\;\varphi(p)\;\tau^1\wedge\tau^2$\\

$+\;\;(p(\Gamma_{\mathbb{C}}(\tau^1,\bar{\tau}^2))\;\psi_1(p)+p(\Gamma_{\mathbb{C}}(\tau^2,\bar{\tau}^2))\;\psi_2(p))\;(\tau^1\wedge\tau^2)\otimes\bar{\tau}^1$\\

$+\;\;(-p(\Gamma_{\mathbb{C}}(\tau^1,\bar{\tau}^1))\;\psi_1(p)-p(\Gamma_{\mathbb{C}}(\tau^2,\bar{\tau}^1))\;\psi_2(p))\;(\tau^1\wedge\tau^2)\otimes\bar{\tau}^2$\\

$+\;\;F(p)\;(\tau^1\wedge\tau^2)\otimes(\bar{\tau}^1\wedge\bar{\tau}^2)\;\;+\;\;(p(\Gamma_{\mathbb{C}}(\tau^1,\bar{\tau}^1))\;p(\Gamma_{\mathbb{C}}(\tau^2,\bar{\tau}^2))-p(\Gamma_{\mathbb{C}}(\tau^1,\bar{\tau}^2))\;p(\Gamma_{\mathbb{C}}(\tau^2,\bar{\tau}^1)))\;F(p)$\\

$+\;\;(p(\Gamma_{\mathbb{C}}(\tau^1,\bar{\tau}^1))\;p(\Gamma_{\mathbb{C}}(\tau^2,\bar{\tau}^2))-p(\Gamma_{\mathbb{C}}(\tau^1,\bar{\tau}^2))\;p(\Gamma_{\mathbb{C}}(\tau^2,\bar{\tau}^1)))\;\psi_1(p)\;\tau^1$\\

$+\;\;(p(\Gamma_{\mathbb{C}}(\tau^1,\bar{\tau}^1))\;p(\Gamma_{\mathbb{C}}(\tau^2,\bar{\tau}^2))-p(\Gamma_{\mathbb{C}}(\tau^1,\bar{\tau}^2))\;p(\Gamma_{\mathbb{C}}(\tau^2,\bar{\tau}^1)))\;\psi_2(p)\;\tau^2$\\

$+\;\;(p(\Gamma_{\mathbb{C}}(\tau^1,\bar{\tau}^1))\;p(\Gamma_{\mathbb{C}}(\tau^2,\bar{\tau}^2))-p(\Gamma_{\mathbb{C}}(\tau^1,\bar{\tau}^2))\;p(\Gamma_{\mathbb{C}}(\tau^2,\bar{\tau}^1)))\;\varphi(p)\;\tau^1\wedge\tau^2$\\

Now\\
 $\Gamma_{\mathbb{C}}(\tau^1,\bar{\tau}^1)=e_0+e_1\quad$,$\quad\Gamma_{\mathbb{C}}(\tau^1,\bar{\tau}^2)=e_2-ie_3\quad$\\
 $\quad\Gamma_{\mathbb{C}}(\tau^2,\bar{\tau}^1)=e_2+ie_3\quad$,$\quad\Gamma_{\mathbb{C}}(\tau^2,\bar{\tau}^2)=e_0-e_1$\\
As a result,\\
 $p(\Gamma_{\mathbb{C}}(\tau^1,\bar{\tau}^1))\;p(\Gamma_{\mathbb{C}}(\tau^2,\bar{\tau}^2))-p(\Gamma_{\mathbb{C}}(\tau^1,\bar{\tau}^2)=(p_0+p_1)(p_0-p_1)-(p_2-ip_3)(p_2+ip_3)$\\
 $=(p_0)^2-(p_1)^2-(p_2)^2-(p_3)^2=\|p\|^2$.\\
Taking this into account, and conjugating $f$, we obtain:\\

$\bar{f}(p)\;\;=\;\;\|p\|^2\;\bar{\varphi}(p)$\\

$+\;\;(p(\Gamma_{\mathbb{C}}(\tau^2,\bar{\tau}^1))\;\bar{\psi}_1(p)+p(\Gamma_{\mathbb{C}}(\tau^2,\bar{\tau}^2))\;\bar{\psi}_2(p))\;\tau^1$\\

$+\;\;(-p(\Gamma_{\mathbb{C}}(\tau^1,\bar{\tau}^1))\;\bar{\psi}_1(p)-p(\Gamma_{\mathbb{C}}(\tau^1,\bar{\tau}^2))\;\bar{\psi}_2(p))\;\tau^2$\\

$+\;\;\bar{F}(p)\;\tau^1\wedge\tau^2\;\;+\;\;(-p(\Gamma_{\mathbb{C}}(\tau^2,\bar{\tau}^2))\;\bar{\varphi}(p))\;\tau^1\otimes\bar{\tau}^1\;\;+\;\;(p(\Gamma_{\mathbb{C}}(\tau^1,\bar{\tau}^2))\;\bar{\varphi}(p))\;\tau^2\otimes\bar{\tau}^1$\\

$+\;\;(p(\Gamma_{\mathbb{C}}(\tau^2,\bar{\tau}^1))\;\bar{\varphi}(p))\;\tau^1\otimes\bar{\tau}^2\;\;+\;\;(-p(\Gamma_{\mathbb{C}}(\tau^1,\bar{\tau}^1)\;\bar{\varphi}(p))\;\tau^2\otimes\bar{\tau}^2$\\

$+\;\;\bar{\psi}_1(p)\;(\tau^1\wedge\tau^2)\otimes\bar{\tau}^1\;\;+\;\;\bar{\psi}_2(p)\;(\tau^1\wedge\tau^2)\otimes\bar{\tau}^2\;\;+\;\;\bar{\varphi}(p)\;(\tau^1\wedge\tau^2)\otimes(\bar{\tau}^1\wedge\bar{\tau}^2)$\\

\vspace{4mm}

and\\

$\zeta_{d^2}(p)(f(p))\;=\;\;\|p\|^2\;\varphi(p)\;\tau^1\wedge\tau^2$\\

$+\;\;(p(\Gamma_{\mathbb{C}}(\tau^1,\bar{\tau}^2))\;\psi_1(p)+p(\Gamma_{\mathbb{C}}(\tau^2,\bar{\tau}^2))\;\psi_2(p))\;(\tau^1\wedge\tau^2)\otimes\bar{\tau}^1$\\

$+\;\;(-p(\Gamma_{\mathbb{C}}(\tau^1,\bar{\tau}^1))\;\psi_1(p)-p(\Gamma_{\mathbb{C}}(\tau^2,\bar{\tau}^1))\;\psi_2(p))\;(\tau^1\wedge\tau^2)\otimes\bar{\tau}^2$\\

$+\;\;F(p)\;(\tau^1\wedge\tau^2)\otimes(\bar{\tau}^1\wedge\bar{\tau}^2)\;\;+\;\;\|p\|^2\;F(p)$\\

$+\;\;\|p\|^2\;\psi_1(p)\;\tau^1\;\;+\;\;\|p\|^2\;\psi_2(p)\;\tau^2\;\;+\;\;\|p\|^2\;\varphi(p)\;\tau^1\wedge\tau^2$\\

Finally, imposing $$\zeta_{d^2}(p)(f(p))=m\bar{f}(p)$$ implies\\

$m\bar{\varphi}(p)=F(p)\quad$,$\quad m\bar{F}(p)=\;\|p\|^2\;\varphi(p)$\\

$m\bar{\psi}_1(p)=p(\Gamma_{\mathbb{C}}(\tau^1,\bar{\tau}^2))\;\psi_1(p)+p(\Gamma_{\mathbb{C}}(\tau^2,\bar{\tau}^2))\;\psi_2(p)$\\

$m\bar{\psi}_2(p)=-p(\Gamma_{\mathbb{C}}(\tau^1,\bar{\tau}^1))\;\psi_1(p)-p(\Gamma_{\mathbb{C}}(\tau^2,\bar{\tau}^1))\;\psi_2(p)$\\

In particular, $$(\|p\|^2-m^2)\;\varphi(p)=0$$

\section{Super Fourier transform} \label{super FT}

In \cite{deb}, a Fourier transform is defined in superspace, using a kernel
that transforms under the group $\hbox{Sp}(2n,\mathbb{R})$ rather than the
orthogonal group (here, $n$ is the odd dimension). Inspired from this, we use
the standard supermetric on $M_{cs}$ to define a natural version of the Fourier
transform for Minkowski superspacetime, taking superfunctions in
$\mathcal{C}^{\infty}_c(\mathring{M})[\theta^a,\bar{\theta}^a]$ to
superfunctions in $\mathcal{C}^{\infty}(V^*)[\tau^a,\bar{\tau}^a]$. Once an
expression for the super Fourier transform of a superfunction is obtained, we
see that the purely odd part of the transform coincides with the Hodge
isomorphism defined by the invariant symplectic structure on the spinors. From
this, it is easy to check that the super Fourier transform has natural
properties such as exchanging the odd derivative
$\displaystyle\frac{\partial}{\partial\theta^1}$ with exterior multiplication
by $i\tau^2$, and multiplication by $\theta^1$ with the contraction
$\displaystyle -i\frac{\partial}{\partial\tau^2}$. The $1\leftrightarrow 2$ exchange is not surprising
since the super Fourier transform is defined via a symplectic structure. Then,
we apply the exchange properties to prove that the supersymmetric differential
operators corresponding to the supersymmetric symbols
$\zeta_{\bar{d}_{\bar{\tau}^{a}}}$ and $\zeta_{d_{\tau^a}}$ constructed in the
preceding section are nothing but the supertranslation-invariant odd vector
fields $D_a$ and $\overline{D}_a$ defined in section \ref{superspacetime}. From
this, we obtain the super Poincar\'e equivariant differential equation
selecting
the massive irreducible unitary representation of superspin 0.\\

We define the super Fourier transform of a (compactly supported) superfunction\\ $f\in \mathcal{C}^{\infty}_c(\mathring{M},\bigwedge^{\bullet}S_{\mathbb{C}}^*)\simeq \mathcal{C}^{\infty}_c(\mathring{M})[\theta^a,\bar{\theta}^a]$ as follows: it is the element $\star\widehat{f}\in \mathcal{C}^{\infty}(V^*)[\tau^a,\bar{\tau}^a]$ defined by:
$$\star\widehat{f}:=\int_{M_{cs}} e^{-i(\langle p,x\rangle+\varepsilon_+(\tau,\theta)+\varepsilon_-(\bar{\tau},\bar{\theta}))}\;f\;dx\;d\theta\;d\bar{\theta}$$

If we define the bosonic Fourier transform of $f$ to be given by:
$$\widehat{f}(p):=\int_{\mathring{M}} e^{-i\langle p,x\rangle}\;f(x)\;dx$$
then $$\star\widehat{f}(p)=\int e^{-i(\varepsilon_+(\tau,\theta)+\varepsilon_-(\bar{\tau},\bar{\theta}))}\;\widehat{f}(p)\;d\theta\;d\bar{\theta}$$

Let $\;\;f(x)\;\;=\;\;\varphi(x)\;\;+\;\;\psi_1(x)\;\theta^1\;\;+\;\;\psi_2(x)\;\theta^2\;\;+\;\;\eta_1(x)\;\bar{\theta}^1\;\;+\;\;\eta_2(x)\;\bar{\theta}^2$\\

$+\;\;F(x)\;\theta^1\wedge\theta^2\;\;+\;\;G(x)\;\bar{\theta}^1\wedge\bar{\theta}^2$\\

$+\;\;A_{11}(x)\;\theta^1\otimes\bar{\theta}^1\;\;+\;\;A_{12}(x)\;\theta^1\otimes\bar{\theta}^2\;\;+\;\;A_{21}(x)\;\theta^2\otimes\bar{\theta}^1\;\;+\;\;A_{22}(x)\;\theta^2\otimes\bar{\theta}^2$\\

$+\;\;\lambda_1(x)\;(\theta^{1}\wedge\theta^{2})\otimes\bar{\theta}^1\;\;+\;\;\lambda_2(x)\;(\theta^{1}\wedge\theta^{2})\otimes\bar{\theta}^2$\\

$+\;\;\mu_1(x)\;\theta^1\otimes(\bar{\theta}^1\wedge\bar{\theta}^2)\;\;+\;\;\mu_2(x)\;\theta^2\otimes(\bar{\theta}^1\wedge\bar{\theta}^2)\;\;+\;\;H(x)\;(\theta^1\wedge\theta^2)\otimes(\bar{\theta}^1\wedge\bar{\theta}^2)$\\

be the expression of a generic superfunction $f:\mathring{M}\longrightarrow \bigwedge^{\bullet}S_{\mathbb{C}}^*$.\\

Then $\;\;\widehat{f}(p)\;\;=\;\;\widehat{\varphi}(p)\;\;+\;\;\widehat{\psi}_1(p)\;\theta^1\;\;+\;\;\widehat{\psi}_2(p)\;\theta^2\;\;+\;\;\widehat{\eta}_1(p)\;\bar{\theta}^1\;\;+\;\;\widehat{\eta}_2(p)\;\bar{\theta}^2$\\

$+\;\;\widehat{F}(p)\;\theta^1\wedge\theta^2\;\;+\;\;\widehat{G}(p)\;\bar{\theta}^1\wedge\bar{\theta}^2$\\

$+\;\;\widehat{A}_{11}(p)\;\theta^1\otimes\bar{\theta}^1\;\;+\;\;\widehat{A}_{12}(p)\;\theta^1\otimes\bar{\theta}^2\;\;+\;\;\widehat{A}_{21}(p)\;\theta^2\otimes\bar{\theta}^1\;\;+\;\;\widehat{A}_{22}(p)\;\theta^2\otimes\bar{\theta}^2$\\

$+\;\;\widehat{\lambda}_1(p)\;(\theta^{1}\wedge\theta^{2})\otimes\bar{\theta}^1\;\;+\;\;\widehat{\lambda}_2(p)\;(\theta^{1}\wedge\theta^{2})\otimes\bar{\theta}^2$\\

$+\;\;\widehat{\mu}_1(p)\;\theta^1\otimes(\bar{\theta}^1\wedge\bar{\theta}^2)\;\;+\;\;\widehat{\mu}_2(p)\;\theta^2\otimes(\bar{\theta}^1\wedge\bar{\theta}^2)\;\;+\;\;\widehat{H}(p)\;(\theta^1\wedge\theta^2)\otimes(\bar{\theta}^1\wedge\bar{\theta}^2)$\\

In order to derive an expression for the super Fourier transform $\star\widehat{f}(p)$, we first expand the exponential:

$e^{-i(\varepsilon_+(\tau,\theta)+\varepsilon_-(\bar{\tau},\bar{\theta}))}=e^{-i\varepsilon_+(\tau,\theta)}e^{-i\varepsilon_-(\bar{\tau},\bar{\theta})}=e^{-i(\tau^1\theta^2-\tau^2\theta^1)}e^{-i(\bar{\tau}^1\bar{\theta}^2-\bar{\tau}^2\bar{\theta}^1)}$\\

$=(1-i\tau^1\theta^2+i\tau^2\theta^1+\tau^1\tau^2\theta^1\theta^2)\;(1-i\bar{\tau}^1\bar{\theta}^2+i\bar{\tau}^2\bar{\theta}^1+\bar{\tau}^1\bar{\tau}^2\bar{\theta}^1\bar{\theta}^2)$\\

Then, we multiply the result by the expansion of $\widehat{f}(p)$, keeping only the coefficients of $(\theta^1\wedge\theta^2)\otimes(\bar{\theta}^1\wedge\bar{\theta}^2)$:\\

$e^{-i(\varepsilon_+(\tau,\theta)+\varepsilon_-(\bar{\tau},\bar{\theta}))}\;\widehat{f}(p)\;\;=\;\;\left(\widehat{H}(p)\;\;+\;\;i\widehat{\mu}_1(p)\;\tau^1\;\;+\;\;i\widehat{\mu}_2(p)\;\tau^2\;\;+\;\;i\widehat{\lambda}_1(p)\;\bar{\tau}^1\;\;+\;\;i\widehat{\lambda}_2(p)\;\bar{\tau}^2\right.$\\

$+\;\;\widehat{G}(p)\;\tau^1\wedge\tau^2\;\;+\;\;\widehat{F}(p)\;\bar{\tau}^1\wedge\bar{\tau}^2$\\

$-\;\;\widehat{A}_{11}(p)\;\tau^1\otimes\bar{\tau}^1\;\;-\;\;\widehat{A}_{12}(p)\;\tau^1\otimes\bar{\tau}^2\;\;-\;\;\widehat{A}_{21}(p)\;\tau^2\otimes\bar{\tau}^1\;\;-\;\;\widehat{A}_{22}(p)\;\tau^2\otimes\bar{\tau}^2$\\

$+\;\;i\widehat{\eta}_1(p)\;(\tau^{1}\wedge\tau^{2})\otimes\bar{\tau}^1\;\;+\;\;i\widehat{\eta}_2(p)\;(\tau^{1}\wedge\tau^{2})\otimes\bar{\tau}^2$\\

$\left.+\;\;i\widehat{\psi}_1(p)\;\tau^1\otimes(\bar{\tau}^1\wedge\bar{\tau}^2)\;\;+\;\;i\widehat{\psi}_2(p)\;\tau^2\otimes(\bar{\tau}^1\wedge\bar{\tau}^2)\;\;+\;\;\widehat{\varphi}(p)\;(\tau^1\wedge\tau^2)\otimes(\bar{\tau}^1\wedge\bar{\tau}^2)\right)$\\

$\quad(\theta^1\wedge\theta^2)\otimes(\bar{\theta}^1\wedge\bar{\theta}^2)\;\;+\;\;\hbox{lower order terms in }\theta,\bar{\theta}$\\

Finally, we perform a Berezin integration. Thus, we have the following proposition.\\

\begin{prop} \label{super FT expression}
The super Fourier transform of a superfunction $f:\mathring{M}\longrightarrow \bigwedge^{\bullet}S_{\mathbb{C}}^*$ is given by:\\

$\star\widehat{f}(p)\;\;=\;\;\widehat{H}(p)\;\;+\;\;i\widehat{\mu}_1(p)\;\tau^1\;\;+\;\;i\widehat{\mu}_2(p)\;\tau^2\;\;+\;\;i\widehat{\lambda}_1(p)\;\bar{\tau}^1\;\;+\;\;i\widehat{\lambda}_2(p)\;\bar{\tau}^2$\\

$+\;\;\widehat{G}(p)\;\tau^1\wedge\tau^2\;\;+\;\;\widehat{F}(p)\;\bar{\tau}^1\wedge\bar{\tau}^2$\\

$-\;\;\widehat{A}_{11}(p)\;\tau^1\otimes\bar{\tau}^1\;\;-\;\;\widehat{A}_{12}(p)\;\tau^1\otimes\bar{\tau}^2\;\;-\;\;\widehat{A}_{21}(p)\;\tau^2\otimes\bar{\tau}^1\;\;-\;\;\widehat{A}_{22}(p)\;\tau^2\otimes\bar{\tau}^2$\\

$+\;\;i\widehat{\eta}_1(p)\;(\tau^{1}\wedge\tau^{2})\otimes\bar{\tau}^1\;\;+\;\;i\widehat{\eta}_2(p)\;(\tau^{1}\wedge\tau^{2})\otimes\bar{\tau}^2$\\

$+\;\;i\widehat{\psi}_1(p)\;\tau^1\otimes(\bar{\tau}^1\wedge\bar{\tau}^2)\;\;+\;\;i\widehat{\psi}_2(p)\;\tau^2\otimes(\bar{\tau}^1\wedge\bar{\tau}^2)\;\;+\;\;\widehat{\varphi}(p)\;(\tau^1\wedge\tau^2)\otimes(\bar{\tau}^1\wedge\bar{\tau}^2)$\\
\end{prop}

Comparing this with the expression of $\widehat{f}(p)$, we have the following theorem:\\ 

\begin{thm} \label{odd super Fourier equals Hodge star}
The purely odd super Fourier transform coincides with the Hodge dual (with respect to the symplectic form $\varepsilon$ on $S_{\mathbb{C}}$).\\
\end{thm}

\begin{prop}\
\begin{enumerate}
\item $\displaystyle \star(\widehat{\frac{\partial f}{\partial\theta^a}})=i\varepsilon_{ab}\tau^b(\star\widehat{f})\quad$,$\quad\displaystyle\star(\widehat{\frac{\partial f}{\partial\bar{\theta}^{\dot{a}}}})=i\varepsilon_{\dot{a}\dot{b}}\bar{\tau}^{\dot{b}}(\star\widehat{f})$
\item $\displaystyle \star(\widehat{\theta^af})=-i\varepsilon^{ab}\frac{\partial}{\partial\tau^b}(\star\widehat{f})\quad$,$\quad\displaystyle\star(\widehat{\bar{\theta}^{\dot{a}}f})=-i\varepsilon^{\dot{a}\dot{b}}\frac{\partial}{\partial\bar{\tau}^{\dot{b}}}(\star\widehat{f})$
\end{enumerate}
\end{prop}

\noindent {\bf Proof~:}
Follow directly from \ref{super FT expression}. $\hfill\square$\\

Recall that for every superfunction $f:\mathring{M}\longrightarrow W$,\\ we have $\;\;\displaystyle D_{a}f=\frac{\partial f}{\partial\theta^{a}}-i\Gamma_{a\dot{b}}^{\mu}\bar{\theta}^{\dot{b}}\frac{\partial f}{\partial x^{\mu}}\;\;$ and $\;\;\displaystyle\overline{D}_{\dot{a}}f=\frac{\partial f}{\partial\bar{\theta}^{\dot{a}}}-i\Gamma_{b\dot{a}}^{\mu}\theta^b\frac{\partial f}{\partial x^{\mu}}$.\\

\begin{prop}\
$\displaystyle\star\widehat{D_{a}f}(p)=i\varepsilon_{ab}\zeta_{d_{\tau^{b}}}(p)(\star\widehat{f}(p))\quad$,$\quad\displaystyle\star\widehat{\overline{D}_{\dot{a}}f}(p)=i\varepsilon_{\dot{a}\dot{b}}\zeta_{\bar{d}_{\bar{\tau}^{\dot{b}}}}(p)(\star\widehat{f}(p))$
\end{prop}

\noindent {\bf Proof~:}
We prove only the second equality, the proof being similar for the first.\\
$\displaystyle\star\widehat{\overline{D}_{\dot{a}}f}=\star(\widehat{\frac{\partial f}{\partial\bar{\theta}^{\dot{a}}}})-i\Gamma_{b\dot{a}}^{\mu}\;\star({\widehat{\theta^b\frac{\partial f}{\partial x^{\mu}}}})$\\

$\displaystyle=i\varepsilon_{\dot{a}\dot{b}}\bar{\tau}^{\dot{b}}(\star\widehat{f})-\Gamma_{b\dot{a}}^{\mu}\varepsilon^{bc}\frac{\partial}{\partial \tau^{c}}(\star({\widehat{\frac{\partial f}{\partial x^{\mu}}}}))\displaystyle=i\varepsilon_{\dot{a}\dot{b}}\bar{\tau}^{\dot{b}}(\star\widehat{f})-i\Gamma_{b\dot{a}}^{\mu}p_{\mu}\varepsilon^{bc}\frac{\partial}{\partial \tau^{c}}(\star\widehat{f})$\\

$\displaystyle=i\varepsilon_{\dot{a}\dot{b}}\left(\bar{\tau}^{\dot{b}}(\star\widehat{f})+\Gamma_{b\dot{a}}^{\mu}p_{\mu}\varepsilon^{bc}\varepsilon^{\dot{a}\dot{b}}\frac{\partial}{\partial \tau^{c}}(\star\widehat{f})\right)\displaystyle=i\varepsilon_{\dot{a}\dot{b}}\left(\bar{\tau}^{\dot{b}}(\star\widehat{f})+\Gamma^{\mu c\dot{b}}p_{\mu}\frac{\partial}{\partial \tau^{c}}(\star\widehat{f})\right)$\\

$\displaystyle=i\varepsilon_{\dot{a}\dot{b}}\left(\bar{\tau}^{\dot{b}}(\star\widehat{f})+p(\Gamma_{\mathbb{C}}(\tau^c,\bar{\tau}^{\dot{b}}))\frac{\partial}{\partial \tau^{c}}(\star\widehat{f})\right)\displaystyle=i\varepsilon_{\dot{a}\dot{b}}\left((\hbox{\upshape{Id}}\otimes e_{\bar{\tau}^{\dot{b}}})+(\zeta_{i_{\bar{\tau}^{\dot{b}}}}(p)\otimes\hbox{\upshape{Id}})\right)(\star\widehat{f})$\\

$\displaystyle=i\varepsilon_{\dot{a}\dot{b}}\zeta_{\bar{d}_{\bar{\tau}^{\dot{b}}}}(p)(\star\widehat{f})$ $\hfill\square$\\

Recall that for every superfunction $f:\mathring{M}\longrightarrow W$,\\ we have
$\;\;\displaystyle D^2f=\varepsilon^{a_1a_2}D_{a_1}D_{a_2}f\;\;$ and $\;\;\displaystyle \overline{D}^2f=\varepsilon^{a_1a_2}\overline{D}_{a_1}\overline{D}_{a_2}f$.\\

\begin{prop}\
$\displaystyle\star\widehat{D^2f}(p)=-\zeta_{d^2}(p)(\star\widehat{f}(p))\quad$,$\quad\displaystyle\star\widehat{\overline{D}^2f}(p)=-\zeta_{\bar{d}^2}(p)(\star\widehat{f}(p))$
\end{prop}

\noindent {\bf Proof~:}
We prove only the first equality, the proof being similar for the second.\\
$\displaystyle\star\widehat{D^2f}=\varepsilon^{a_1a_2}\;\star(\widehat{D_{a_1}D_{a_2}f})$\\

$\displaystyle=\varepsilon^{a_1a_2}\;i\varepsilon_{a_1b_1}\zeta_{d_{\tau^{b_1}}}(p)(\star\widehat{D_{a_2}f})=\varepsilon^{a_1a_2}\;i\varepsilon_{a_1b_1}\zeta_{d_{\tau^{b_1}}}(p)\;i\varepsilon_{a_2 b_2}\zeta_{d_{\tau^{b_2}}}(p)(\star\widehat{f})$\\

$=-\varepsilon^{a_1a_2}\varepsilon_{a_1b_1}\varepsilon_{a_2 b_2}\;\zeta_{d_{\tau^{b_1}}}(p)\;\zeta_{d_{\tau^{b_2}}}(p)(\star\widehat{f})=-\varepsilon_{b_1b_2}\;\zeta_{d_{\tau^{b_1}}}(p)\;\zeta_{d_{\tau^{b_2}}}(p)(\star\widehat{f})$\\

$=-\zeta_{d^2}(p)(\star\widehat{f})$ $\hfill\square$\\

We deduce immediately from the above the following proposition.\\

\begin{prop}
The equations\\

 $\zeta_{\bar{d}_{\bar{\tau}^{\dot{1}}}}(p)(\star\widehat{f}(p))=0\;\;$, $\;\;\zeta_{\bar{d}_{\bar{\tau}^{\dot{2}}}}(p)(\star\widehat{f}(p))=0\;\;$ and $\;\;\zeta_{d^2}(p)(\star\widehat{f}(p))=m\;\overline{\star\widehat{f}}(p)$\\

  are equivalent to:\\

   $\overline{D}_{\dot{1}}f=0\;\;$, $\;\;\overline{D}_{\dot{2}}f=0\;\;$ and $\;\;D^2f=-m\bar{f}$.\\
\end{prop}

The following theorem summarizes the main result that we have obtained: a realization of the irreducible unitary representation of the super-Poincar\'e group of mass $m$ and superspin 0 in terms of partial differential equations involving superfunctions in the Berezin-Kostant-Leites sense (resp. \emph{ordinary} ({\it i.e.} non-Grassmannian) complex-valued functions on spacetime).\\

\begin{thm} \label{super equations}
Let $M_{cs}$ be the linear cs supermanifold associated to the super vector space $V_{\mathbb{C}}\oplus S_{\mathbb{C}}$ (where $V$ is a four-dimensional Lorentzian vector space, and $S_{\mathbb{C}}$ the corresponding four-dimensional complex space of Dirac spinors). The irreducible unitary representation of the super-Poincar\'e group of mass $m$ and superspin 0 can be realized as the sub-super vector space of $\mathcal{O}_{M_{cs}}(\mathring{M})=\mathcal{C}^{\infty}(\mathring{M},\bigwedge^{\bullet}S_{\mathbb{C}}^*)$ made of the superfunctions satisfying the differential equations:\\
 
   $\overline{D}_{\dot{1}}f=0\;\;$, $\;\;\overline{D}_{\dot{2}}f=0\;\;$ and $\;\;D^2f=-m\bar{f}$.\\
   
In components, this representation space corresponds to:

 $\{(\varphi,\psi)\in \mathcal{C}^{\infty}(\mathring{M},\mathbb{C})\times \mathcal{C}^{\infty}(\mathring{M},S_+^*)\;|\;\left\{ \begin{array}{rcl}
(\square+m^2)\varphi &=&0 \\
i\Gamma_{ab}^{\mu}\partial_{\mu}\bar{\psi}^b+m\psi_a &=& 0 \\
\end{array} \right.\}\,.$
\end{thm}

\section{Link with the superfield-theoretic approach} \label{superfunctions vs superfields}

We have obtained at the end of the preceding section supersymmetric differential equations
corresponding to the massive irreducible unitary representations of superspin 0 of the
super Poincar\'e group. These equations involve superfunctions, whose components are
ordinary complex-valued functions on spacetime (we are working in the Berezin-Kostant-Leites
category of supermanifolds). In particular, these equations reduce to a Klein-Gordon equation,
and a Dirac equation which involves ordinary spinor fields with complex-valued components.
This is in contrast with the physics literature, where the spinor fields occurring in
supersymmetric theories have always anticommuting Grassmann-valued components.
In fact, one can proceed differently in order to realize the representations: one can consider
{\it a priori} a suitable action functional for superfields on Minkowski superspacetime, and
then obtain differential equations selecting the representation as the Euler-Lagrange equations
corresponding to that action functional. This Lagrangian field-theoretic approach involves
the differential geometry of the underlying supermanifold (here Minkowski superspacetime),
in order to carry out the calculus of variations, and is most conveniently dealt with by
applying the functor of points. This is what is implicitly done in the physics literature,
and it leads naturally to odd Grassmannian spinor fields. One can obtain in this way the
Wess-Zumino equations for massive chiral superfields ({\it cf.} \cite{wb} for instance).
In this section, we view the solutions of these equations as a functor; it turns out that
this functor is representable, precisely by the solutions of our supersymmetric equations
(that we have obtained otherwise from momentum space via super Fourier transform). \\

The generalized supermanifold of superfields is $\mathcal{F}:=\underline{\hbox{Hom}}(M_{cs},\mathbb{C})$.
It is by definition the contravariant functor from $\mathbf{sMan}_{fd}^{cs}$ to $\mathbf{Set}$ given by
$$\mathcal{F}(B)=\hbox{Hom}(B\times M_{cs},\mathbb{C})$$ for all complex supermanifolds $B$.
One would like to think of $\mathcal{F}$ as some kind of infinite-dimensional supermanifold $(\mathcal{|F|},\mathcal{O}_{\mathcal{F}})$. \\

If $\Phi_{geom}=(\check{\Phi},\Phi^{\sharp}):B\times M_{cs}\longrightarrow \mathbb{C}$ is a superfield, then $\Phi^{\sharp}:\mathcal{C}^{\infty}(\mathbb{C},\mathbb{C})\longrightarrow \mathcal{O}_B(|B|)\;\hat{\otimes}\;\mathcal{O}_{M_{cs}}(\mathring{M})$ is entirely determined by $\Phi^{\sharp}(\hbox{\upshape{Id}}_{\mathbb{C}})\in (\mathcal{O}_B(|B|)\;\hat{\otimes}\;\mathcal{O}_{M_{cs}}(\mathring{M}))_0$. Thus, the information about $\Phi_{geom}$ is fully contained in $\Phi^{\sharp}(\hbox{\upshape{Id}}_{\mathbb{C}})\in (\bigwedge^{\bullet}S_{\mathbb{C}}^*\otimes\mathcal{C}^{\infty}(\mathring{M},\mathbb{C})\;\hat{\otimes}\;\mathcal{O}_B(|B|))_0$.\\

Using $x^{\mu},\theta^a,\bar{\theta}^b$ as coordinates on $M_{cs}$, we may write, for any $f\in\mathcal{C}^{\infty}(\mathbb{C},\mathbb{C})$,
$$\Phi^{\sharp}(f)=\varphi^{\sharp}(f)+\theta^a(\psi_a)^{\sharp}(f)+\bar{\theta}^{a}(\eta_a)^{\sharp}(f)+\theta^1\theta^2 F^{\sharp}(f)+\bar{\theta}^{1}\bar{\theta}^{2}G^{\sharp}(f)+i\Gamma^{\mu}_{ab}\theta^a\bar{\theta}^{b}(A_{\mu})^{\sharp}(f)$$
$$+\theta^1\theta^2\bar{\theta}^{a}(\lambda_a)^{\sharp}(f)+\bar{\theta}^{1}\bar{\theta}^{2}\theta^a(\mu_a)^{\sharp}(f)+\theta^1\theta^2\bar{\theta}^{1}\bar{\theta}^{2}H^{\sharp}(f)$$

where $\varphi^{\sharp}(f)\in \mathcal{C}^{\infty}(\mathring{M},\mathbb{C})\;\hat{\otimes}\;\mathcal{O}_B(|B|)_0$, $\;\; (\psi_a)^{\sharp}(f)\in \mathcal{C}^{\infty}(\mathring{M},\mathbb{C})\;\hat{\otimes}\;\mathcal{O}_B(|B|)_1$ $\;\;\hbox{(while }\theta^a(\psi_a)^{\sharp}(f)\in S_+^*\otimes\mathcal{C}^{\infty}(\mathring{M},\mathbb{C})\;\hat{\otimes}\;\mathcal{O}_B(|B|)_1$), etc...\\

If $g\in\mathcal{C}^{\infty}(\mathbb{C},\mathbb{C})$ is another function, writing $\Phi^{\sharp}(fg)=\Phi^{\sharp}(f)\;\Phi^{\sharp}(g)$ will give\\ $\varphi^{\sharp}(fg)=\varphi^{\sharp}(f)\;\varphi^{\sharp}(g)$, so $\varphi^{\sharp}$ corresponds indeed to a morphism $\varphi_{geom}:B\times\mathring{M}\longrightarrow \mathbb{C}$. \\

But at order 1, $\Phi^{\sharp}(fg)=\Phi^{\sharp}(f)\;\Phi^{\sharp}(g)$ will give $(\psi_a)^{\sharp}(fg)=(\psi_a)^{\sharp}(f)\;\varphi^{\sharp}(g)+\varphi^{\sharp}(f)\;(\psi_a)^{\sharp}(g)$ and same for $(\eta_a)^{\sharp}$. So $(\psi_a)^{\sharp}$ and $(\eta_a)^{\sharp}$ are derivations, and not pull-backs of morphisms. Consequently, $\psi_{geom}:B\times\mathring{M}\longrightarrow \hbox{L}(\{0\}\oplus S_+^*)$ and $\eta_{geom}:B\times\mathring{M}\longrightarrow \hbox{L}(\{0\}\oplus S_-^*)$ should be considered as odd vector fields along $\varphi_{geom}$.\\

To the superfield $\Phi_{geom}$, we can associate a map $\Phi:M_{cs}(B)\longrightarrow\mathcal{O}_B(|B|)_0$ in the following way: let $\beta\in M_{cs}(B)=\hbox{Hom}(B,M_{cs})$. Then $\beta^{\sharp}_{\mathring{M}}:\bigwedge^{\bullet}S_{\mathbb{C}}^*\otimes\mathcal{C}^{\infty}(\mathring{M},\mathbb{C})\longrightarrow\mathcal{O}_B(|B|)$ can be extended into $\beta^{\sharp}_{\mathring{M}}\;\hat{\otimes}\;\hbox{\upshape{Id}}_{\mathcal{O}_B(|B|)}:\bigwedge^{\bullet}S_{\mathbb{C}}^*\otimes\mathcal{C}^{\infty}(\mathring{M},\mathbb{C})\;\hat{\otimes}\;\mathcal{O}_B(|B|)\longrightarrow\mathcal{O}_B(|B|)\;\hat{\otimes}\;\mathcal{O}_B(|B|)$. On the other hand, we have a canonical map $\Delta^{\sharp}_{|B|\times |B|}:\mathcal{O}_B(|B|)\;\hat{\otimes}\;\mathcal{O}_B(|B|)\longrightarrow \mathcal{O}_B(|B|)$. Composing with $\Phi_{\mathbb{C}}^{\sharp}:\mathcal{C}^{\infty}(\mathbb{C},\mathbb{C})\longrightarrow (\bigwedge^{\bullet}S_{\mathbb{C}}^*\otimes\mathcal{C}^{\infty}(\mathring{M},\mathbb{C})\;\hat{\otimes}\;\mathcal{O}_B(|B|))_0$, we obtain a map $\Delta^{\sharp}_{|B|\times |B|}\circ (\beta^{\sharp}_{\mathring{M}}\;\hat{\otimes}\;\hbox{\upshape{Id}}_{\mathcal{O}_B(|B|)})\circ \Phi_{\mathbb{C}}^{\sharp}:\mathcal{C}^{\infty}(\mathbb{C},\mathbb{C})\longrightarrow \mathcal{O}_B(|B|)_0$.\\
Now set $\Phi(\beta):=(\Delta^{\sharp}_{|B|\times |B|}\circ (\beta^{\sharp}_{\mathring{M}}\;\hat{\otimes}\;\hbox{\upshape{Id}}_{\mathcal{O}_B(|B|)})\circ \Phi_{\mathbb{C}}^{\sharp})(\hbox{\upshape{Id}}_{\mathbb{C}})$. Then $\Phi(\beta)\in \mathcal{O}_B(|B|)_0$. \\

Here is alternative way to define $\Phi(\beta)$, equivalent but slightly simpler. We have the diagonal morphism $\Delta:B\longrightarrow B\times B$, as well as the morphism $\hbox{\upshape{Id}}_B\times \beta:B\times B\longrightarrow B\times M_{cs}$. Composing with $\Phi_{geom}:B\times M_{cs}\longrightarrow \mathbb{C}$, we obtain an element $\Phi_{geom}\circ (\hbox{\upshape{Id}}_B\times \beta)\circ \Delta\in \hbox{Hom}(B,\mathbb{C})$. Now set $\Phi(\beta)=(\Phi_{geom}\circ (\hbox{\upshape{Id}}_B\times \beta)\circ \Delta)^{\sharp}(\hbox{\upshape{Id}}_{\mathbb{C}})$.\\

Let $y^{\mu}:=\beta^{\sharp}_{\mathring{M}}(x^{\mu})\in \mathcal{O}_B(|B|)_0$, $\xi^a:=\beta^{\sharp}_{\mathring{M}}(\theta^a)\in \mathcal{O}_B(|B|)_1$, and $\bar{\xi}^b:=\beta^{\sharp}_{\mathring{M}}(\bar{\theta}^b)\in \mathcal{O}_B(|B|)_1$. We may write $\beta=(y^{\mu},\xi^a,\bar{\xi}^b)$ or $\beta=(y,\xi,\bar{\xi})$.\\

Recall that $$\Phi^{\sharp}(\hbox{\upshape{Id}}_{\mathbb{C}})=\varphi^{\sharp}(\hbox{\upshape{Id}}_{\mathbb{C}})+\theta^a(\psi_a)^{\sharp}(\hbox{\upshape{Id}}_{\mathbb{C}})+\bar{\theta}^{a}(\eta_a)^{\sharp}(\hbox{\upshape{Id}}_{\mathbb{C}})+\theta^1\theta^2 F^{\sharp}(\hbox{\upshape{Id}}_{\mathbb{C}})+\bar{\theta}^{1}\bar{\theta}^{2}G^{\sharp}(\hbox{\upshape{Id}}_{\mathbb{C}})+i\Gamma^{\mu}_{ab}\theta^a\bar{\theta}^{b}(A_{\mu})^{\sharp}(\hbox{\upshape{Id}}_{\mathbb{C}})$$
$$+\theta^1\theta^2\bar{\theta}^{a}(\lambda_a)^{\sharp}(\hbox{\upshape{Id}}_{\mathbb{C}})+\bar{\theta}^{1}\bar{\theta}^{2}\theta^a(\mu_a)^{\sharp}(\hbox{\upshape{Id}}_{\mathbb{C}})+\theta^1\theta^2\bar{\theta}^{1}\bar{\theta}^{2}H^{\sharp}(\hbox{\upshape{Id}}_{\mathbb{C}})$$

Applying $\Delta^{\sharp}_{|B|\times |B|}\circ (\beta^{\sharp}_{\mathring{M}}\;\hat{\otimes}\;\hbox{\upshape{Id}}_{\mathcal{O}_B(|B|)})$ to both sides of this equality, we obtain the following expression for $\Phi:M_{cs}(B)\longrightarrow\mathcal{O}_B(|B|)_0$:
$$\Phi(y,\xi,\bar{\xi})=\varphi(y)+\xi^a\psi_a(y)+ \bar{\xi}^{a}\eta_a(y)+\xi^1\xi^2 F(y)+\bar{\xi}^{1}\bar{\xi}^{2}G(y)+i\Gamma^{\mu}_{ab}\xi^a\bar{\xi}^{b}A_{\mu}(y)$$
$$+\xi^1\xi^2\bar{\xi}^{a}\lambda_a(y)+\bar{\xi}^{1}\bar{\xi}^{2}\xi^a\mu_a(y)+\xi^1\xi^2\bar{\xi}^{1}\bar{\xi}^{2}H(y)$$
where $\varphi(y)\in\mathcal{O}_B(|B|)_0$, $\xi^a\in\mathcal{O}_B(|B|)_1$ and $\psi_a(y)\in \mathcal{O}_B(|B|)_1$ (so $\xi^a\psi_a(y)\in\mathcal{O}_B(|B|)_0$), etc...\\

\begin{defi}
Let $\Phi:M_{cs}\longrightarrow\mathcal{O}_B(|B|)_0$ be a scalar superfield. We say that $\Phi$ is {\bf chiral} (resp. {\bf antichiral}) if $\overline{D}_1\Phi=\overline{D}_2\Phi=0$ (resp. $D_1\Phi=D_2\Phi=0$).\\
\end{defi}

For any chiral superfield, let
$$\mathcal{A}(\Phi)=\int_{M_{cs}(B)}\bar{\Phi}\Phi\;d^2\xi\;d^2\bar{\xi}\;d^4 y+\int_{M_{cs}^+(B)}\frac{1}{2}m\Phi^2\;d^2\xi\;d^4 y$$

\begin{prop} The superfield equation corresponding to the above action functional is $$-\overline{D}^2\bar{\Phi}+m\Phi=0$$
\end{prop}

Let $\mathcal{E}(B):=\{\Phi:M_{cs}(B)\longrightarrow \mathcal{O}_B(|B|)_0\;|\;-\overline{D}^2\bar{\Phi}+m\Phi=0\}$\\
$\simeq \{(\varphi,\psi)\in \hbox{\upshape{Map}}(\mathring{M}(B),\mathcal{O}_B(|B|)_0)\times \hbox{\upshape{Map}}(\mathring{M}(B),(\mathcal{O}_B(|B|)_1)^2)\;|\; \left\{ \begin{array}{rcl}
(\square+m^2)\varphi &=&0 \\
i\Gamma_{ab}^{\mu}\partial_{\mu}\bar{\psi}^b+m\psi_a &=& 0 \\
\end{array} \right.\}$\\

\begin{prop}\
$\mathcal{E}(B)$ is $\hbox{\upshape{S}}\Pi(V)(B)$-invariant.\\
\end{prop}

$\mathcal{E}$ is the solution functor of the superfield equation. It is a generalized supermanifold on which $\hbox{S}\Pi(V)$ acts. We will see that $\mathcal{E}$ is representable by a super-vector space $E$ which gives rise to the super-Hilbert space of $1$-superparticle states of mass $m$ and superspin 0.\\

The action of the super Lie group $\hbox{S}\Pi(V)$ on the generalized supermanifold $\underline{\hbox{Hom}}(M_{cs},\mathbb{C})$ is obtained from the representation of $\hbox{S}\Pi(V)$ on the super-vector space $\mathcal{O}_{M_{cs}}(\mathring{M})$. Viewing $\hbox{S}\Pi(V)$ as a Harish-Chandra pair $(\Pi(V),\mathfrak{s}\pi(V))$, this representation is equivalent to a pair $(\rho,\eta)$ where $\rho:\Pi(V)\longrightarrow\hbox{Aut}(\mathcal{O}_{M_{cs}}(\mathring{M}))$ is a morphism of Lie groups, and $\eta:\mathfrak{s}\pi(V)\longrightarrow\mathfrak{gl}(\mathcal{O}_{M_{cs}}(\mathring{M}))$ is a morphism of super Lie algebras, and we have seen that $\eta(e_{\mu})=P_{\mu}$, $\eta(f_a)=Q_a$ and $\eta(\bar{f}_b)=\overline{Q}_b$.\\

Recall that $\;\mathcal{O}_{M_{cs}}(\mathring{M})\;=\;\mathcal{C}^{\infty}(\mathring{M},\mathbb{C})\otimes\bigwedge^{\bullet}S_{\mathbb{C}}^*\;\simeq\;\mathcal{C}^{\infty}(\mathring{M},\mathbb{C})\otimes\bigwedge^{\bullet}S_{+}^*\otimes\bigwedge^{\bullet}S_{-}^*$.\\

There are two interesting sub-super vector spaces in $\mathcal{O}_{M_{cs}}(\mathring{M})$, the subspace of chiral (resp. antichiral) superfunctions (and it is not difficult to see that each of them is $\hbox{\upshape{S}}\Pi(V)$-invariant):\\ 

 $\mathcal{O}_{M_{cs}}(\mathring{M})_{chiral}:=\{f\in \mathcal{O}_{M_{cs}}(\mathring{M})\;|\;\overline{D}_1f=\overline{D}_2f=0\}\;\simeq \mathcal{C}^{\infty}(\mathring{M},\mathbb{C})\otimes\bigwedge^{\bullet}S_{+}^*$\\

 $\mathcal{O}_{M_{cs}}(\mathring{M})_{antichiral}:=\{f\in \mathcal{O}_{M_{cs}}(\mathring{M})\;|\;D_1f=D_2f=0\}\;\simeq \mathcal{C}^{\infty}(\mathring{M},\mathbb{C})\otimes\bigwedge^{\bullet}S_{-}^*$\\

Note that any $f\in \mathcal{O}_{M_{cs}}(\mathring{M})_{chiral}$ can be written as follows:
$$f=\varphi+\theta^a\psi_a+\theta^1\theta^2 F-\Gamma_{ab}^{\mu}\theta^a\bar{\theta}^b\partial_{\mu} \varphi-\Gamma_{ab}^{\mu}\theta^c\theta^a\bar{\theta}^b\partial_{\mu}\psi_c+\theta^1\theta^2\bar{\theta}^1\bar{\theta}^2\square\varphi$$
where $\varphi\in\mathcal{C}^{\infty}(\mathring{M},\mathbb{C})$, $\psi\in\mathcal{C}^{\infty}(\mathring{M},S_{+}^*)$ and $F\in \mathcal{C}^{\infty}(\mathring{M},\mathbb{C})$. \\

Then $\bar{f}\in \mathcal{O}_{M_{cs}}(\mathring{M})_{antichiral}$ and is given by:
$$\bar{f}=\bar{\varphi}+\bar{\theta}^a\bar{\psi}_a+\bar{\theta}^1\bar{\theta}^2 \bar{F}+\Gamma_{ab}^{\mu}\theta^a\bar{\theta}^b\partial_{\mu} \bar{\varphi}-\Gamma_{ba}^{\mu}\bar{\theta}^c\bar{\theta}^a\theta^b\partial_{\mu}\bar{\psi}_c+\theta^1\theta^2\bar{\theta}^1\bar{\theta}^2\square\bar{\varphi}$$\

\begin{prop}
Let $E:=\{f\in \mathcal{O}_{M_{cs}}(\mathring{M})_{chiral}\;|\;-\overline{D}^2 \bar{f}+mf=0\}$. Then 
\begin{enumerate}
\item
$E\simeq \{(\varphi,\psi)\in \mathcal{C}^{\infty}(\mathring{M},\mathbb{C})\times \mathcal{C}^{\infty}(\mathring{M},S_+^*)\;|\;\left\{ \begin{array}{rcl}
(\square+m^2)\varphi &=&0 \\
i\Gamma_{ab}^{\mu}\partial_{\mu}\bar{\psi}^b+m\psi_a &=& 0 \\
\end{array} \right.\}$
\item $E$ is $\hbox{\upshape{S}}\Pi(V)$-invariant.\\
\end{enumerate}
\end{prop}

The link between our equations and the Wess-Zumino equations for massive chiral superfields is made precise by the following result:\\

\begin{thm} \label{representability}
Let $\mathcal{E}$ be the solution functor of the Wess-Zumino equations for massive chiral superfields (that is, the functor from the category of supermanifolds to the category of sets defined by $\mathcal{E}(B):=\{\Phi:M_{cs}(B)\longrightarrow \mathcal{O}_B(|B|)_0\;|\;-\overline{D}^2\bar{\Phi}+m\Phi=0\}$). Then $\mathcal{E}$ is representable by the super vector space $E=\{f\in \mathcal{O}_{M_{cs}}(\mathring{M})_{chiral}\;|\;-\overline{D}^2 \bar{f}+mf=0\}$. In other words, there is a natural isomorphism of functors:
$$\mathcal{E}\simeq\mathcal{L}E$$
where $\mathcal{L}E(B):=(E_0\hat{\otimes}\mathcal{O}_B(|B|)_0)\oplus (E_1\hat{\otimes}\mathcal{O}_B(|B|)_1)$ for every supermanifold $B$. Moreover, this isomorphism is $\hbox{\upshape{S}}\Pi(V)$-equivariant.\\
\end{thm}

\end{document}